\documentclass[usenatbib,fleqn]{mnras}

\pdfoutput=1

\usepackage{amsmath}
\usepackage{amssymb}
\usepackage{blindtext}
\usepackage{bm}
\usepackage{color}
\usepackage{graphicx}
\usepackage{hyperref}
\usepackage[all]{hypcap}
\usepackage{longtable}
\usepackage{multicol}
\usepackage{multirow}
\usepackage{natbib}
\usepackage{txfonts}
\usepackage{wasysym}
\usepackage[capitalize,nameinlink]{cleveref}

\creflabelformat{equation}{#2#1#3}
\crefrangelabelformat{equation}{#3#1#4 to #5#2#6}

\hypersetup{colorlinks,
  linkcolor=red,
  filecolor=red,
  urlcolor=magenta,
  citecolor=blue}

\usepackage{array}
\newcolumntype{L}[1]{>{\raggedright\let\newline\\\arraybackslash\hspace{0pt}}m{#1}}
\newcolumntype{C}[1]{>{\centering\let\newline\\\arraybackslash\hspace{0pt}}m{#1}}
\newcolumntype{R}[1]{>{\raggedleft\let\newline\\\arraybackslash\hspace{0pt}}m{#1}}

\def\avgrhstar{\langle r_\mathrm{h}^\star\rangle}
\def\ct{c_\mathrm{t}}
\def\ESD{\Delta\Sigma}
\def\ezgal{\textsc{EzGal}}

\def\galfit{\textsc{galfit}}
\def\gammat{\gamma_\mathrm{t}}
\def\LCDM{$\Lambda$CDM}

\def\megacam{MegaCam}
\def\meneacs{MENeaCS}
\def\Mh{M_\mathrm{h}}
\def\Msub{m_\mathrm{bg}}
\def\Msun{\mathrm{M}_\odot}
\def\Mstar{m_\star}
\def\Nbody{N-body}

\def\percent{ per cent}
\def\rmag{m_\mathrm{phot}}
\def\Rsat{R_\mathrm{sat}}
\def\rs{r_\mathrm{s}}
\def\size{s_\mathrm{eff}}
\def\slope{0.95\pm0.10}
\def\sqdeg{sq.\ deg.}
\def\thetals{\theta_\mathrm{ls}}

\title[Satellite galaxy-galaxy lensing in low-$z$ clusters]
{The galaxy-subhalo connection in low-redshift galaxy clusters from weak gravitational lensing}

\author[C.\ Sif\'on et al.]{Crist\'obal~Sif\'on$^{1,2}$,
                            Ricardo~Herbonnet$^2$,
                            Henk~Hoekstra$^2$,
                            Remco~F.~J.~van~der~Burg$^3$
                            \newauthor
                            and
                            Massimo~Viola$^2$
\\
$^1$Department of Astrophysical Sciences, Peyton Hall, Princeton University, Princeton, NJ 08544, USA\\
$^2$Leiden Observatory, Leiden University, PO Box 9513, NL-2300 RA Leiden, Netherlands\\
$^3$Laboratoire AIM, IRFU/Service d'Astrophysique - CEA/DSM - CNRS - Universit\'e Paris Diderot, B\^{a}t. 709, CEA-Saclay, 91191 Gif-sur-Yvette Cedex, France
}

\pubyear{2018}

\begin{document}
\label{firstpage}
\pagerange{\pageref{firstpage}--\pageref{lastpage}}

\maketitle

\begin{abstract}
We measure the gravitational lensing signal around satellite galaxies in a sample of galaxy clusters at $z<0.15$ by combining high-quality imaging data from the Canada-France-Hawaii Telescope with a large sample of spectroscopically-confirmed cluster members. We use extensive image simulations to assess the accuracy of shape measurements of faint, background sources in the vicinity of bright satellite galaxies. We find a small but significant bias, as light from the lenses makes the shapes of background galaxies appear radially aligned with the lens. We account for this bias by applying a correction that depends on both lens size and magnitude. We also correct for contamination of the source sample by cluster members. We use a physically-motivated definition of subhalo mass, namely the mass bound to the subhalo, $m_\mathrm{bg}$, similar to definitions used by common subhalo finders in numerical simulations. Binning the satellites by stellar mass we provide a direct measurement of the subhalo-to-stellar-mass relation, $\log\Msub/\Msun = (11.54\pm0.05) + (\slope)\log[\Mstar/(2\times10^{10}\Msun)]$. This best-fitting relation implies that, at a stellar mass $m_\star\sim3\times10^{10}\,\Msun$, subhalo masses are roughly 50\percent\ of those of central galaxies, and this fraction decreases at higher stellar masses. We find some evidence for a sharp change in the total-to-stellar mass ratio around the clusters' scale radius, which could be interpreted as galaxies within the scale radius having suffered more strongly from tidal stripping, but remain cautious regarding this interpretation.
\end{abstract}



\section{Introduction}\label{s:intro}

According to the hierarchical structure formation paradigm, galaxy clusters grow by the continuous accretion of smaller galaxy groups and individual galaxies. Initially, each of these systems is hosted by their own dark matter halo, but as a galaxy falls into a larger structure, tidal interactions transfer mass from the infalling galaxy to the new host. The galaxy then becomes a satellite and its dark matter halo, a subhalo.
 
Detailed studies on the statistics of subhaloes from numerical \Nbody\ simulations have revealed that subhaloes are severely affected by their host haloes. Dynamical friction makes more massive subhaloes sink towards the centre faster, while tidal stripping removes mass preferentially from the outskirts of massive subhaloes closer to the centre. These two effects combined destroy the most massive subhaloes soon after infall \citep[e.g.,][]{tormen98,taffoni03}, a result exaggerated in simulations with limited resolution \citep[e.g.,][]{klypin99,taylor05,han16}. Tidal stripping makes subhaloes more concentrated than field haloes of the same mass \citep[e.g.,][]{ghigna98,springel08,moline17}, and counterbalances the spatial segregation induced by dynamical friction \citep{vdbosch16}.

One of the most fundamental questions is how these subhaloes are linked to the satellite galaxies they host, which are what we observe in the real Universe. Taking \Nbody\ simulations at face value results in serious inconsistencies with observations, the most famous of which are known as the ``missing satellites'' \citep{klypin99,moore99} and ``too big to fail'' \citep{boylan11} problems. It has since become clear that these problems may arise because baryonic physics has a strong influence on the small-scale distribution of matter. Energetic feedback from supernovae at the low-mass end, and active galactic nuclei at the high-mass end, of the galaxy population affect the ability of dark matter (sub)haloes to form stars and retain them. In addition, the excess mass in the centre of galaxies (compared to dark matter-only simulations) can modify each subhalo's susceptibility to tidal stripping \citep[e.g.,][]{zolotov12}. 

Despite these difficulties, given the current technical challenges of generating cosmological high-resolution hydrodynamical simulations (in which galaxies form self-consistently), \Nbody\ simulations remain a valuable tool to try to understand the evolution of galaxies and (sub)haloes. In order for them to be applied to real observations, however, one must post-process these simulations in some way that relates subhaloes to galaxies, taking into account the aforementioned complexities (and others). For instance, semi-analytic models contain either physical or phenomenological recipes whether or not to form galaxies in certain dark matter haloes based on the mass and assembly history of haloes \citep[e.g.,][]{bower06,lacey16}. A different method involves halo occupation distributions (HODs), which assume that the average number of galaxies in a halo depends only on host halo mass. Because they provide an analytical framework to connect galaxies and dark matter haloes, HODs are commonly used to interpret galaxy-galaxy lensing and galaxy clustering measurements through a conditional stellar mass (or luminosity) function \citep[e.g.,][]{seljak00,peacock00,mandelbaum06_ggl,cacciato09,vdbosch13}.

One of the key aspects of these prescriptions is the stellar-to-halo mass relation. While many studies have constrained the stellar-to-halo mass relation of central galaxies \citep[e.g.,][]{hoekstra05,heymans06_ggl,mandelbaum06_ggl,mandelbaum16,more11,vanuitert11,vanuitert16,leauthaud12,velander14,coupon15,zu15}, this is not the case for satellite galaxies, whose \emph{sub}halo-to-stellar mass relation (SHSMR) remains essentially unexplored, and the constraints so far are largely limited to indirect measurements. \cite{rodriguez12} used abundance matching (the assumption that galaxies rank-ordered by stellar mass can be uniquely mapped to [sub]haloes rank-ordered by total mass) to infer the SHSMR using the satellite galaxy stellar mass function, and \cite{rodriguez13} extended these results using galaxy clustering measurements. They showed that the SHSMR is significantly different from the central stellar-to-total mass relation, and that assuming an average relation when studying a mixed population can lead to biased results \citep[see also][]{yang09}.

Instead, only stellar dynamics and gravitational lensing provide direct ways to probe the total gravitational potential of a galaxy. However, the quantitative connection between stellar velocity dispersion and halo mass is not straightforward \citep[e.g.,][]{li13_massproxy,old15}, and only gravitational lensing provides a direct measurement of the total surface mass density \citep{fahlman94,clowe98}. Using deep Hubble Space Telescope (HST) observations, \cite{natarajan98,natarajan02,natarajan07,natarajan09} measured the weak (and also sometimes strong) lensing effect of galaxies in six clusters at $z=0.2-0.6$. After fitting a truncated density profile to the ensemble signal using a maximum likelihood approach, they concluded that galaxies in clusters are strongly truncated with respect to field galaxies. Using data for clusters at $z\sim0.2$ observed with the CFH12k instrument on the Canada-Hawaii-France Telescope (CFHT), \cite{limousin07} arrived at a similar conclusion. \cite{halkola07} and \cite{suyu10} used strong lensing measurements of a single cluster and a small galaxy group, respectively, and also found evidence for strong truncation of the density profiles of satellite galaxies, and other strong lensing studies have reached similar conclusions \citep[e.g.,][]{eichner13,monna15,monna17_macs}.
Likewise, \cite{okabe14} analyzed the weak lensing signal of subhaloes in the Coma cluster and found that, while group-scale subhaloes show (mild) evidence of sharp truncations at  radii $r<200$ kpc, stacked weak lensing measurements of satellite galaxies show no signs of truncation.
Similarly, \cite{pastormira11} found no evidence of truncation of subhaloes in the Millenium simulation \citep{springel05_millenium}; their density profiles are instead fully consistent with a NFW profile. It is unclear whether the differences between these studies are due to different galaxy (and cluster) samples, different modelling assumptions or, in the case of \cite{pastormira11}, due to the lack of baryonic physics in the simulations.

In addition, recent combinations of large weak lensing surveys with high-purity galaxy group catalogues have allowed direct measurements of the average subhalo masses associated with satellite galaxies using weak galaxy-galaxy lensing \citep{li14,li16,sifon15_kids,niemiec17}. Like the studies cited above, these studies did not focus on the SHSMR but on the segregation of subhaloes by mass within galaxy groups, by measuring subhalo masses at different group-centric distances. The observational results are consistent, within their large errorbars, with the mild segregation of dark matter subhaloes seen in numerical simulations \citep{han16,vdbosch16}. However, it is not clear whether results based on subhaloes in N-body simulations can be directly compared to observations. In fact, \cite{vdbosch17} has shown that the statistics of subhaloes inferred from N-body simulations are problematic even to this day, because of severe numerical destruction of subhaloes.

In this work, we present weak gravitational lensing measurements of the total mass of satellite galaxies in 48 massive galaxy clusters at $z<0.15$. Our images were taken with the \megacam\ instrument on the Canada-France-Hawaii Telescope (CFHT), which has a field of view of 1 \sqdeg, allowing us to focus on very low redshift clusters and take advantage of the $<0.8''$ seeing (corresponding to 1.48 kpc at $z=0.1$) typical of our observations. We can therefore probe the lensing signal close to the galaxies themselves, at a physical scale equivalent to what can be probed in a cluster at $z\sim0.5$ with HST, but out to the clusters' virial radii. In addition, the low-redshift clusters we use have extensive spectroscopic observations available from various data sets, compiled by \cite{sifon15_cccp}, so we do not need to rely on uncertain photometric identification of cluster members.

This paper is organized as follows. We summarize the galaxy-galaxy lensing formalism in \Cref{s:ggl}. We describe our data set in \Cref{s:data}, taking a close look at the source catalogue and the shapes of background sources in \Cref{s:calibration}. We present our modelling of the satellite lensing signal in \Cref{s:model}, and discuss the connection between mass and light in satellite galaxies, in the form of the subhalo-to-stellar mass relation and subhalo mass segregation, in \Cref{s:shsmr,s:segregation}, respectively. Finally, we summarize in \Cref{s:conclusions}.

We adopt a flat $\Lambda$ cold dark matter (\LCDM) cosmology with $\Omega_\mathrm{m}=0.315$, based on the latest results from cosmic microwave background observations by \cite{planck15xiii}, and $H_0=70\,\mathrm{km\,s^{-1}Mpc^{-1}}$. In this cosmology, $10\arcsec=\{9.8,18.4,26.1\}\,{\rm kpc}$ at $z=\{0.05,0.1,0.15\}$. As usual, stellar and (sub)halo masses depend on the Hubble constant as $m_\star\sim1/H_0^2$ and $m\sim1/H_0$, respectively.


\section{Weak galaxy-galaxy lensing}\label{s:ggl}

Gravitational lensing distorts the images of background (``source'') galaxies as their light passes near a matter overdensity along the line-of-sight. This produces a distortion in the shape of the background source, called \emph{shear}, and a \emph{magnification} effect on the source's size (and consequently its brightness). The shear field around a massive object aligns the images of background sources around it in the tangential direction. Therefore, starting from a measurement of the shear of an object in a cartesian frame with components $(\gamma_1,\gamma_2)$ (see \Cref{s:shapes}), it is customary to parametrize the shear as
\begin{equation}
  \begin{pmatrix}
    \gammat \\
    \gamma_\times
  \end{pmatrix}
  = 
    \begin{pmatrix}
      -\cos\,2\phi & -\sin\,2\phi \\
      \sin\,2\phi & -\cos\,2\phi
    \end{pmatrix}
  \begin{pmatrix}
    \gamma_1 \\
    \gamma_2
  \end{pmatrix}
  ,
\end{equation}
where $\phi$ is the azimuthal angle of the lens-source vector, $\gammat$ measures the ellipticity in the tangential ($\gammat>0$) and radial ($\gammat<0$) directions and $\gamma_\times$ measures the ellipticity in directions $45^\circ$ from the tangent. Because of parity symmetry, we expect $\langle\gamma_\times\rangle=0$ for an ensemble of lenses \citep{schneider03} and therefore $\gamma_\times$ serves as a test for systematic effects.

The shear is related to the excess surface mass density (ESD), $\ESD$, via
\begin{equation}
 \ESD(R) \equiv \bar\Sigma(<R) - \bar\Sigma(R) = \gammat \Sigma_\mathrm{c},
\end{equation}
where $\bar\Sigma(<R)$ and $\bar\Sigma(R)$ are the average surface mass density within a radius\footnote{As a convention, we denote three-dimensional distances with lower case $r$ and two-dimensional distances (that is, projected on the sky) with upper case $R$.} $R$ and within a thin annulus at distance $R$ from the lens. The critical surface density, $\Sigma_\mathrm{c}$, is a geometrical factor that accounts for the lensing efficiency,
\begin{equation}\label{eq:sigmac}
 \Sigma_\mathrm{c} = \frac{c^2}{4\pi G}\frac{D_\mathrm{s}}{D_\mathrm{l}D_\mathrm{ls}},
\end{equation}
where, $D_\mathrm{l}$, $D_\mathrm{s}$, and $D_\mathrm{ls}$ are the angular diameter distances to the lens, to the source and between the lens and the source, respectively. The ESD for each bin in lens-source separation is then
\begin{equation}\label{eq:esd}
 \ESD = \frac{\sum_i w_i \Sigma_{\mathrm{c},i} \gamma_{{\rm t},i}}{\sum_i w_i},
\end{equation}
where the sums run over all lens-source pairs in a given bin and the weight of each source galaxy is given by
\begin{equation}\label{eq:weight}
 w_i = \frac1{\langle\epsilon_\mathrm{int}^2\rangle + (\sigma_{\gamma,i})^2}.
\end{equation}
Here, $\sigma_\gamma$ is the measurement uncertainty in $\gammat$, which results from the quadrature sum of statistical uncertainties due to shot noise in the images (see \Cref{s:shapes}) and from uncertainties in the modelling of a measurement bias detailed in \Cref{s:ct,ap:ct}.\footnote{In practice, the latter is negligible in most cases.} We set the intrinsic root-mean-square galaxy ellipticity, $\langle\epsilon_\mathrm{int}^2\rangle^{1/2} = (\langle\epsilon^2\rangle - \langle\sigma_{\gamma,i}^2\rangle)^{1/2} = 0.25$ \citep[e.g.,][]{hoekstra00,schrabback18}, where $\langle\epsilon^2\rangle=0.104$ is the observed rms ellipticity. In \Cref{eq:esd}, we use a single value for $\Sigma_\mathrm{c}$ for all satellites in each cluster (see \Cref{s:beta}).

In fact, the weak lensing observable is the \emph{reduced} shear, $g\equiv\gamma/(1-\kappa)$ (where $\kappa=\Sigma/\Sigma_\mathrm{c}$ is the lensing convergence), but in the weak limit $\kappa\ll1$ so that $g\approx\gamma$. However, close to the centres of galaxy clusters the convergence becomes significant, so this approximation is not accurate anymore. To account for this, the lensing model presented in \Cref{s:model} is corrected using
\begin{equation}
 g(R) = \frac{\gamma(R)}{1 - \bar\Sigma(R)/\Sigma_\mathrm{c}}
    = \frac{\ESD(R)/\Sigma_\mathrm{c}}{1 - \bar\Sigma(R)/\Sigma_\mathrm{c}}\,.
\end{equation}

Because the gravitational potential of satellites in a cluster is traced by the same background source galaxies, data points in the ESD are correlated. Following equations 13--17 of \cite{viola15}, we can re-arrange \Cref{eq:esd} to reflect the contribution from each \emph{source} galaxy. The data covariance of measurements in a single cluster can then be written as
\begin{equation}\label{eq:cov}
 \mathbfss{C}_{mnij} = \Sigma_\mathrm{c}^2\langle\epsilon^2\rangle
   \frac{\sum_s \left(C_{si,m}C_{sj,n} + S_{si,m}S_{sj,n}\right)}
        {\left(\sum_s Z_{si,m}\right)\left(\sum_s Z_{sj,n}\right)},
\end{equation}
where index pairs $m,n$ and $i,j$ run over the observable bins (e.g., stellar mass) and lens-source separation, $R$, respectively, and $C$, $S$ and $Z$ are sums over the lenses:
\begin{equation}\label{eq:Cs}
\begin{split}
 C_{si} &= -\sum_l w_{ls} \cos 2\phi_{ls} \,, \\
 S_{si} &= -\sum_l w_{ls} \sin 2\phi_{ls} \,, \\
 Z_{si} &= \sum_l w_{ls} \,,
\end{split}
\end{equation}
where we explicitly allow for the possibility that the source weight, $w$, may be different for each lens-source pair (as opposed to a unique weight per source). This is indeed the case when we consider the corrections to the shape measurements from lens contamination discussed in \Cref{s:ct,ap:ct}, although in practice differences are negligible. As implied by \Cref{eq:cov}, we assign the same $\Sigma_\mathrm{c}$ to all galaxies that are part of the same cluster. The total ESD is then the inverse-covariance--weighted sum of the ESDs of individual clusters.
 
In addition to the data covariance there are, in principle, contributions to the measurement uncertainty from sample variance and from distant large scale structure. By comparing \Cref{eq:cov} to uncertainties estimated by bootstrap resampling, \cite{sifon15_kids} have shown that the contribution from sample variance is less than 10 per cent for satelite galaxy-galaxy lensing measurements when limited to small lens-source separations ($R\lesssim2$ Mpc). Since the signal from satellites themselves is limited to $R\lesssim300\,{\rm kpc}$ \citep[\Cref{f:esd_spec_rs}; see also][]{sifon15_kids}, in this work we ignore the sample variance contribution to the lensing covariance. Similarly, the distant large scale-structure introduces correlations preferentially on large scales when the signal is averaged around a few positions, as first shown in \cite{hoekstra01}. This adds noise and is relevant for individual cluster mass estimates \citep[e.g.,][]{okabe14}. Here, we stack the signal at relatively small scales around many different positions (all the lenses) and the large-scale structure contribution is suppressed accordingly. It is therefore reasonable to ignore this contribution as well.

Measurements are independent between clusters. We therefore combine measurements from different clusters, $X_i \equiv \Sigma_{\mathrm{c},i}\mathbf{\gamma}_i$, as:
\begin{equation}
  \langle X \rangle
   = \mathbfss{C}_\mathrm{tot}
     \left[\sum_i \mathbfss{C}_i^{-1} \Sigma_{\mathrm{c},i}\mathbf{\gamma}_i \right]
\end{equation}
where $\mathbfss{C}_i^{-1}$ is the inverse covariance matrix of measurements of the $i$-th cluster, and
\begin{equation}
  \mathbfss{C}_\mathrm{tot} = \left[ \sum_i \mathbfss{C}_i^{-1} \right]^{-1}
\end{equation}
is the covariance of the average measurements entering our analysis (\Cref{s:likelihood}).

Finally, we note that the formalism described above is valid under spherical symmetry, and for a smooth cluster stack. For an rms cluster ellipticity of roughly 0.3 \citep[e.g.,][]{vanuitert17}, the ellipticity of our stack is $0.3/\sqrt{48}\sim0.04$, close enough to circular for our purposes. Substructure from individual clusters will naturally smooth out during the stacking process as well. We therefore consider said assumptions of our formalism to be valid within the precision of our measurements.


\section{Data set}\label{s:data}

In this section we describe the lens and source galaxy samples we use in our analysis. In the next section, we make a detailed assessment of the shape measurement and quality cuts on the source sample using extensive image simulations.

\subsection{Cluster and lens galaxy samples}\label{s:lenses}

The Multi-Epoch Nearby Cluster Survey \citep[\meneacs,][]{sand12} is a targeted survey of 57 galaxy clusters in the redshift range $0.05\lesssim z \lesssim0.15$ observed in the $g$ and $r$ bands with \megacam\ on CFHT. We only use the 48 clusters affected by $r$-band Galactic extinctions $A_r\leq0.2$ mag, since we find that larger extinctions bias the source number counts and the correction for cluster member contamination (\Cref{s:calibration}). The image processing and photometry are described in detail in \cite{vdburg13}; most images have seeing $<0.8''$. We list our sample of 48 clusters in \Cref{t:clusters}. \cite{sifon15_cccp} compiled a large sample of spectroscopic redshift measurements in the direction of 46 of these clusters, identifying a total of 7945 spectroscopic members. Since, \cite{rines16} have published additional spectroscopic redshifts for galaxies in 12 \meneacs\ clusters, six of which are included in \cite{sifon15_cccp} but for which the observations of \cite{rines16} represent a significant increase in the number of member galaxies. We select cluster members in these 12 clusters in an identical way as \cite{sifon15_cccp}. The median dynamical mass of MENeaCS clusters is $M_{200}\sim6\times10^{14}\,\Msun$ \citep{sifon15_cccp}.

From the member catalogue of \cite{sifon15_cccp}, we exclude all brightest cluster galaxies (BCGs), and refer to all other galaxies as satellites. Because the shapes of background galaxies near these members are very likely to be contaminated by light from the BCG, we also exclude all satellite galaxies within 10$\arcsec$ of the BCGs to avoid severe contamination from extended light. Finally, we impose a luminosity limit $L_\mathrm{sat}<\min(2L^\star,0.5L_\mathrm{BCG})$ (where $L^\star(z)$ is the $r$-band luminosity corresponding to the characteristic magnitude, $m^\star_\mathrm{phot}(z)$ of the \cite{schechter76} function, fit to red satellite galaxies in redMaPPer galaxy clusters over the redshift range $0.05<z<0.7$ \citep{rykoff14}).\footnote{Equation 9 of \citet{rykoff14} provides a fitting function for the $i$-band $m^\star_\mathrm{phot}(z)$, which we convert to $r$-band magnitudes assuming a quiescent spectrum, appropriate for the majority of our satellites, using \ezgal\ \citep[\url{http://www.baryons.org/ezgal/},][]{mancone12}.} We use SExtractor's \textsc{mag\_auto} \citep{bertin96} as our estimates of galaxy magnitudes. We choose the maximum possible luminosity,    $2L^\star$, because the BCGs in our sample have $L_\mathrm{BCG}\gtrsim3L^\star$, so this ensures we do not include central galaxies of massive (sub)structures that could, for instance, have recently merged with the cluster. In addition, we only include satellites within 2 Mpc of the BCG. At larger distances, contamination by fore- and background galaxies becomes an increasingly larger problem. Our final spectroscopic sample consists of 5414 satellites in 45 clusters.

In addition, we include red sequence galaxies in all \meneacs\ clusters in low Galactic extinction regions in order to improve our statistics. We measure the red sequence by fitting a straight line to the colour-magnitude relation of red galaxies in each cluster using a maximum likelihood approach, based on the methodology of \cite{hao09}. Following \cite{sifon15_cccp}, we include only red sequence galaxies brighter than $M_\mathrm{r}=-19$ and within 1 Mpc of the BCG.\footnote{Here, $M_\mathrm{r}$ is the $k+e$--corrected absolute magnitude in the $r$-band, calculated with \ezgal\ using a passively evolving Charlot \& Bruzual \citep[2007, unpublished, see][]{bruzual03} model with formation redshift $z_\mathrm{f}=5$.} When we include red sequence galaxies, we also use the six clusters without spectroscopic cluster members. Therefore our combined spectroscopic plus red sequence sample includes 7909 cluster members in 48 clusters (including three clusters without spectroscopic data). Throughout, we refer to the spectroscopic and spectroscopic plus red sequence samples as `spec' and `spec+RS', respectively.

For the purpose of estimating stellar masses and photometric redshifts, the original \meneacs\ observations in $g$ and $r$ were complemented by $u$- and $i$-band observations with the Wide-Field Camera on the Isaac Newton Telescope in La Palma \citep[except for a few clusters with archival \megacam\ data in either of these bands, see][for details]{vdburg15}. Stellar masses were estimated by \cite{vdburg15} by fitting each galaxy's spectral energy distribution using \textsc{fast} \citep{kriek09} assuming a \cite{chabrier03} initial mass function.

\begin{table}
\centering
\caption{Cluster sample.
Clusters with a $u$ and/or $i$ superscript were observed with INT/WFC in the respective 
band(s); all other observations were performed with CFHT/Megacam.
}
\label{t:clusters}
\begin{tabular}{lcrcc}
\hline\hline
Cluster &     R.A.     & \multicolumn{1}{c}{Decl.} & Redshift &    $r$-band   \\
        & (hh:mm:ss.s) & \multicolumn{1}{c}{(dd:mm:ss)} &     & seeing ($''$) \\[0.5ex]
\hline
Abell 7$^u$$^i$ & 00:11:45.3 & 32:24:57 & 0.106 & 0.60 \\
Abell 21$^u$$^i$ & 00:20:37.0 & 28:39:33 & 0.095 & 0.63 \\
Abell 85$^u$$^i$ & 00:41:50.4 & $-$09:18:11 & 0.055 & 0.62 \\
Abell 119$^i$ & 00:56:16.1 & $-$01:15:19 & 0.044 & 0.65 \\
Abell 133$^u$$^i$ & 01:02:41.7 & $-$21:52:55 & 0.057 & 0.68 \\
Abell 646$^u$$^i$ & 08:22:09.5 & 47:05:53 & 0.129 & 0.69 \\
Abell 655 & 08:25:29.0 & 47:08:01 & 0.127 & 0.65 \\
Abell 754$^i$ & 09:08:32.4 & $-$09:37:47 & 0.054 & 0.74 \\
Abell 780$^u$$^i$ & 09:18:05.7 & $-$12:05:44 & 0.054 & 0.80 \\
Abell 795$^i$ & 09:24:05.3 & 14:10:22 & 0.136 & 0.72 \\
Abell 961$^u$$^i$ & 10:16:22.9 & 33:38:18 & 0.124 & 0.71 \\
Abell 990$^u$$^i$ & 10:23:39.9 & 49:08:39 & 0.144 & 0.78 \\
Abell 1033$^u$$^i$ & 10:31:44.3 & 35:02:29 & 0.126 & 0.65 \\
Abell 1068$^u$$^i$ & 10:40:44.5 & 39:57:11 & 0.138 & 0.61 \\
Abell 1132$^u$$^i$ & 10:58:23.6 & 56:47:42 & 0.136 & 0.68 \\
Abell 1285$^u$$^i$ & 11:30:23.8 & $-$14:34:52 & 0.106 & 0.82 \\
Abell 1361$^u$$^i$ & 11:43:39.6 & 46:21:21 & 0.117 & 0.61 \\
Abell 1413$^i$ & 11:55:18.0 & 23:24:18 & 0.143 & 0.66 \\
Abell 1650$^u$$^i$ & 12:58:41.5 & $-$01:45:41 & 0.084 & 0.76 \\
Abell 1651$^u$$^i$ & 12:59:22.5 & $-$04:11:46 & 0.085 & 0.91 \\
Abell 1781$^u$$^i$ & 13:44:52.5 & 29:46:16 & 0.062 & 0.73 \\
Abell 1795$^u$$^i$ & 13:48:52.5 & 26:35:35 & 0.062 & 0.68 \\
Abell 1927$^u$$^i$ & 14:31:06.8 & 25:38:02 & 0.095 & 0.62 \\
Abell 1991$^i$ & 14:54:31.5 & 18:38:33 & 0.059 & 0.67 \\
Abell 2029$^u$$^i$ & 15:10:56.1 & 05:44:41 & 0.077 & 0.65 \\
Abell 2033$^u$$^i$ & 15:11:26.5 & 06:20:57 & 0.082 & 0.61 \\
Abell 2050$^u$$^i$ & 15:16:17.9 & 00:05:21 & 0.118 & 0.62 \\
Abell 2055$^u$$^i$ & 15:18:45.7 & 06:13:56 & 0.102 & 0.61 \\
Abell 2064$^u$$^i$ & 15:20:52.2 & 48:39:39 & 0.108 & 0.69 \\
Abell 2065$^u$$^i$ & 15:22:29.2 & 27:42:28 & 0.073 & 0.66 \\
Abell 2069$^u$$^i$ & 15:24:07.5 & 29:53:20 & 0.116 & 0.62 \\
Abell 2142$^u$$^i$ & 15:58:20.0 & 27:14:00 & 0.091 & 0.62 \\
Abell 2420$^u$$^i$ & 22:10:18.8 & $-$12:10:14 & 0.085 & 0.67 \\
Abell 2426$^u$$^i$ & 22:14:31.6 & $-$10:22:26 & 0.098 & 0.73 \\
Abell 2440$^u$$^i$ & 22:23:56.9 & $-$01:35:00 & 0.091 & 0.70 \\
Abell 2443$^u$$^i$ & 22:26:07.9 & 17:21:24 & 0.108 & 0.62 \\
Abell 2495$^u$$^i$ & 22:50:19.7 & 10:54:13 & 0.078 & 0.61 \\
Abell 2597$^u$$^i$ & 23:25:19.7 & $-$12:07:27 & 0.085 & 0.67 \\
Abell 2627$^u$$^i$ & 23:36:42.1 & 23:55:29 & 0.126 & 0.64 \\
Abell 2670$^i$ & 23:54:13.7 & $-$10:25:08 & 0.076 & 0.77 \\
Abell 2703$^u$$^i$ & 00:05:23.9 & 16:13:09 & 0.114 & 0.60 \\
MKW3S$^u$$^i$ & 15:21:51.8 & 07:42:32 & 0.045 & 0.65 \\
RX J0736$^u$$^i$ & 07:36:38.1 & 39:24:53 & 0.118 & 0.70 \\
RX J2344$^u$$^i$ & 23:44:18.2 & $-$04:22:49 & 0.079 & 0.70 \\
ZwCl 1023$^u$$^i$ & 10:25:58.0 & 12:41:09 & 0.143 & 0.72 \\
ZwCl 1215$^u$$^i$ & 12:17:41.1 & 03:39:21 & 0.075 & 0.86 \\

\hline
\end{tabular}
\end{table}

\begin{table*}
 \centering
\caption{Number of galaxies and average properties of stellar mass and cluster-centric distance bins used in \Cref{s:shsmr,s:segregation}. Sub-columns correspond to the values of the fiducial spectroscopic-plus-red-sequence and the spectroscopic-only samples. 
}
\label{t:bins}
\begin{tabular}{l c c | c c | c c | c c}
\hline\hline
Binning & Bin & \multirow{3}{*}{Range} & \multicolumn{2}{c|}{$N_\mathrm{sat}$} & 
\multicolumn{2}{c|}{$\langle\Rsat/{\rm Mpc}\rangle$} & 
\multicolumn{2}{c}{$\log\langle\Mstar/\Msun\rangle$} \\
\cline{4-9}
observable & label &  & spec+RS & spec & spec+RS & spec & spec+RS & spec \\[0.5ex]
\hline
\multirow{4}{*}{$\log(\Mstar/\Msun)$}
 & M1 &   $[9.0-9.8)$ &      2144 &      1010 & 0.66 & 0.88 & \,\,\,9.51 & \,\,\,9.51 \\
 & M2 &  $[9.8-10.2)$ &      2017 &      1315 & 0.67 & 0.87 &      10.01 &      10.03 \\
 & M3 & $[10.2-10.5)$ &      1387 &      1146 & 0.80 & 0.91 &      10.36 &      10.35 \\
 & M4 & $[10.5-10.9)$ &      1178 &      1052 & 0.83 & 0.89 &      10.67 &      10.67 \\
 & M5 & $[10.9-11.2]$ & \,\,\,278 & \,\,\,265 & 0.93 & 0.98 &      11.01 &      11.01 \\
\hline
\multirow{4}{*}{$\Rsat\,({\rm Mpc})$}
  & D1 & $[0.1-0.35)$ & 1346 & \,\,\,664 & 0.23 & 0.23 & \,\,\,9.97 & 10.20 \\
  & D2 & $[0.35-0.7)$ & 1934 &      1139 & 0.52 & 0.52 &      10.03 & 10.20 \\
  & D3 &  $[0.7-1.2)$ & 1994 &      1397 & 0.90 & 0.94 &      10.07 & 10.22 \\
  & D4 &  $[1.2-2.0)$ & 1550 &      1529 & 1.55 & 1.55 &      10.24 & 10.25 \\
\hline
\end{tabular}
\end{table*}

\begin{figure*}
  \centerline{\includegraphics[width=2.2in]{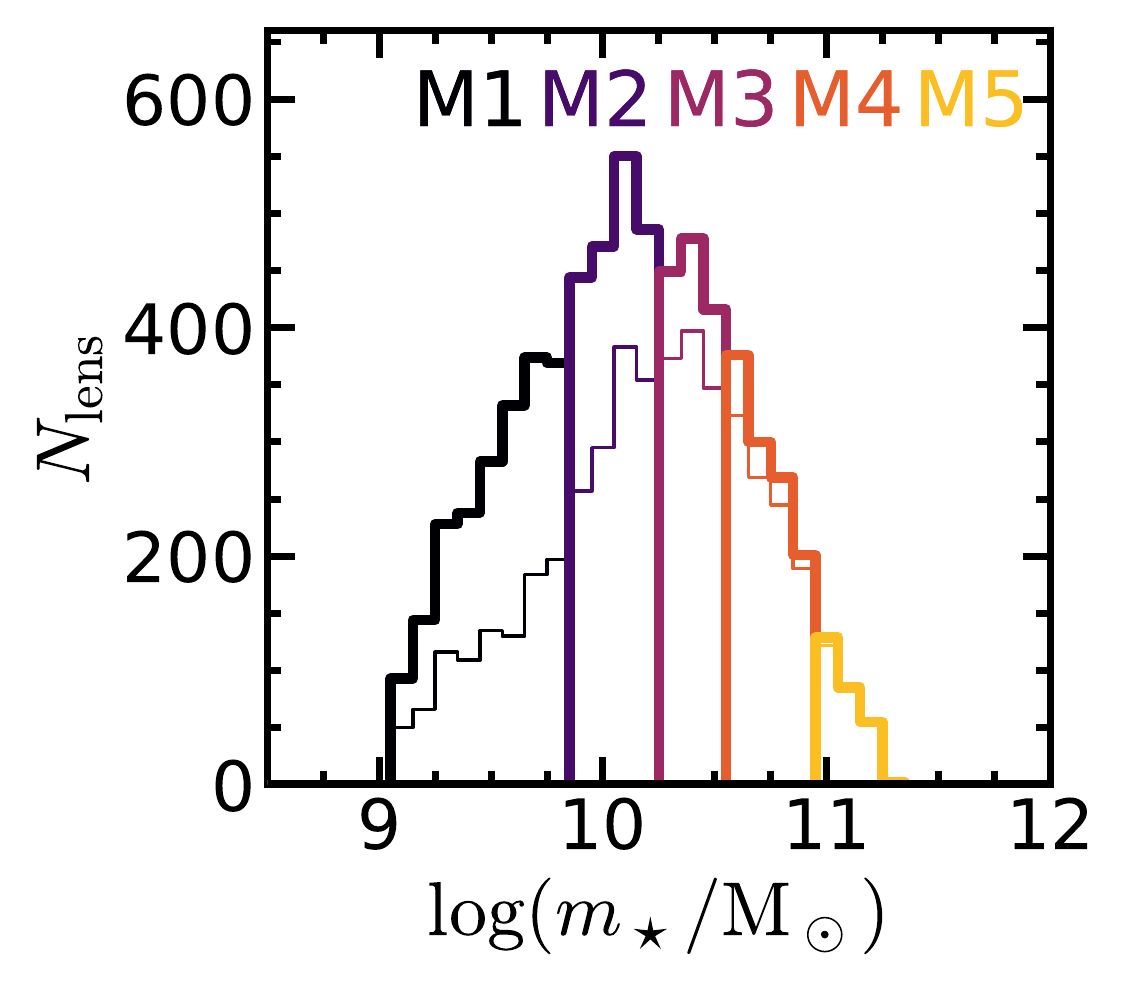}
              \includegraphics[width=2.2in]{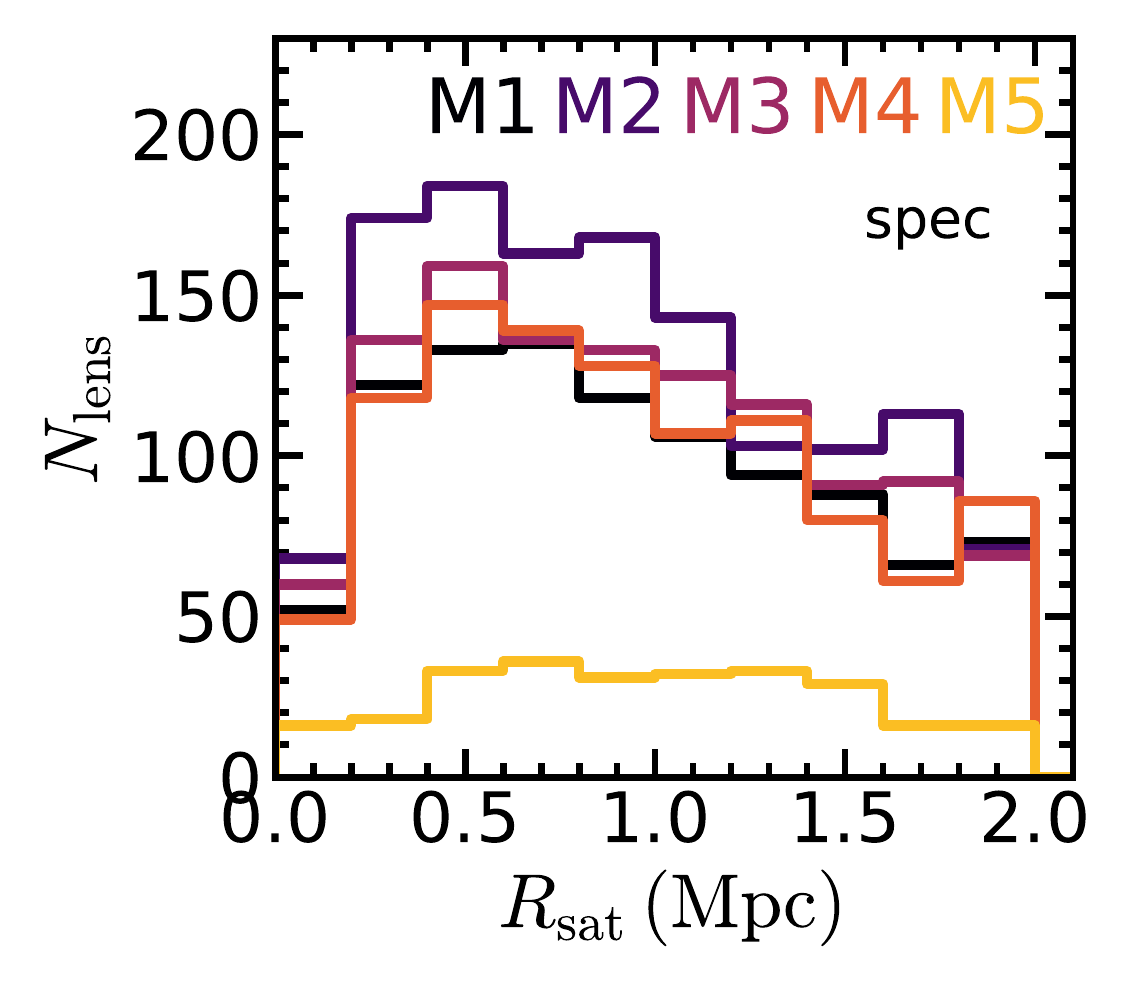}
              \includegraphics[width=2.2in]{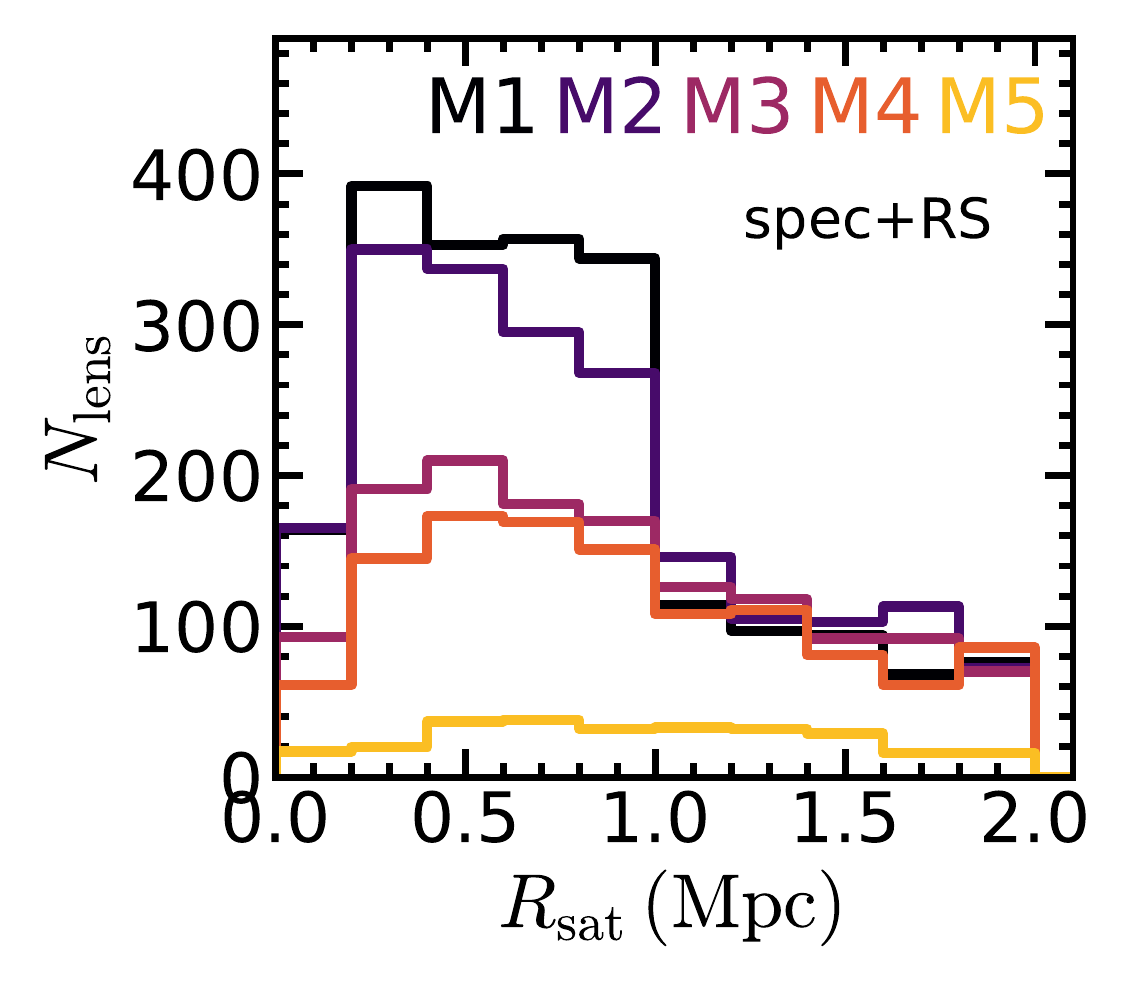}}
  \centerline{\includegraphics[width=2.2in]{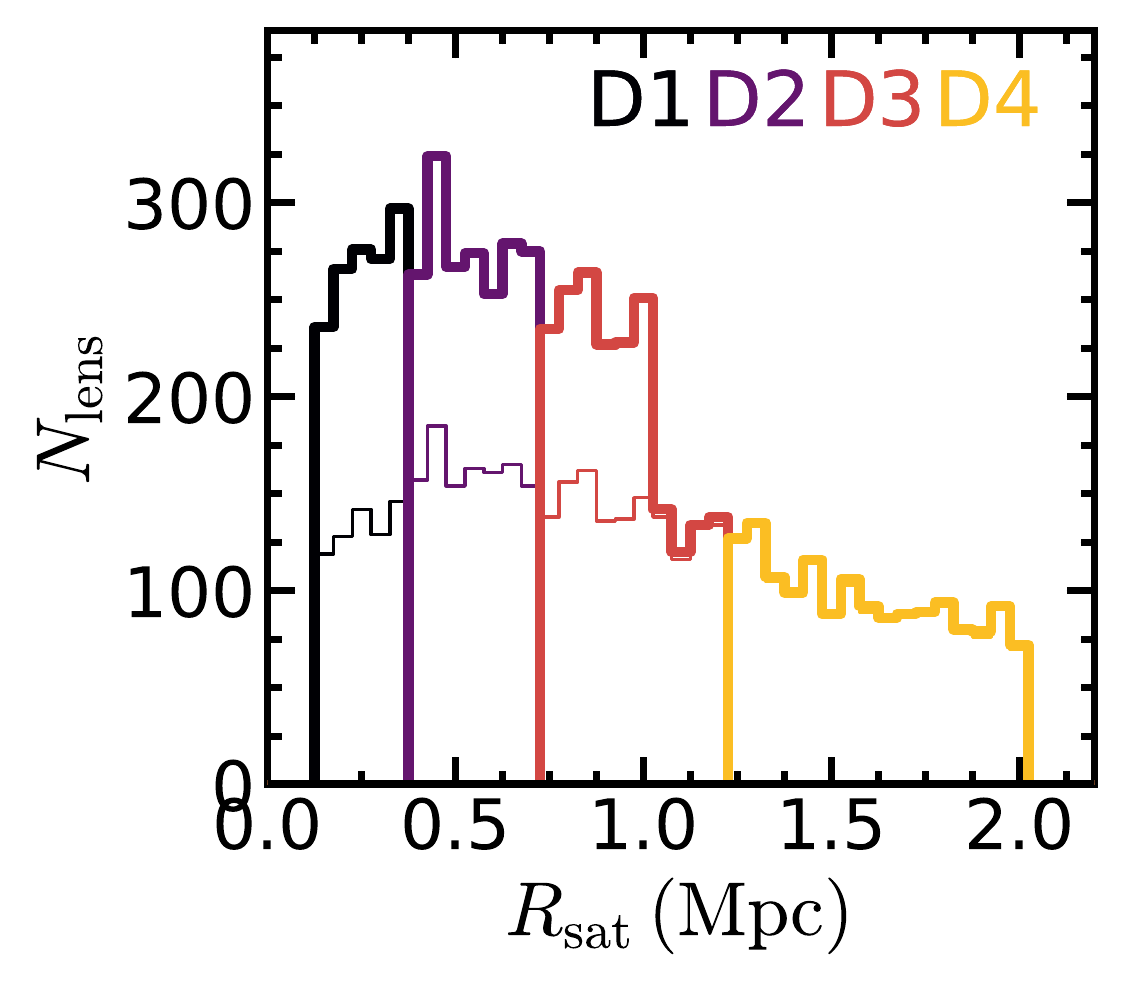}
              \includegraphics[width=2.2in]{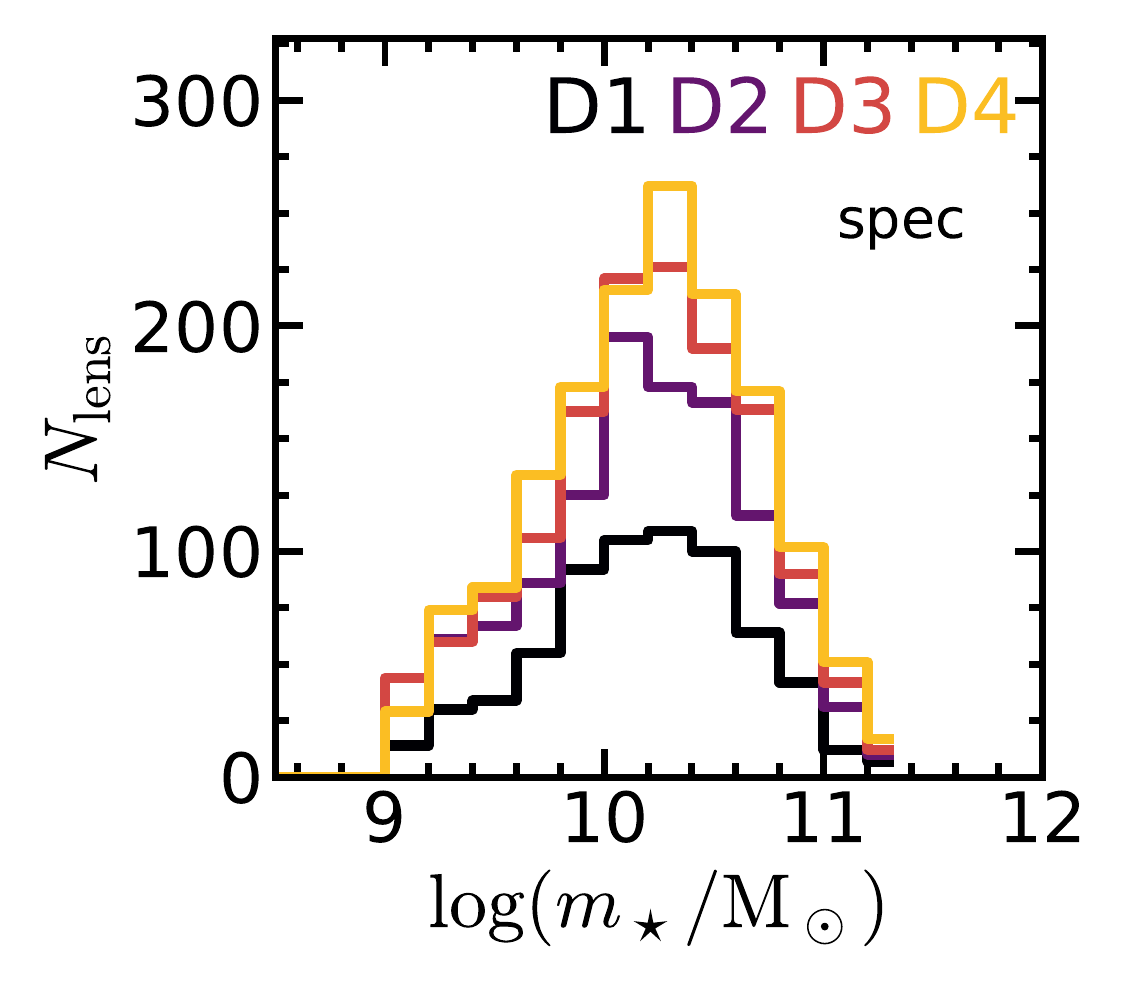}
              \includegraphics[width=2.2in]{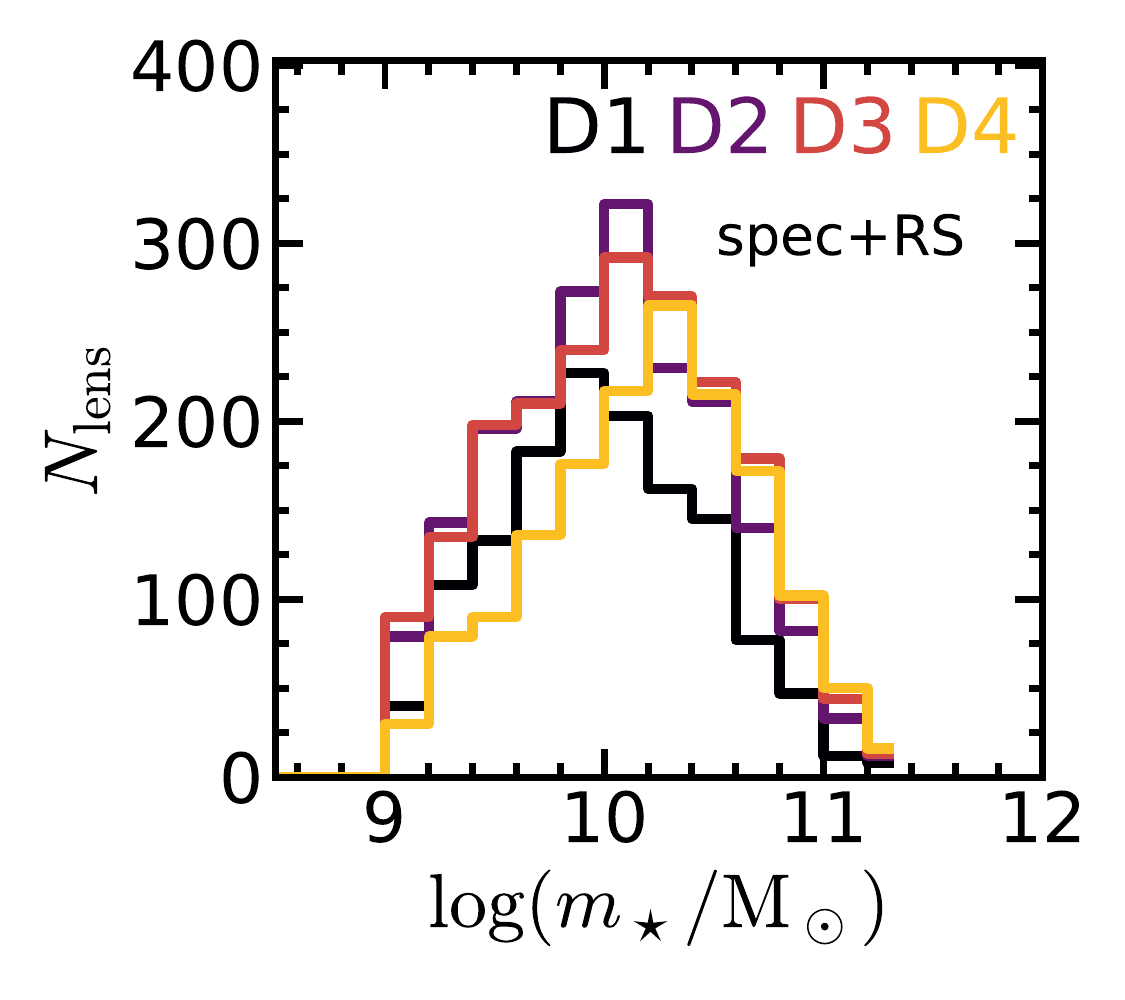}}
\caption{The five stellar mass bins used in \Cref{s:shsmr} (top) and the five cluster-centric distance bins used in \Cref{s:segregation} (bottom). Left panels show histograms for spectroscopic (`spec' sample, thin lines) and spectroscopic plus red sequence (`spec+RS' sample, thick lines) members. Middle and right panels show distribution of different stellar mass bins in cluster-centric distance (top) and of different cluster-centric radius bins in stellar mass (bottom), for the spec and spec+RS samples, respectively. Note the different vertical scales in each panel.}
\label{f:histograms}
\end{figure*}

In order to characterize the connection between satellite galaxies and their host subhaloes, we split the sample by stellar mass (\Cref{s:shsmr}) and cluster-centric distance (\Cref{s:segregation}), each time splitting the sample in five bins. We show the stellar mass and cluster-centric distributions of the resulting subsamples in \Cref{f:histograms}, and list the average values in \Cref{t:bins}.

\subsection{Source galaxy sample}
\label{s:sources}

We construct the source catalogues in an identical manner to \cite{hoekstra15}, except for one additional flag to remove galaxies whose shape is significantly biased by the presence of a nearby bright object. This step is discussed in detail in \Cref{s:delmag}. The biases in the shape measurements of the sources, depending on how the source sample is defined, have been characterized in great detail by \cite{hoekstra15}. Although the study of \cite{hoekstra15} refers to a different cluster sample, both samples have been observed with the same instrument under very similar conditions of high image quality, so we can safely take the analysis of \cite{hoekstra15} as a reference for our study.

Specifically, we select only sources with $r$-band magnitudes\footnote{We denote $r$-band magnitudes with $\rmag$ in order to avoid confusion with subhalo masses, which we denote with lower case $m$ and subscripts depending on the definition (see \Cref{s:subhalo}).} $20<\rmag<24.5$, with sizes $r_\mathrm{h}<5$ pix and an additional constraint on $\delta \rmag$, the difference in estimated magnitude before and after the local background subtraction used for shape measurements (see \Cref{s:delmag}). Compared to \cite{hoekstra15}, who used $22<\rmag<25$, we choose different magnitude limits (i) at the bright end because our cluster sample is at lower redshift and therefore cluster members tend to be brighter, and (ii) at the faint end because our data are slightly shallower, complicating the shape measurements of very faint sources. The magnitudes $\rmag$ have been corrected for Galactic extinction using the \cite{schlafly11} recalibration of the \cite{schlegel98} infrared-based dust map. 

Unlike most cluster lensing studies \citep[e.g.,][]{hoekstra12,applegate14,umetsu14}, we do not apply a colour cut to our source sample, since this only reduces contamination by $\sim$30\percent\ for $z\sim0.2$ clusters \citep{hoekstra07}. In fact, one of the advantages of using low-redshift clusters is that contamination by cluster members is significantly lower than at higher redshifts, since cluster members are spread over a larger area on the sky. Instead of applying colour cuts to reduce contamination, we follow \cite{hoekstra15} and correct for contamination in the source sample by applying a `boost factor' to the measured lensing signal to account for the dilution by cluster members \citep[e.g.,][]{mandelbaum05_errors}. We discuss this and other corrections to the shape measurements, along with the source redshift distribution, in \Cref{s:calibration}.

\subsection{Shape measurements}
\label{s:shapes}

To measure the galaxy-galaxy lensing signal we must accurately infer the shear field around the lenses by measuring the shapes of as many background galaxies as possible. For most of the sources this is a difficult procedure as they are faint and of sizes comparable to the image resolution, quantified by the point spread function (PSF). Blurring by the PSF and noise lead to a multiplicative bias, $\mu$, while an anisotropic PSF introduces an additive bias, $c$ \citep[e.g.,][]{heymans06_step}. The measured (or observed) shear is therefore related to the true shear by
\begin{equation}\label{eq:gammaobs}
  \gamma^{\rm obs}(\thetals\vert\Rsat) =
      \left(1+\mu\right)\gamma^{\rm true}(\thetals) 
\mathcal{B}^{-1}(\Rsat) + c \,,
\end{equation}
where $\thetals$ is the lens-source separation and $\mu$ and $c$ are referred to simply as the multiplicative and additive biases, respectively; $\mathcal{B}(\Rsat)$ is the `boost factor' that corrects for contamination by cluster members, described in \Cref{s:boost}. Note that $\mu$, $c$ and $\mathcal{B}(\Rsat)$ all depend on both the dataset and the shape measurement method.

We measure galaxy shapes by calculating the moments of galaxy images using the KSB method \citep{kaiser95,luppino97}, incorporating the modifications by \cite{hoekstra98,hoekstra00}. The PSF is measured from the shapes of stars in the image and interpolation between stars is used to estimate the PSF for each galaxy. \cite{hoekstra15} used extensive image simulations to assess the performance of KSB depending on the observing conditions and background source ellipticity, magnitude and size distributions. We adopt the size-- and signal-to-noise--dependent multiplicative bias correction obtained by \cite{hoekstra15}. Instead of correcting each source's measured shape, we apply an average correction to each data point (which is an average over thousands of sources), since the latter is more robust to uncertainties in the intrinsic ellipticity distribution \citep{hoekstra15}. In the next section we take a detailed look at possible sources of bias in our shape measurements.

Due to lensing, sources are magnified as well as sheared, and this may alter the inferred source density, affecting the boost correction discussed in \Cref{s:boost}. The increase in flux boosts the number counts relative to an unlensed area of the sky, but the decrease in effective area works in the opposite direction. The net effect depends on the intrinsic distribution of source galaxies as a function of magnitude, and cancels out for a slope $d\log N_\mathrm{source}/d\rmag=0.40$ \citep{mellier99}. In fact, this slope is 0.38--0.40 for the MegaCam $r$-band data \citep{hoekstra15}, so we can safely ignore magnification in our analysis.

In order to account for the measurement uncertainties in defining the quality of our lensing data, throughout this work we use the source \emph{weight} density. We define the weight density,
\begin{equation}
  \xi_\mathrm{s}\equiv(1/A)\sum_i w_i \,,
\end{equation}
as the sum of the shape measurement weights (\Cref{eq:weight}) per square arcminute.


\section{Source sample and shear calibration}\label{s:calibration}

We now explore the impact of cluster galaxies in our analysis, as they contaminate our source sample and in some cases bias shape measurements through blending of their light with that of source galaxies. In order to assess the impact of cluster galaxies in the shear measurement pipeline, we use dedicated sets of image simulations. We extend the image simulations produced by \cite{hoekstra15} by introducing simulated cluster galaxies into the images of source galaxies. We create two sets of image simulations with different cluster galaxies to investigate different features of the analysis pipeline, as described in the following sections.

The image simulation pipeline of \citet{hoekstra15} creates mock images of the MegaCam instrument with randomly placed source galaxies. In short, these simulated galaxies have properties based on \galfit\ \citep{peng02} measurements of galaxies in the GEMS survey \citep{rix04}. The modulus of the ellipticity is drawn from a Rayleigh distribution with a width of 0.25 and truncated at 0.9, and galaxies are assigned random position angles.  \Cref{f:magsize} shows the distribution of magnitudes and sizes measured with \galfit\ of \meneacs\ cluster galaxies \citep[from][]{sifon15_cccp}. We use these measurements to simulate lens galaxies which we add to the simulations of source galaxies. The surface brightness profiles of galaxies are drawn, assuming their light follows \cite{sersic68}  radial profiles, using the \textsc{GalSim} software \citep{rowe15}.

\subsection{Sensitivity to background subtraction}\label{s:delmag}

Before discussing the impact of cluster galaxies in the source sample and shape measurements, we describe a bias pertaining to the shape measurement pipeline itself. The pipeline proceeds in two steps: the first is to detect sources using a global background estimation, while the second is to measure the shapes of these detected objects. In the second step, a local background level is determined by measuring the root mean square brightness in an annulus with inner and outer radii of 16 and 32 pixels respectively, after masking all detected objects. This annulus is split into four quadrants. The background is modelled by fitting a plane through them, and is then subtracted from the image. This background subtraction works well in general, but when light from nearby objects is not properly accounted for, it significantly modifies the estimated magnitude of the test galaxy. Since the simulations do not have a diffuse background component, a proper background subtraction would leave the galaxy magnitude untouched. Therefore, changes in the magnitude pre- and post-background subtraction in the simulations, which we denote $\delta \rmag \equiv m_\mathrm{postbg}-m_\mathrm{prebg}$, mean that the shape measurement process is not robust for that particular galaxy. As our sources are in close proximity to bright satellite galaxies, this feature is potentially detrimental to our shear measurements. The cluster image simulations indeed contain a population of sources with large values of $\delta \rmag$, which is absent in the simulations without cluster galaxies. Comparing the simulations with- and without cluster galaxies we determined an empirical relation to flag any galaxies severely affected by the local background subtraction. We discard all source galaxies with
\begin{equation}\label{eq:delmag}
  \delta \rmag < -49.04 - 7.00\rmag + 0.333\rmag^2 - 0.0053\rmag^3 \, ,
\end{equation}
since these galaxies are outliers in the $\delta\rmag-\rmag$ plane. Inspecting the images of the galaxies thus discarded in the real data, we find that they are mostly located either near bright, saturated stars (but these galaxies would be discarded in subsequent steps by masking stellar spikes and ghosts), or close to big galaxies with resolved spiral arms or other features, that make the plane approximation of the background a bad description of the local background. We have verified that the calibration of the shape measurements by \cite{hoekstra15} remains unchanged when discarding these galaxies (which were included in their sample); this is essentially because \Cref{eq:delmag} is independent of galaxy shape. Typically, an additional 10--12\percent\ of sources in the data are flagged by \Cref{eq:delmag}.

\subsection{Additive shear bias}\label{s:ct}

\begin{figure}
 \centerline{\includegraphics[width=3.2in]{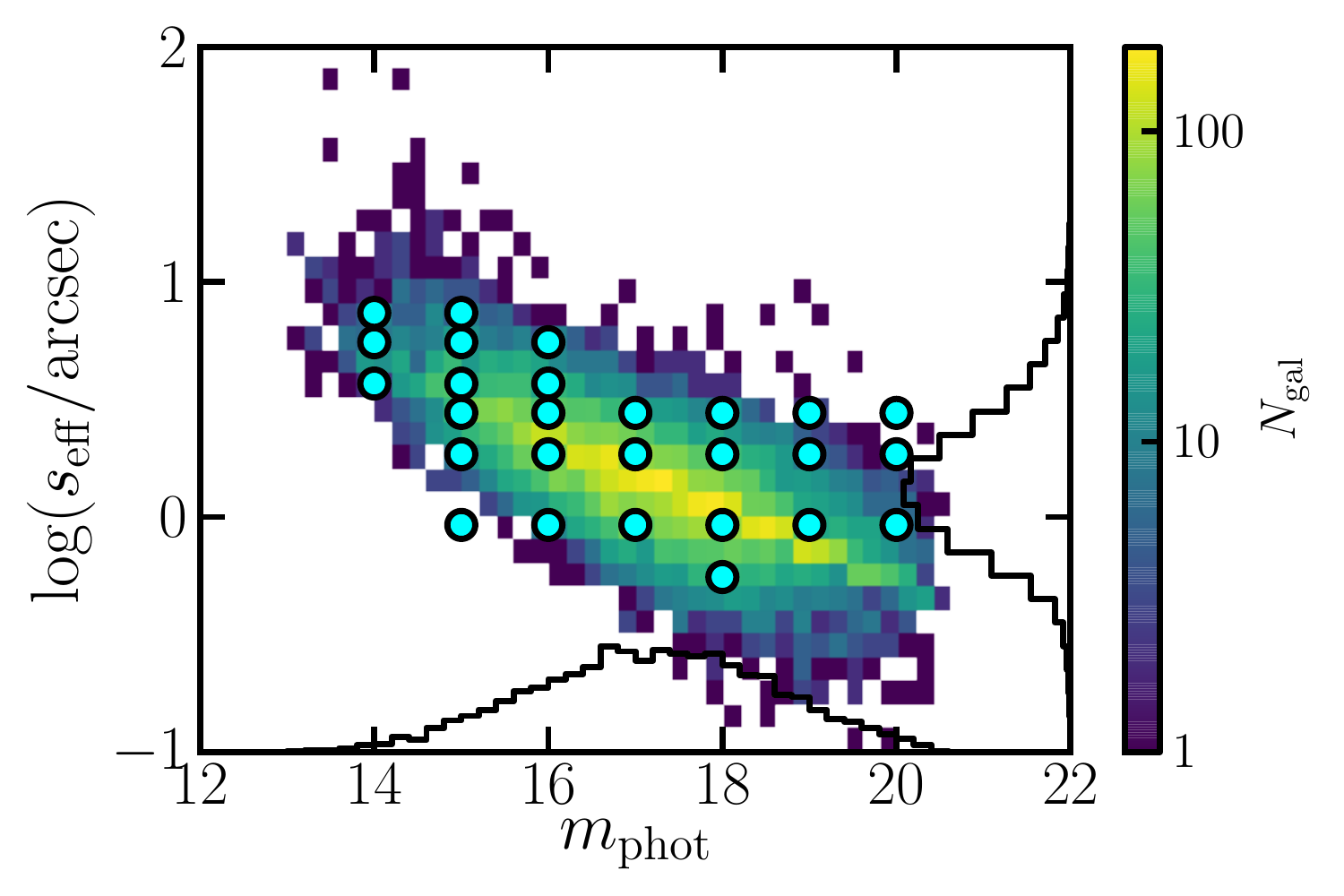}}
\caption{Distribution of magnitudes and effective radii (as measured by GALFIT) of satellites in the \meneacs\ spec+RS sample,. The logarithmic color scale shows the number of galaxies per two-dimensional bin, while black histograms show the one-dimensional distributions. Cyan circles show the coordinates used in the grid image simulations used to determine the additive bias on the shape measurements in \Cref{s:ct}. Note that galaxies at the bottom-right corner of the distribution, not covered by the simulations, are faint and small and therefore can be safely assumed to produce no obscuration (see \Cref{ap:ct}).}
\label{f:magsize}
\end{figure}

In galaxy-galaxy lensing (and equivalently cluster lensing), source shapes are azimuthally averaged around the lenses. This  washes out any spatial PSF anisotropy, and the additive bias $c$ in \Cref{eq:gammaobs} can be neglected. (In other words, additive biases in $\gamma_1$ and $\gamma_2$ vanish when projected onto $\gammat$.) However, our measurements are focused on the immediate surroundings (tens of arcseconds to few arcminutes) of thousands of luminous lenses, such that galaxy light may bias the shape measurements of fainter background sources. Given that the light profile always decreases radially, the azimuthal averaging can in fact introduce an additive bias in $\gammat$ (as opposed to $\gamma_{1,2}$) by biasing the background subtraction along the radial direction. We refer to this additive bias in $\gammat$ as $\ct$ hereafter.

We expect the bias to depend on the size and magnitude of cluster galaxies and therefore create image simulations to determine this relation. We selected a set of magnitudes and sizes representative of the full sample of cluster members (shown as cyan circles in \Cref{f:magsize}) and simulated lens galaxies with those properties. In order to accurately estimate $\ct$, we simulate large numbers of galaxies with the same magnitude and size, placed in a regular grid in the image simulations, separated by at least 60$\arcsec$ to avoid overlap between the lenses. We refer to these simulations as `grid image simulations'. The PSF in these simulations is circular with a full width at half maximum of 0.$\!\arcsec$67. We generate a large number of grid image simulations spanning a range of lens size and $r$-band magnitude and measure the average shear around these simulated lenses, which is by construction zero in the source-only image simulations.

\begin{figure}
 \centerline{\includegraphics[width=3.2in]{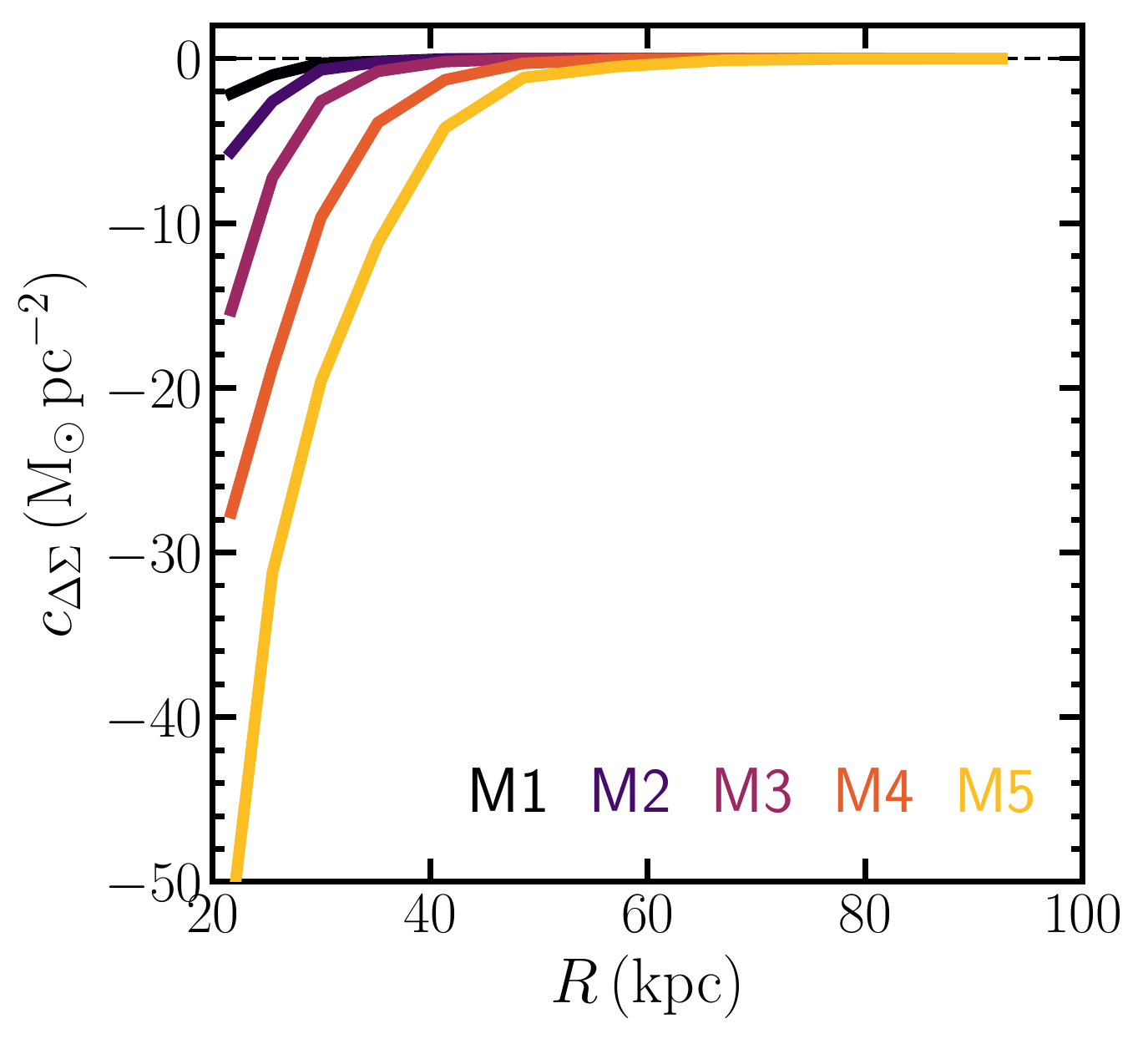}}
\caption{Average tangential additive bias, $c_{\ESD} \equiv \Sigma_\mathrm{c} \ct$, for the five stellar mass bins studied in \Cref{s:shsmr}, from low (M1) to high (M5) stellar mass (see \Cref{t:bins}). Note the smaller extent of the horizontal axis compared to other similar figures.  }
\label{f:ct}
\end{figure}

In \Cref{ap:ct} we show that we can model this (negative) bias as a function of lens-source separation, lens magnitude and size, and we correct the shear measured for each lens-source pair for this bias. For illustration, we show in \Cref{f:ct} the average $\Sigma_\mathrm{c}\ct$ obtained from the image simulations after weighting the results in the simulations by the two-dimensional distribution of real galaxies in $r$-band magnitude and size, when binning \meneacs\ galaxies into five stellar mass bins (see \Cref{s:shsmr}). As expected, the correction is larger for more massive galaxies, which are on average larger. At $R\sim20$ kpc (i.e., the smallest scales probed), the correction is 20--30\percent\ and is negligible at $R\sim50$ kpc. We find that on average $\ct$ is approximately independent of cluster-centric distance, because there is no strong luminosity segregation of galaxies in clusters as massive as those in \meneacs\ \citep[e.g.,][]{roberts15}. For reference, a fraction of order $10^{-6}$ lens-source pairs have $\lvert\ct\rvert>0.01$, which corresponds to the typical shear produced by massive cluster galaxies in our sample. We remove these lens-source pairs from our analysis, since such corrections are most of the time larger than the signal itself, although such a small fraction of lens-source pairs has no effect on our results. We find that lens galaxies with $s_\mathrm{eff}<1 \arcsec$ produce no noticeable obscuration at the scales of interest (for details see \Cref{ap:ct}), and we therefore did not produce simulations for the smallest lenses (\Cref{f:magsize}).

\subsection{Contamination by cluster members}
\label{s:boost}

\begin{figure}
 \centerline{\includegraphics[width=3.4in]{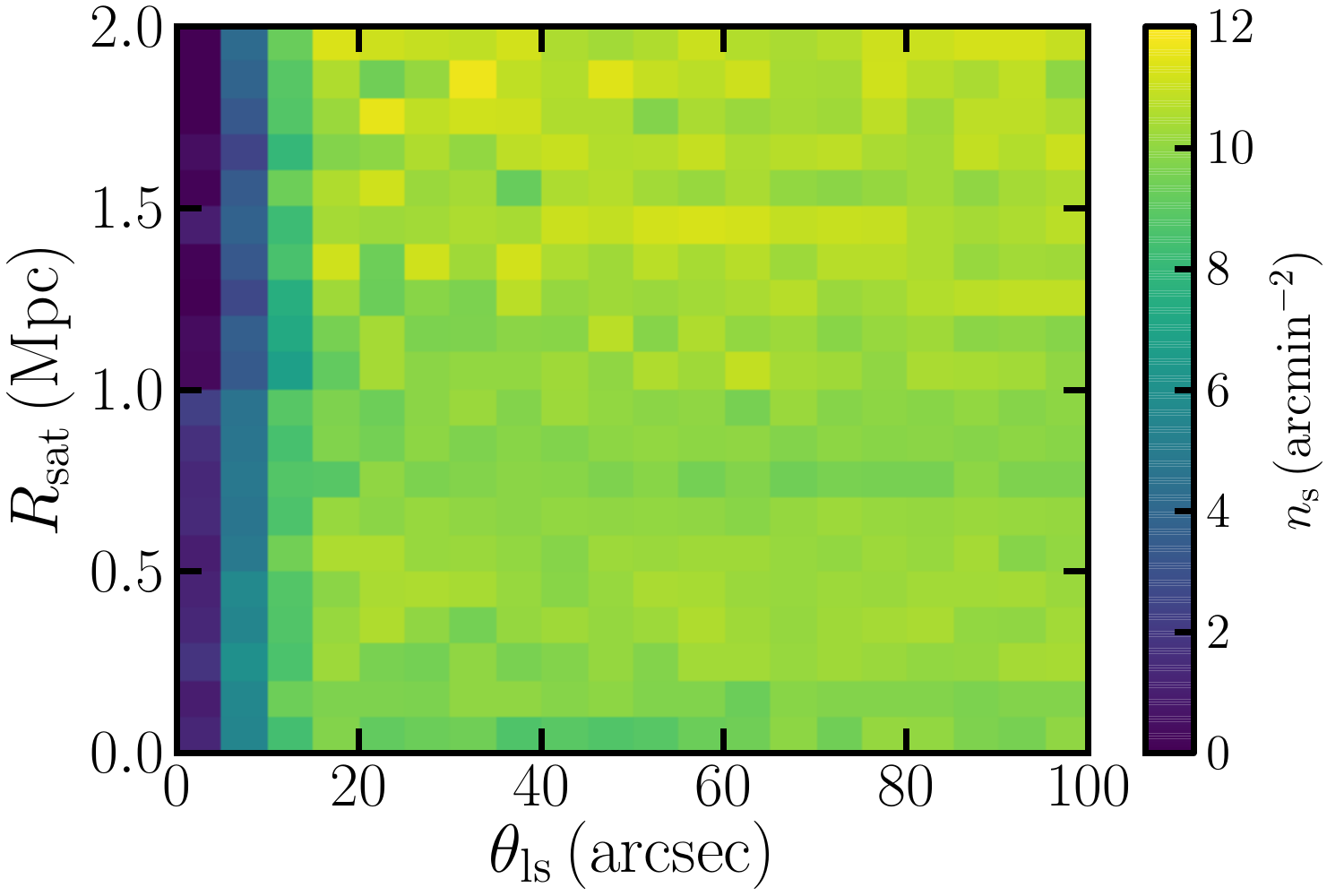}}
\caption{Observed number density of background sources as a function of lens-source separation, $\thetals$, and distance from the lens to the cluster centre, $R_\mathrm{sat}$, for all MENeaCS clusters, after applying all the cuts described in \Cref{s:data}.}
\label{f:nsources_grid}
\end{figure}

In addition to the additive bias discussed above, lens galaxies affect the source density in their vicinity for two reasons: big lenses act as masks on the background source population, while small ones enter the source sample. We refer to these effects as obscuration and contamination, respectively. 

Since on average cluster galaxies are randomly oriented \citep{sifon15_cccp}, contamination by cluster members biases the (positive) lensing signal low; the correction for this effect is usually referred to as the `boost factor' \citep[e.g.,][]{mandelbaum05_errors}. Obscuration, in turn, has two effects: it reduces the statistical power of small-scale measurements, and it complicates the determination of the contamination correction, since the observed source density is affected by obscuration. \Cref{f:nsources_grid} shows the number density of sources as a function of lens-source separation and cluster-centric distance. The obscuration of soure galaxies is evident: the source density decreases rapidly at $\thetals\lesssim10''$, while it remains essentially constant over the rest of the $\thetals-\Rsat$ plane. The effect of contamination is not so readily seen (i.e., the source density is approximately constant for varying $\Rsat$), because of the low redshift of our clusters: cluster galaxies are sufficiently separated on the sky that they do not appreciably boost the source density if obscuration is not accounted for.

\subsubsection{Correcting the observed source density profile for obscuration}\label{s:boost_obsc}

To measure the obscuration by cluster members, we generate a new set of image simulations, in which the spatial distribution of lens galaxies in the observations is reproduced and each lens galaxy is simulated with its measured properties. In this way a realistic simulation of each observed \meneacs\ cluster is created. We refer to these image simulations as `cluster image simulations'.

The cluster image simulations were designed to mimic the data as closely as possible to accurately capture the obscuration produced by \meneacs\ cluster members. We used the image simulation pipeline of \cite{hoekstra15} to create images of the source population with the same seeing and noise level measured from the data for each cluster. We then created images with the same properties, including a foreground cluster. Where available, we used the \galfit\ measurements of \cite{sifon15_cccp} to create surface brightness profiles for cluster members. For cluster members without reliable \galfit\ measurements (which constitute approximately 10\percent\ of the simulated cluster galaxies, and are mostly on the faint end of the population) we draw random values following the distribution of morphological parameters for galaxies with similar magnitude and redshift. Although individual galaxies may not be accurately represented in the simulations, the average obscuration should be well captured. We include all spectroscopic and red-sequence member galaxies down to an apparent magnitude $\rmag=23$ and to $\Rsat=3$ Mpc. As shown by \cite{sifon15_cccp}, the red sequence is severely contaminated at such large distances. As we show below, this `interloper' population can be easily accounted for, since the density of interlopers is not a function of cluster-centric distance.

\begin{figure}
 \centerline{\includegraphics[width=\linewidth]{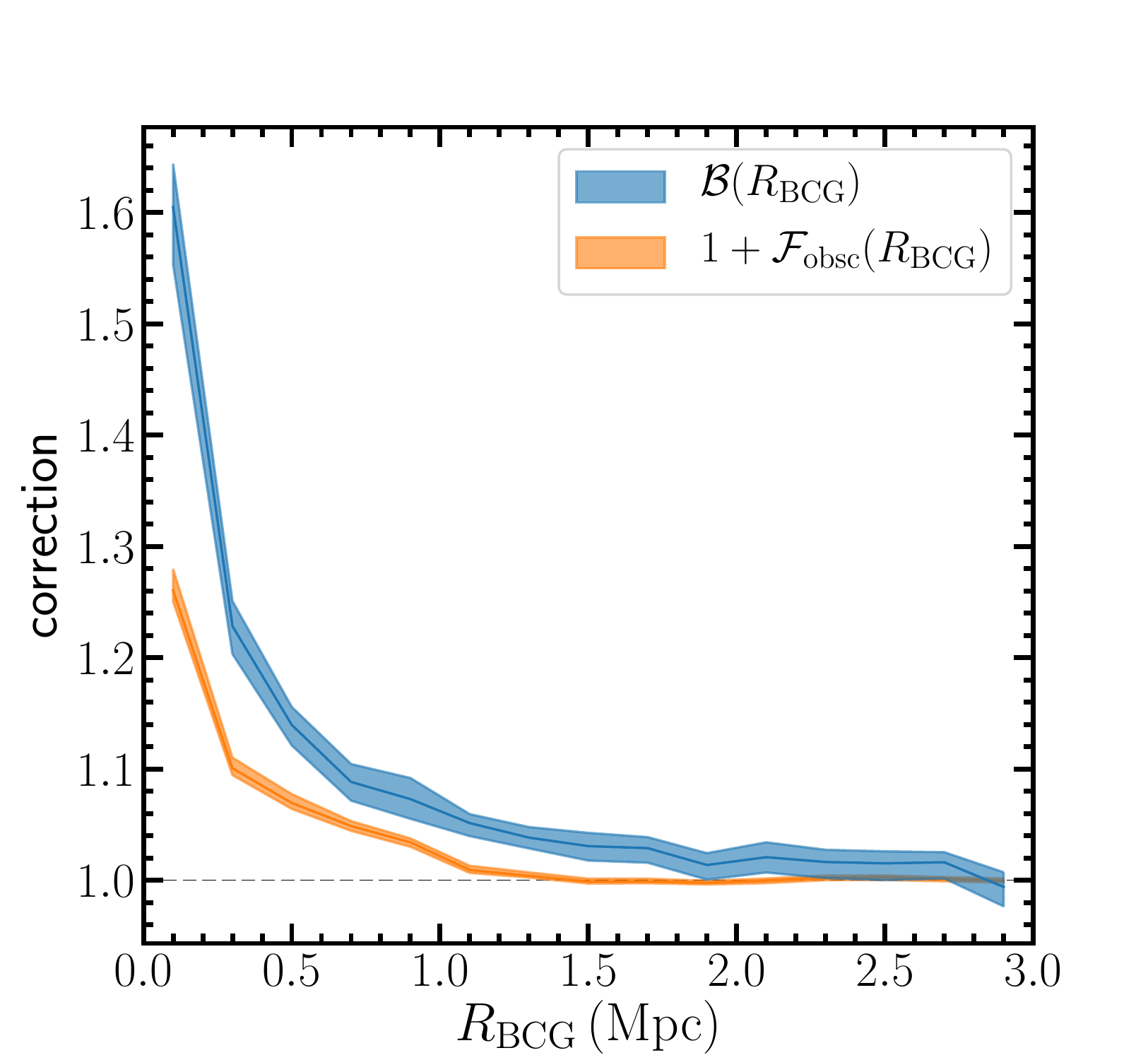}}
\caption{Obscuration correction (orange) and obscuration-corrected contamination correction (i.e., boost factor, blue) as a function of cluster-centric distance. Both quantities are averages over all clusters. The width of each curve shows the uncertainty on the mean correction. 
}
\label{f:boost}
\end{figure}

We use the cluster image simulations to calculate the average obscuration produced by cluster galaxies by measuring the source density as a function of cluster-centric radius. Because in these simulations we reproduce the spatial distribution of cluster galaxies, we can account for the radial dependence of the obscuration, given the number density profile of cluster galaxies. We show in \Cref{f:boost} the average obscuration profile, defined as
\begin{equation}
  \mathcal{F}_\mathrm{obsc}(R_\mathrm{BCG}) \equiv 
        \frac{\xi_\mathrm{s,cluster}(R_\mathrm{BCG})}{\xi_\mathrm{s,background}} \,,
\end{equation}
where $\xi_\mathrm{s}$ is the source weight density, and the subscripts ``cluster'' and ``background'' refer to the image simulations with and without the cluster galaxies, respectively.

In fact, the obscuration at large cluster-centric distance is not exactly zero, but reaches a constant value $\hat{\mathcal{F}}(R_\mathrm{BCG}>1.5\mathrm{Mpc})\simeq0.06$ (where the hat symbol simply denotes a biased measurement of the true $\mathcal{F}(R)$). This is because, to ensure completeness, the image simulations include all red sequence galaxies, which inevitably includes a contaminating population of galaxies that are in fact not part of the cluster, especially at large $R_\mathrm{BCG}$ \citep{sifon15_cccp}. We account for this excess obscuration by contaminating galaxies by simply subtracting the large-scale value of $\hat{\mathcal{F}}(R_\mathrm{BCG})$, which results in the curve shown in \Cref{f:boost}.

\subsubsection{Boost correction}
\label{s:blank}

Because the source sample is both obscured and contaminated by cluster galaxies, we need an external measurement of the reference source density. Furthermore, because the bulk of our sample is at $z<0.1$, the Megacam field of view is not enough to estimate cluster-free source densities---our images only reach $\Rsat\sim3$ Mpc at $z=0.1$. Therefore, we retrieved data for 41 blank fields from the Megacam archive \citep{gwyn08}, which provides an area of approximately 33 \sqdeg\ after manual masking. These blank fields contain no galaxy clusters and have noise and seeing properties at least as good as the MENeaCS data. We construct the source sample and shape catalogue exactly as described above, after degrading the blank field observations to the typical noise level of MENeaCS data (see Herbonnet et al.\ in prep.).

\begin{figure}
 \centerline{\includegraphics[width=0.95\linewidth]{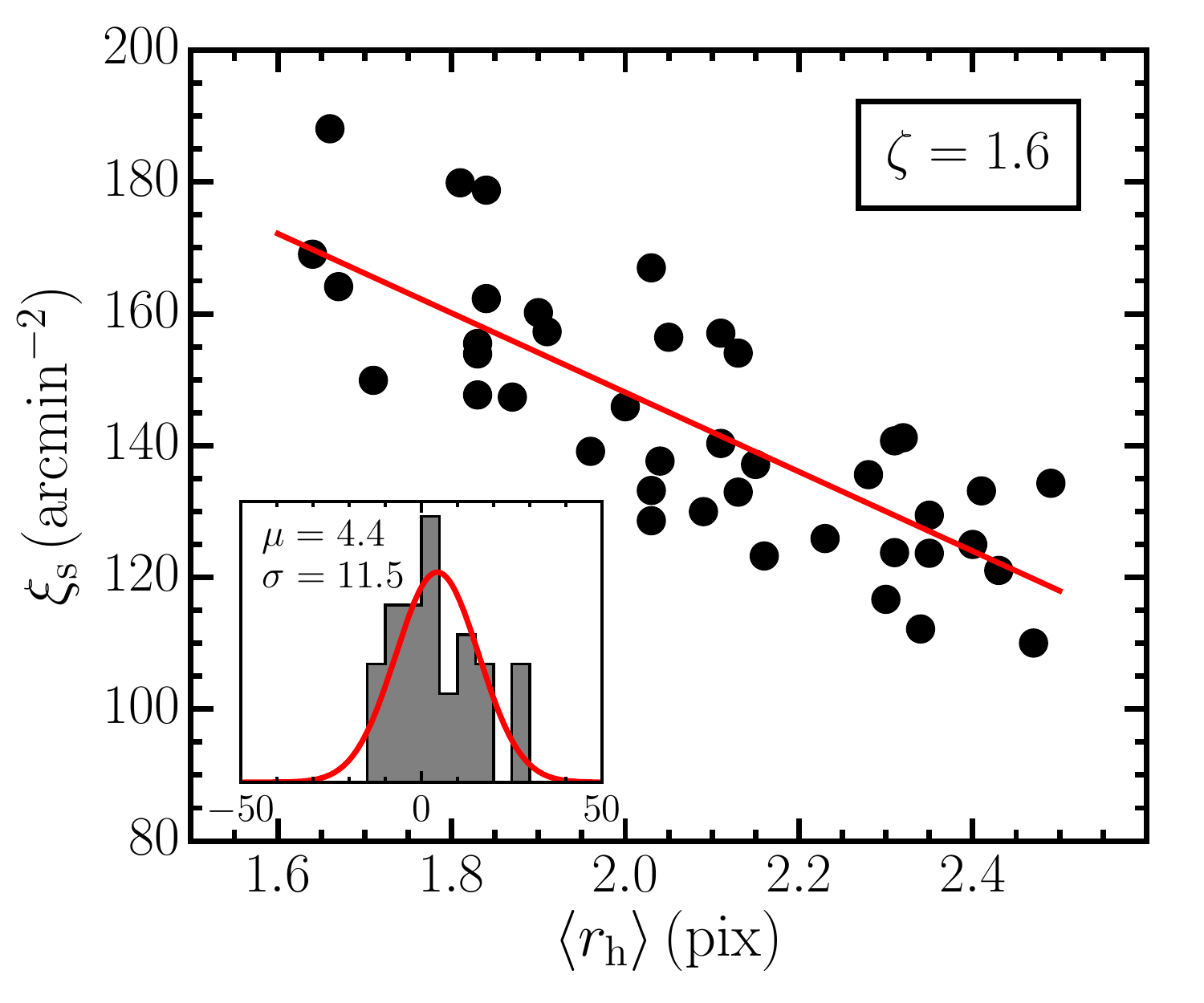}}
\caption{Total source weight density as a function of half-light radius of stars in each of the blank Megacam fields, with the background level artificially increased to 1.6 counts per pixel (the mean noise level in MENeaCS) and assuming no Galactic extinction for illustration. The red solid line shows the best-fitting function described by \Cref{eq:blank}. The inset shows a histogram of the residuals in $\xi_\mathrm{s}$ about the best-fit, with the best-fit Gaussian distribution in red, and the legend reports the mean ($\mu$) and standard deviation ($\sigma$) of this distribution.
}
\label{f:blank}
\end{figure}

As described in Herbonnet et al.\ (in prep.), we fit the the blank field source weight densities, $\xi_\mathrm{s,blank}$, as a linear combination of the image quality (quantified by the average half-light radius of stars, $\avgrhstar$), the background noise level, $\zeta$, and the Galactic extinction in the $r$-band, $A_r$,
\begin{equation}\label{eq:blank}
 \langle\xi_\mathrm{s,blank}\rangle(\avgrhstar,\zeta,A_r) = 
     p_1\avgrhstar + p_2\zeta + p_3 A_r + p_4\, ,
\end{equation}
where $\zeta$, $\avgrhstar$ and $A_r$ are in units of counts per pixel, pixels, and magnitudes respectively, and $p_i = (-68.4, -40.6, -122.8, 364.2)$  are the best-fit parameters. The blank field measurements are well described by a normal distribution around \Cref{eq:blank}, with a constant scatter of 12 weight-units per sq.\ arcmin, as shown in \Cref{f:blank}. We adopt the noise-, extinction-, and seeing-dependent source density measured in the blank fields as the background level for each of the MENeaCS clusters. We have checked that at the high redshift end of our sample, the source densities at the outskirts of clusters ($R_\mathrm{BCG}\gtrsim3$ Mpc) are consistent with the expectations from the blank fields. The limiting factor to the precision of the blank field source density prediction is the number of blank fields. For the available 41 fields, the relative uncertainty in the blank field prediction is 1.0\percent, which is precise enough for our analysis.

Having computed the obscuration from the image simulations and the contamination by comparing with blank fields, we now calculate the boost correction appropriate to our dataset. Given a source's $R_\mathrm{BCG}$, we calculate its corrected (or `true') shear through \Cref{eq:gammaobs}, where the boost correction is 
\begin{equation}\label{eq:boost}
  \mathcal{B}(R_\mathrm{BCG})
    = \left\langle 
          \frac{\xi_\mathrm{s,data}(R_\mathrm{BCG})}{\xi_\mathrm{s,blank}}
      \right\rangle
      \left\langle1 - \mathcal{F}_\mathrm{obsc}(R_\mathrm{BCG})\right\rangle^{-1}\,,
\end{equation}
where all quantities are averaged over all clusters, weighted by the number of lens-source pairs in each cluster to match the weighting applied to the average ESD. \Cref{eq:boost} assumes that faint cluster galaxies (which enter the source sample) do not cluster strongly with the bright cluster members constituting our lens sample; this small-scale cluster would introduce a dependence of $\mathcal{B}$ on $\thetals$. For reference, \cite{fang16} showed that there is an excess of galaxies in the vicinity of cluster members, but at the level of a few galaxies per cluster, which would have no impact on our results. In fact, we find no evidence of small-scale clustering in our sample of red sequence galaxies.

\subsection{More details on obscuration by cluster members}
\label{s:obscuration}

\begin{figure}
 \centerline{\includegraphics[width=\linewidth]{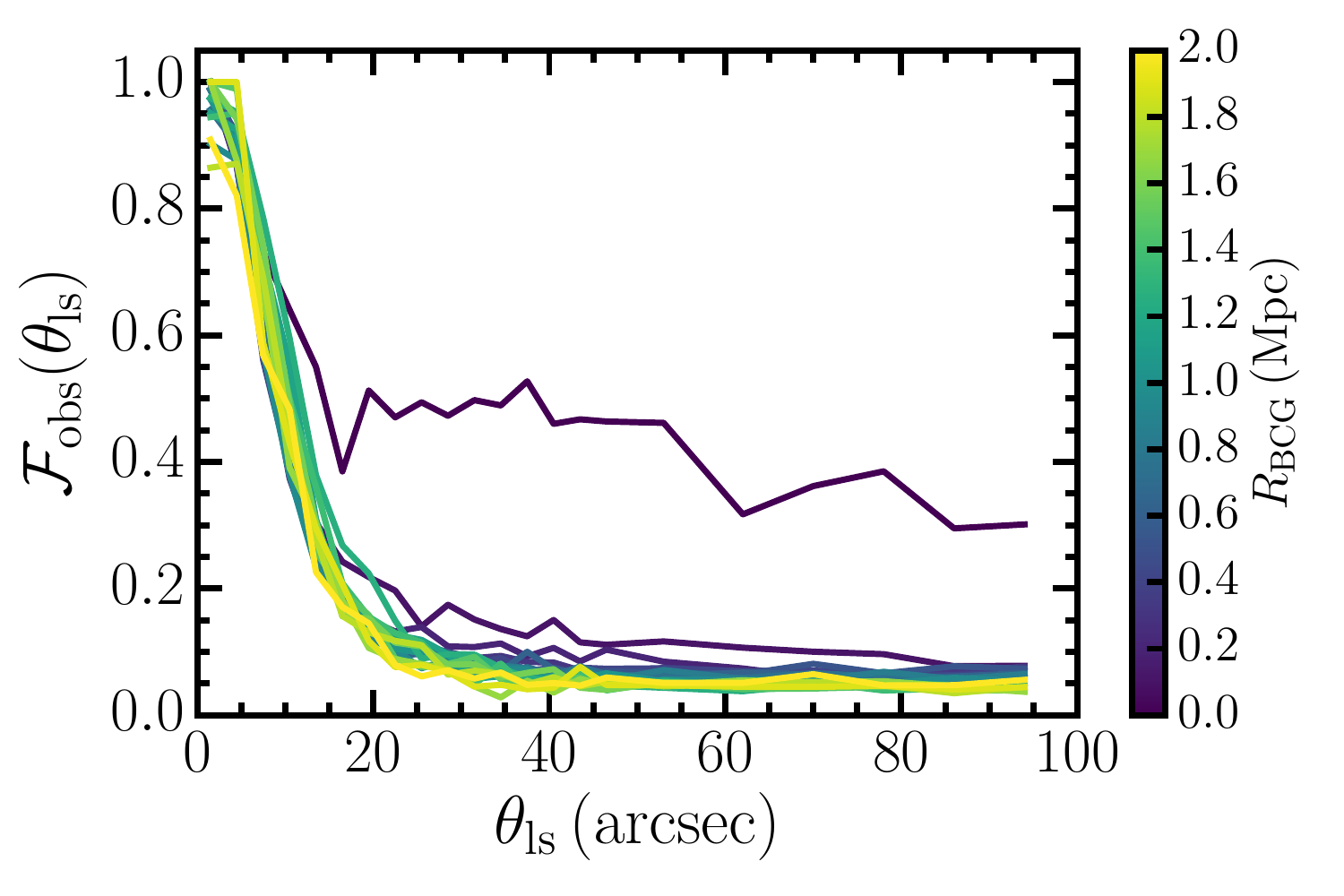}}
\caption{Obscuration profile measured in the image simulations as a function of lens-source separation, $\thetals$, in bins of cluster-centric distance, $R_\mathrm{BCG}$, averaged over all \meneacs\ clusters.}
\label{f:obscuration}
\end{figure}

In the previous section we calculated the average obscuration produced by cluster members as a function of cluster-centric distance, $R_\mathrm{BCG}$, in order to properly estimate the boost correction. In this section, we look closer at the obscuration by cluster galaxies individually rather than collectively as part of the cluster.

We calculate obscuration profiles around galaxies, $\mathcal{F}_\mathrm{obsc}(\theta_\mathrm{ls})$, in bins of cluster centric distance, $R_\mathrm{BCG}$; we show these profiles in \Cref{f:obscuration}. Because of the high \emph{lens} density at small $R_\mathrm{BCG}$, the obscuration drops only down to roughly 0.45 up to $\thetals\sim50''$, decreasing slowly at larger separations. However, the effect of neighbouring lenses is negligible at $R_\mathrm{BCG}\gtrsim200$ kpc. Note that these obscuration profiles do not affect the calculation of the boost factor, because as mentioned in the preceding section the density of cluster galaxies does not depend on $\thetals$. (Integration of this set of curves over $\thetals$ gives rise to \Cref{eq:boost}). Instead,  the steep rise in the obscuration below $\thetals\approx20$ arcsec fundamentally limits the scales accessible in this study. Pushing to smaller scales would require subtraction of the light profiles of lens galaxies, an avenue we will explore in future work.

\subsection{Source redshift distribution}\label{s:beta}

The measurement of the ESD is averaged over each lens source pair in the source population so that redshifts for individual sources are required. However, we lack the deep colour information to estimate reliable photometric redshifts for individual source galaxies. Instead, we can use an average lensing efficiency $\langle \beta \rangle = \langle \mathrm{max}[0,D_\mathrm{ls}/D_\mathrm{s} ]\rangle$ for the entire source population, which can be inferred from a representative field with a reliable redshift distribution, as a proxy for the cluster background \citep[see, e.g.,][]{hoekstra15}.

We take as a reference the COSMOS2015 catalogue \citep{laigle16}, which contains photometric redshift estimates for galaxies in the 2 square degrees COSMOS field. This catalogue is deep enough to cover our magnitude range and contains near infrared measurements that help break degeneracies in photometric redshift estimation. The COSMOS field was also targeted by a deep observation with the CFHT, from which there exists a lensing catalogue. The matched lensing-photometric redshift catalogue allows us to apply the same quality cuts on the redshift distribution as have been applied to the lensing data, which could otherwise bias the results \citep{gruen17}. The overlapping area is only 1 square degree, which raises concerns that it might be unrepresentative for our cluster fields. However, we have checked our results by calculating the redshift distribution of sources in the 4 sq.\ deg.\ CFHTLS deep survey. We confirm that the uncertainties on our mean lensing efficiency, $\langle \beta \rangle$, including cosmic variance, are less than 2\percent. Such precision is sufficient for our analysis. For more details see Herbonnet et al.\ (in prep.).

The assumption of using only the average value $\langle \beta \rangle$ and ignoring the width of the distribution introduces a bias into our measurement of $\Delta\Sigma$ \citep{hoekstra00}. However, for our low redshift clusters the effect is expected to be very small. With our photometric redshift catalogue and equation 7 from \cite{hoekstra00} we estimate that this bias is at most 1+0.06$\kappa$ (where $\kappa$ is the lensing convergence, and $\kappa\ll1$ in the weak lensing regime). This is a negligible bias compared to other sources of uncertainty and we therefore ignore it in the rest of our analysis.

\subsection{Resulting lensing signal}

\begin{figure*}
 \centerline{\includegraphics[width=\textwidth]{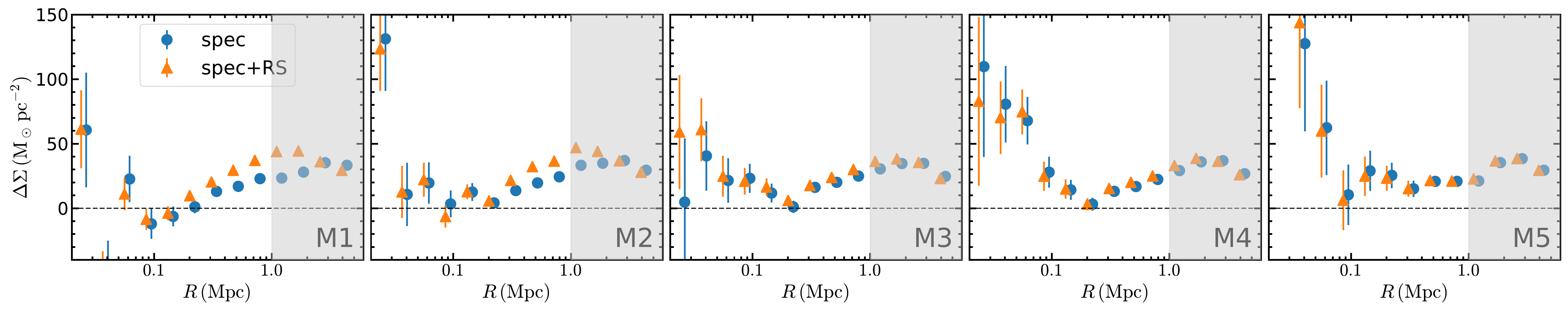}}
\caption{Excess surface mass density (ESD) of satellite galaxies binned by stellar mass. Blue circles and orange triangles show the ESD of the spectroscopic and spectroscopic-plus-red sequence samples, respectively. Errorbars are the square roots of the diagonal terms of the covariance matrices. The dashed horizontal line shows $\ESD=0$ for reference. In our analysis we only use data points up to 1 Mpc, shown over a white background.}
\label{f:esd_spec_rs}
\end{figure*}

\Cref{f:esd_spec_rs} shows the resulting lensing signal from satellites in \meneacs\ clusters, corrected for both $\ct(\thetals)$ and $\mathcal{B}(R)$. We make the distinction in the arguments of both corrections because the former is applied to each lens-source pair, while the latter is applied as an average correction after stacking all lenses in each bin. We compare the ESDs of the five bins in satellite stellar mass for the spec and spec+RS samples. There are two differences in the signal measured for both samples. Firstly, the signal from the spec+RS sample is slightly lower than the signal from the spec sample at the smallest scales. This is expected, as in general the more massive galaxies have been targeted in the spectroscopic observations; this is reflected also in the average stellar masses listed in \Cref{t:bins}. Secondly, the spec+RS signal is larger at intermediate scales, which is a reflection of the fact that spectroscopic observations tend to be incomplete at the dense centres of clusters, so the average cluster-centric distance of the spec+RS sample is lower. We base our analysis on the spec+RS sample, which is a more complete lens sample.

At intermediate scales, $0.3\lesssim R/{\rm Mpc}\lesssim2$, the two samples produce different signals. In particular, the signal from the spec+RS sample is higher. This is a consequence of the fact that we only include red sequence galaxies out to 1 Mpc, so the spec+RS sample is on average closer to the cluster centre than the spec sample. Therefore, the peak of the host cluster signal happens at smaller $R$. Beyond the peak the two signals are consistent, because all galaxies come from the same clusters. See Figure 3 of \cite{sifon15_kids} for a graphical representation. We account for the measured radial distribution of satellites in our modelling below.


\section{Satellite galaxy-galaxy lensing model}\label{s:model}

We interpret the galaxy-galaxy lensing signal produced by subhaloes using the formalism introduced by \citeauthor{yang06} \citep[\citeyear{yang06}, see also][]{li13_ggl}, and applied to observations by \cite{li14,li16,sifon15_kids,sifon18_udg} and \cite{niemiec17}. This formalism assumes that measurements are averages over a large number of satellites \emph{and} clusters, such that the stacked cluster is (to a sufficient approximation) point-symmetric around its centre and well-described by a given parametrization of the density profile. A similar method was introduced by \cite{pastormira11}, which however does not rely on such parametrization by virtue of subtracting the signal at the opposite point in the host cluster. A different approach is to perform a maximum likelihood reconstruction of the lensing potential of cluster galaxies accounting for the cluster potential, which should either be well known a priori \citep[e.g.,][]{natarajan97,geiger98} or modelled simultaneously with the cluster galaxies \citep{limousin05}. This method has been applied in several observational studies \citep[e.g.,][]{natarajan98,natarajan09,limousin07}. We discuss results from the literature using either method after presenting our analysis, in \Cref{s:segregation}. In the following we describe our modelling of the satellite galaxy-galaxy lensing signal.

The ESD measured around a satellite galaxy is a combination of the contributions from the subhalo (including the galaxy itself) at small scales, and that from the host halo at larger scales,
\begin{equation}\label{eq:esd_tot}
 \ESD_\mathrm{sat}(R) =
      \ESD_\star(R\vert\Mstar) + \ESD_\mathrm{sub}(R\vert\Msub,c_\mathrm{sub}) + 
      \ESD_\mathrm{host}(R\vert\Mh,c_\mathrm{h}),
\end{equation}
where $\ESD_\star$ represents the contribution from baryons in the satellite galaxy, which we model as a point source contribution throughout, such that
\begin{equation}
 \ESD_\star(R \vert \Mstar) = \frac{\Mstar}{\pi R^2}.
\end{equation}
Here, we take $\Mstar$ to be the median stellar mass of all satellites in the corresponding sample (e.g., a given bin in satellite luminosity). In \Cref{eq:esd_tot}, $R$ refers to the lens-source separation in physical units; $\Msub$ is the average subhalo mass (see below) and $c_\mathrm{sub}$ its concentration; and $\Mh$ and $c_\mathrm{h}$ are the average mass and concentration of the host clusters. In the remainder of this section we describe the other two components in \Cref{eq:esd_tot}. Detailed, graphical descriptions of these components can be found in \cite{yang06}, \cite{li13_ggl} and \cite{sifon15_kids}.

\subsection{Host cluster contribution}
\label{s:host}

In numerical simulations, the density profiles of dark matter haloes are well described by a Navarro-Frenk-White \citep[NFW,][]{nfw95} profile,
\begin{equation}\label{eq:nfw}
 \rho_\mathrm{NFW}(r) = \frac{\delta_c\,\rho_\mathrm{m}}{r/\rs\left(1+r/\rs\right)^2},
\end{equation}
where $\rho_\mathrm{m}(z)=3H_0^2(1+z)^3\Omega_\mathrm{m}/(8\pi G)$ is the mean density of the Universe at redshift $z$ and
\begin{equation}\label{eq:delta_c}
 \delta_c = \frac{200}{3}\frac{c^3}{\ln(1+c) - c/(1+c)}.
\end{equation}
The two free parameters, $\rs$ and $c \equiv r_{200}/\rs$, are the scale radius and concentration of the profile, respectively. Stacked weak lensing measurements have shown that this theoretical profile is a good description, on average, of real galaxy clusters as well \citep{oguri12,umetsu16}. We therefore adopt this parametrization for the density profile of the host clusters.

The concentration parameter is typically anti-correlated with mass. This relation, referred to as $c(M)$ hereafter, has been the subject of several studies \citep[e.g.,][]{bullock01,duffy08,maccio08,prada12,dutton14}. Most of these studies parametrize the $c(M)$ relation as a power law with mass (and some with redshift as well), with the mass dependence being typically very weak. Since our sample covers relatively narrow ranges in both quantities (i.e., cluster mass and redshift), the exact function adopted is of relatively little importance. We therefore parametrize the mass-concentration relation as a power law with mass,
\begin{equation}\label{eq:cMhost}
 c_\mathrm{h}(M_\mathrm{200,h}) = a_\mathrm{c} 
\left(\frac{M_\mathrm{200,h}}{10^{15}\Msun}\right)^{b_\mathrm{c}}
\end{equation}
where $M_\mathrm{200,h}$ is the host halo mass within $r_\mathrm{200,h}$, and $a_\mathrm{c}$ and $b_\mathrm{c}$ are free parameters. We follow \cite{sifon15_kids} and account for the observed separations between the satellites and the cluster centre (which we assume to coincide with the BCG) in each observable bin to model the total host halo contribution to \Cref{eq:esd_tot}.

\begin{figure*}
 \centerline{\includegraphics[width=\textwidth]{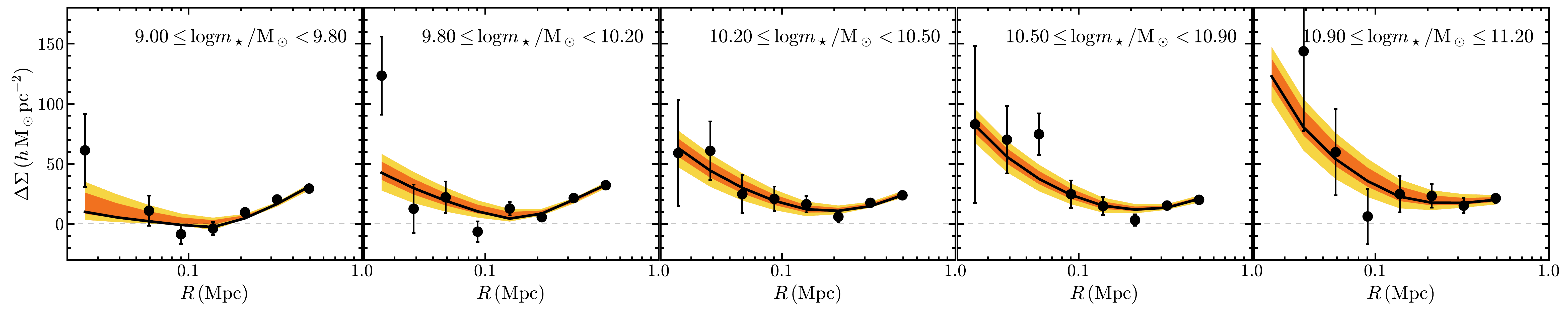}}
\caption{Excess surface mass density of the spec+RS sample, binned by stellar mass as shown in the legends (same as the orange triangles in \Cref{f:esd_spec_rs}). The black line shows the best-fitting model from the MCMC and the orange and yellow regions outline the 68 and 95 per cent credible intervals.}
\label{f:esd_logmstar}
\end{figure*}

\subsection{Subhalo contribution}\label{s:subhalo}

Although in numerical simulations satellite galaxies are heavily stripped by their host cluster, the effect on their density profile is not well established. For instance, \cite{hayashi03} found that, although tidal stripping removes mass in an outside-in fashion, tidal heating causes the subhalo to expand, and the resulting density profile is similar in shape to that of a central galaxy (which has not been subject to tidal stripping). Similarly, \cite{pastormira11} found that the NFW profile is a better fit than truncated profiles for subhaloes in the Millenium Simulation \citep{springel05_millenium}, and that the reduction in mass produced by tidal stripping is simply reflected as a change in the NFW concentration of subhaloes, which have roughly a factor 2--3 higher concentration than host haloes, consistent with the mass-concentration relation for subhaloes derived by \cite{moline17} from N-body simulations. \cite{moline17} further showed that the subhalo concentration depends on cluster-centric distance, with subhaloes closer in having a larger concentration as a result of the stronger stripping.

We therefore assume that the density profile of subhaloes can also be described by an NFW profile. We adopt the subhalo mass-concentration relation derived by \cite{moline17}, which depends on both the subhalo mass and its halo-centric distance,
\begin{equation}\label{eq:cMsub}
 \begin{split}
 c_\mathrm{sub}(m_{200},x) = 
    & 
    c_0\left(1 + \sum_{i=1}^3\left[a_i\log\left(
                 \frac{m_{200}}{10^8\,h^{-1}\Msun}\right)\right]^i \,
       \right) \\
    & \times \left[1 + b\log x\right],
 \end{split}
\end{equation}
where $x \equiv r_\mathrm{sat}/r_\mathrm{h,200}$ (defined in three-dimensional space), $c_0=19.9$, $a_i=\{-0.195,0.089,0.089\}$ and $b=-0.54$.

Note that the quantity $m_{200}$ is used for mathematical convenience only, but is not well defined physically. Instead, we report subhalo masses within the radius at which the subhalo density matches the background density of the cluster at the distance of the subhalo in question (which we denote $r_\mathrm{bg}$), and refer to this mass as $m_\mathrm{bg}$. This radius $r_\mathrm{bg}$ scales roughly with cluster-centric distance as $r_\mathrm{bg} \propto (\Rsat/r_\mathrm{200,h})^{2/3}$ \citep[see also][for a comparison between $m_\mathrm{bg}$ and $m_{200}$]{natarajan07}. The reported subhalo masses are therefore similar to those that would be measured by a subhalo finder based on local overdensities such as \textsc{subfind} \citep{springel01_cluster}, which allows us to compare our results with numerical simulations consistently.

Because the density profile is a steep function of cluster-centric distance, we take the most probable three-dimensional cluster-centric distance, $\langle r_\mathrm{sat} \rangle$, to be equal to the weighted average of the histogram of two-dimensional distances, $\Rsat$:
\begin{equation}
 \langle r_\mathrm{sat} \rangle = 
      \frac{\sum_i n(R_{{\rm sat},i})R_{{\rm sat},i}}
           {\sum_i n(R_{{\rm sat},i})},
\end{equation}
where the index $i$ runs over bins of width $\Delta\Rsat=0.1\,{\rm Mpc}$ (see \Cref{f:histograms}). We use this $\langle r_\mathrm{sat} \rangle$ in \Cref{eq:cMsub}.

\subsection{Fitting procedure}\label{s:likelihood}

We fit the model presented above to the data using the affine-invariant Markov Chain Monte Carlo (MCMC) ensemble sampler \textsc{emcee} \citep{foreman13}. This sampler uses a number of walkers (set here to 500) which move through parameter space depending on the position of all other walkers at a particular step, using a Metropolis Hastings acceptance criterion \citep[see][for a detailed description]{goodman10}. The loss function to be maximized is defined as
\begin{equation}\label{eq:likelihood}
 \begin{split}
 \mathcal{L} = \frac1{(2\pi)^{k^2/2}}
     &\prod_{m=1}^k\prod_{n=1}^k\frac1{\sqrt{{\rm det}(\mathbfss{C}_{mn})}} \\
     &\times
      \exp\left[-\frac1{2}\boldsymbol{(O-E)}^T_m\mathbfss{C}^{-1}_{mn}\boldsymbol{(O-E)}_n\right],
 \end{split}
\end{equation}
where $k=5$ is the number of bins into which the sample is split (in stellar mass or cluster-centric distance bins); $\boldsymbol{O}$ and $\boldsymbol{E}$ are the observational data vector and the corresponding model predictions, respectively; $\mathbfss{C}$ is the covariance matrix; ${\rm det}(\cdot)$ is the determinant operator; and the index pair $(i,j)$ runs over data points in each bin $(m,n)$. As implied by \Cref{eq:likelihood}, we account for the full covariance matrix, including elements both within and between observable bins.

We quote the prior ranges and marginalized posterior central values and 68\percent\ uncertainties for all free parameters in our model in \Cref{t:mcmc}, both when binning by stellar mass and by cluster-centric distance (each discussed in \Cref{s:shsmr,s:segregation}, respectively). Although we quote parameters of host clusters, we treat them as nuisance parameters throughout. Note that priors are defined in linear space, and are only quoted as logarithmic quantities in \Cref{t:mcmc} for convenience. As a result, when poorly constrained by the data, posterior host cluster masses are unrealistically high. For guidance, the values in \Cref{t:mcmc} can be compared to dynamical masses and weak lensing masses reported for the same clusters by \cite{sifon15_cccp} and Herbonnet et al.\ (in prep.), which suggest an average cluster mass $M_\mathrm{200,h}\sim6\times10^{14}\,\Msun$.


\section{The subhalo-to-stellar mass relation}\label{s:shsmr}

\begin{figure*}
 \centerline{\includegraphics[width=0.5\textwidth]{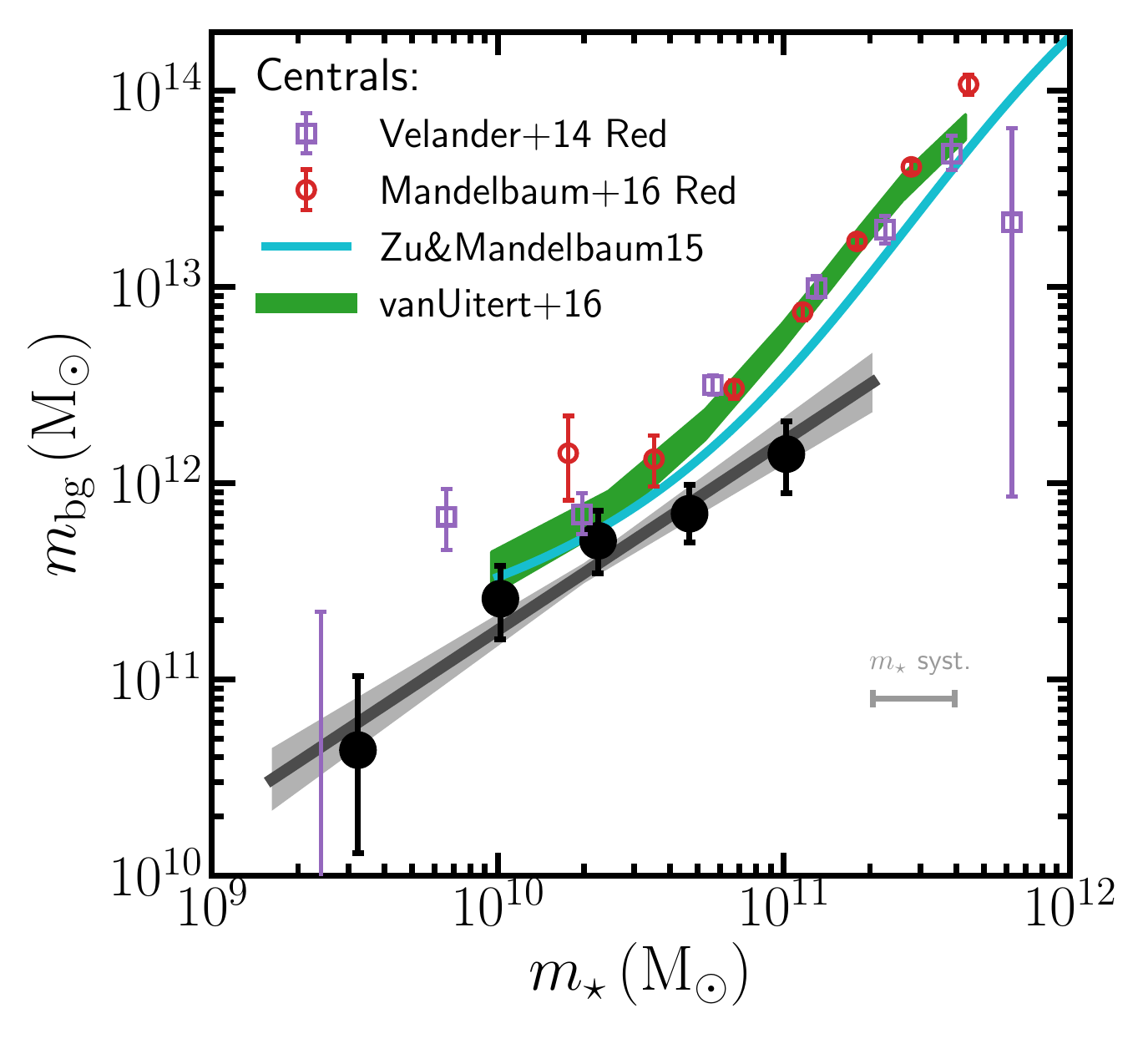}
             \includegraphics[width=0.5\textwidth]{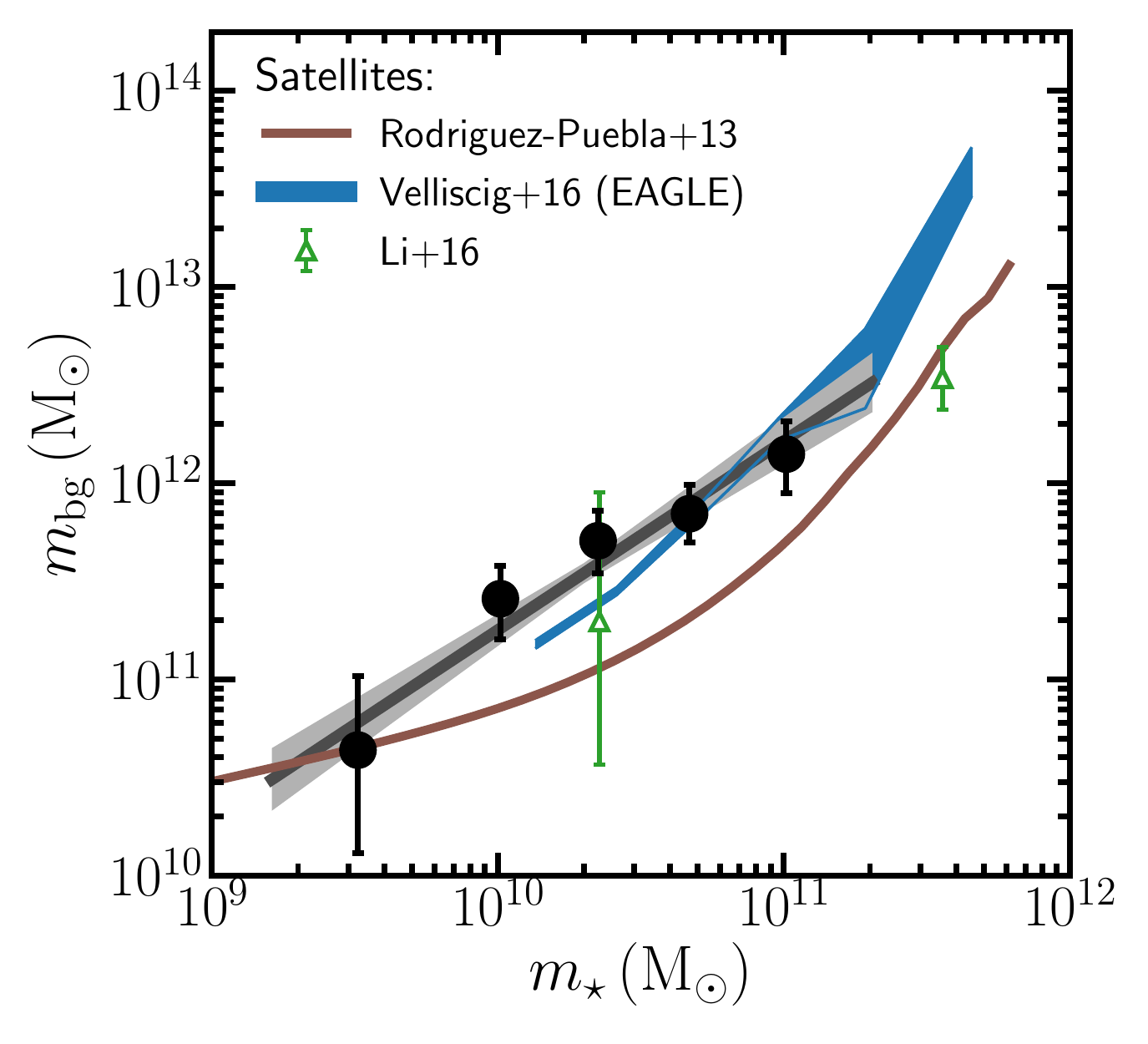}}
\caption{Stellar-to-subhalo mass relation. Big black circles in both panels correspond to the best-fit subhalo masses of spectroscopic plus red sequence satellites, assuming the subhalo mass-concentration relation of \citet{moline17}. The grey line and shaded regions show the best-fit linear relation using the BCES $X_2 \vert X_1$ estimator and the 68 per cent confidence interval on the fit, respectively. Subhalo masses refer to the mass within $r_\mathrm{bg}$ (see \Cref{s:subhalo}). The left panel shows for comparison the stellar-to-halo mass relations of central galaxies (where halo mass refers to $M_\mathrm{200,h}$) from galaxy-galaxy lensing measurements by \citet{vanuitert16} and specifically of \emph{red} central galaxies by \citet{velander14} and \citet{mandelbaum16}, plus the relation from combined lensing and clustering measurements by \citet{zu15}. The grey horizontal errorbar shows our estimate of the systematic uncertainty in stellar masses present in this comparison. The right panel shows measurements of subhalo masses as a function of stellar mass by \citet{li16} (green triangles), in addition to the subhalo-to-stellar mass relations for satellite galaxies in the EAGLE simulation \citep{velliscig17}, and   from abundance matching applied to galaxy clustering measurements by \citet{rodriguez13} as a cyan band (with the width corresponding to the error on the mean) and a brown line, respectively.
}
\label{f:shsmr}
\end{figure*}

We first bin the sample by stellar mass, as shown in the top-left panel of \Cref{f:histograms}. The ESD of the five bins, along with the best-fit model, are shown in \Cref{f:esd_logmstar}. For reference, the total $\chi^2$ is 53.2, with 27 `nominal' degrees of freedom, although we caution that neither of these values can be interpreted in the usual way given i) the non-linear nature of our model, and ii) the existence of priors that limit the model but to which the data are not subject \citep[e.g.,][]{andrae10}.\footnote{For instance, the model is not allowed to produce negative subhalo masses, while data points may well scatter to negative values due to statistical noise. This is especially relevant for low signal-to-noise ratio measurements (notice, for instance, that the second data point in the lowest-stellar mass bin is missing from the plot; its value is $\Delta\Sigma=-53\pm20\,\mathrm{M_\odot\,pc^{-2}}$). Furthermore, our use of uniform priors in mass drives the model away from zero more strongly than a uniform prior in $\log m$, for instance. The models tested in \Cref{ap:models} give similar $\chi^2$/d.o.f.\ values.}

The best-fit masses resulting from this model are shown in both panels of \Cref{f:shsmr}. We fit a power law relation\footnote{Our choice of a single power law to model the SHSMR is motivated only by our limited statistics.} between subhalo and stellar masses using the BCES $X_2 \vert X_1$ estimator, an extension of least squares linear regression which accounts for measurement uncertainties on both variables (although here we neglect uncertainties on the average stellar masses) and intrinsic scatter \citep{akritas96}, and find an approximately linear relation,
\begin{equation}\label{eq:shsmr}
 \frac{\Msub}{\Msun} = 10^{11.54\pm0.05}
                       \left(\frac{\Mstar}{2\times10^{10}\Msun}\right)^{\slope}.
\end{equation}
We remind the reader that this relation applies to the subhalo mass, $\Msub$, within the radius $r_\mathrm{bg}$ where the subhalo density equals the host halo background density. If we replace $\Msub$ with $m_{200}$, the normalization increases by a factor 3.0, while the best-fit slope is $0.97\pm0.10$, indistinguishable from that reported in \Cref{eq:shsmr}. We also tested that varying the concentration of subhaloes by 20\percent\ does not change the slope of the SHSMR.

\subsection{The SHSMR in the context of the total-to-stellar mass relation of central galaxies}

We also show in the left panel of \Cref{f:shsmr} various determinations of the relation between total and stellar mass of central galaxies from the literature \citep{velander14,zu15,mandelbaum16,vanuitert16}, where halo mass refers to $M_\mathrm{200}$.\footnote{We scale all these relations to both the value of $H_0$ and the definition of halo mass---that is, $M_{200}$ defined with respect to the average Universal density---adopted in this paper.} These have all been determined with weak lensing measurements \citep[in combination with measurements of galaxy clustering and the stellar mass function in the cases of][respectively]{zu15,vanuitert16}, and are broadly consistent with each other. Both \cite{velander14} and \cite{mandelbaum16} divided their samples into red and blue centrals, and we only show their results for red galaxies since \meneacs\ satellites are in their great majority red as well. Indeed, small differences between some of these determinations are probably driven by the different galaxy samples used in each study, and are not relevant for the present discussion. In particular, \cite{mandelbaum16} found good agreement with the model of \cite{zu15} once the galaxy samples are matched between the two studies.

Similarly, we assessed consistency of stellar masses by comparing our stellar mass estimates with those from both the MPA-JHU and NYU value-added catalogs  of SDSS galaxies \citep[respectively]{kauffmann03_mstar,blanton05_vagc}---  \citealt{zu15,mandelbaum16} adopted the latter---for the overlapping samples (1500 and 1818 galaxies in the range $10 \leq \log m_\star/\Msun \leq 11$, respectively). We find that our stellar masses are 0.10 and 0.12 dex lower, respectively, and roughly 0.1 dex lower than the \cite{taylor11} stellar masses used by \cite{vanuitert16}, which we estimated by comparing the \cite{taylor11} stellar masses to SDSS stellar masses. Unfortunately, we have no direct way of comparing our stellar masses to those of \cite{velander14}, and a detailed comparison is beyond the scope of this work. We therefore consider 0.12 dex as the nominal systematic uncertainty in stellar masses for the purpose of this comparison, which is roughly the expected value when the initial mass function is kept fixed \citep[all the cited studies adopted the IMF from][]{chabrier03}, and is consistent with numbers found by other authors \citep[e.g.,][]{coupon15,vanuitert16}.

The comparison between the central total-to-stellar mass relation and the satellite SHSMR is however not straightforward. In principle, we may consider in the case of central galaxies that $M_\mathrm{bg}=M_\mathrm{200}$, so at least the mass definitions can be regarded consistent. However, identifying the progenitors of present-day satellites is not an easy task, as there is evidence that most satellites in massive clusters today were part of smaller groups long before entering their current hosts. In the context of the decreased star formation of satellite galaxies, this is usually referred to as `pre-processing' \citep[e.g.,][]{mcgee09,gabor15,haines15}. The impact of this pre-processing on the total mass content of present-day satellites is unknown. Nevertheless, we can make a phenomenological comparison. We find that at $\log m_\mathrm{bg}\lesssim10.3$ the shape of the SHSMR coincides with that of the analogous relation for central galaxies, consistent with the prediction that galaxies lose dark matter more easily than stellar matter \citep[e.g.,][]{chang13_stripping}. We also cautiously note an increased difference in the SHSMR with the relation for centrals at high masses, $m \gtrsim10^{11}\,\Msun$, which may suggest that massive satellites lose dark matter more easily than lower-mass satellites, compared to their stellar mass loss. This may for instance be related to the fact that dynamical friction pulls more massive satellites toward the centre more efficiently (compared to lower mass galaxies), where tidal forces are stronger. If stellar mass is more resistant to tidal stripping then we would expect satellite galaxies with larger stellar masses to have a lower total-to-stellar mass ratio, as observed.

\begin{figure}
  \centerline{\includegraphics[width=\linewidth]{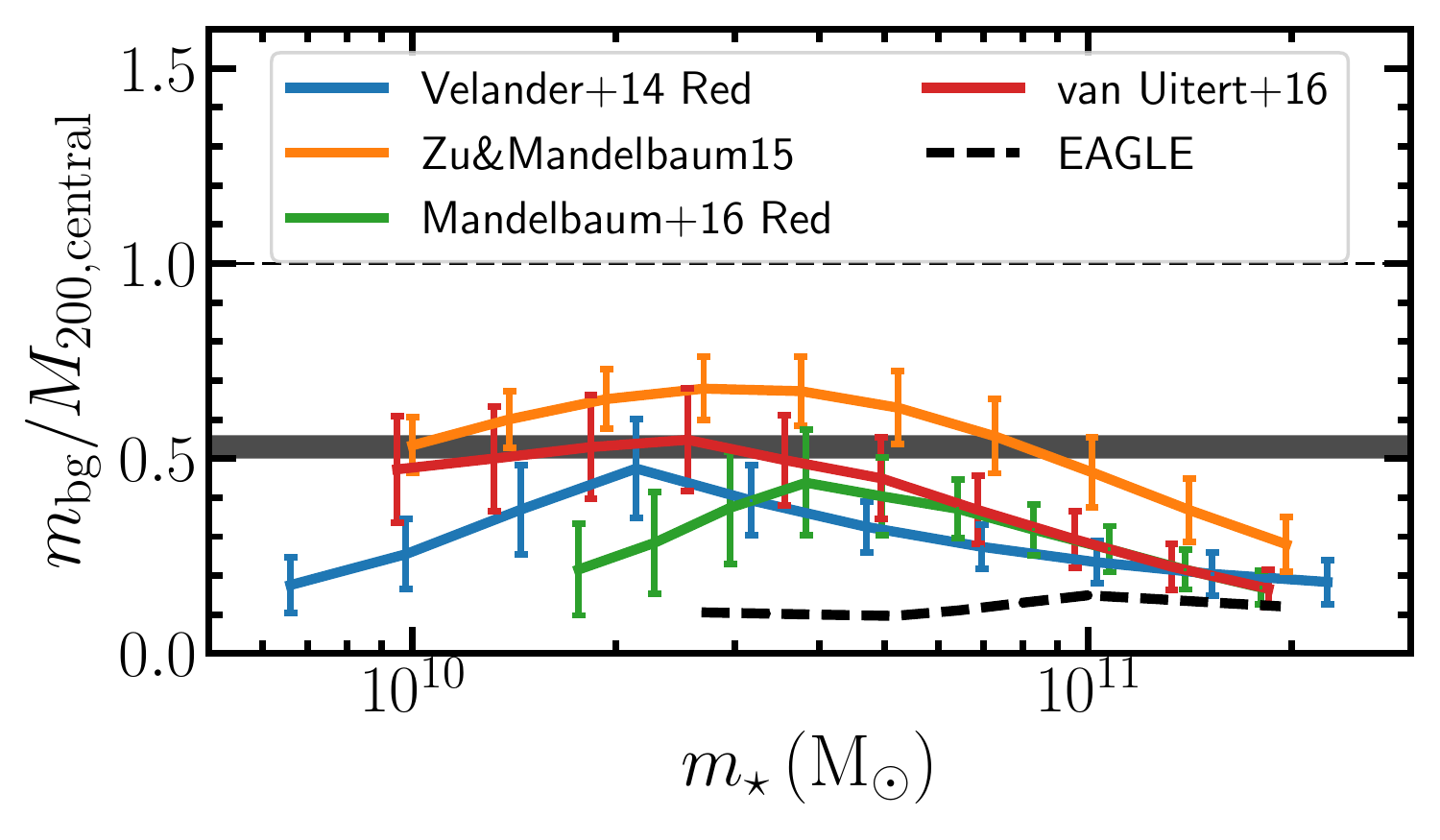}}
\caption{Ratio of subhalo masses measured in this work to central halo masses from different observational studies (solid lines) and of both quantities measured in the EAGLE simulation by \citet{velliscig17} (black dashed line), at fixed stellar mass. Uncertainties are propagated considering uncertainties both on the literature measurements and in the power law fit to our own measurements. The ratios for \citet{velander14} and \citet{mandelbaum16} are interpolated linearly in logarithmic space   from the data points. See \Cref{f:shsmr} for a more detailed description of the different measurements.
The horizontal grey band shows for comparison the range of values for $m_\mathrm{sub}/M_\mathrm{200,central}$ adopted by these studies. Note that the horizontal range is smaller than that shown in \Cref{f:shsmr}.}
\label{f:massratio}
\end{figure}

\cite{velander14} and \cite{mandelbaum16} constrained the total-to-stellar mass relation of central galaxies making use of a mixture of central and satellite galaxies, assuming that subhaloes have lost approximately half of their mass since being accreted onto the clusters. They achieve this by truncating the NFW density profiles of subhaloes at $r_\mathrm{t}=0.4r_{200}$ \citep{mandelbaum05_ggl}, but they do not fit for any parameter relating to the subhalo contribution to their signal except for the fraction of satellite galaxies (which are hosted by subhaloes). We show in \Cref{f:massratio} the ratio between our subhalo masses (more precisely, of the power-law fit to them) and the central halo masses shown in the left panel of \Cref{f:shsmr}. For red galaxies \citep{velander14,mandelbaum16} the ratio reaches a maximum of up to 0.7 at $m_\star=(2-5)\times10^{10}\,\Msun$, but quickly drops for both lower and higher stellar masses, reaching $\sim20$\percent\ at the low- and high-stellar mass ends probed here, although we note that uncertainties are significant at the low mass end (cf.\ \Cref{f:shsmr}). It is apparent that the scatter between lines is larger than individual uncertainties. We attribute this additional scatter partially to potential systematics from different modelling assumptions (both in halo mass and in stellar mass), but also partially to the fact that each study used a different galaxy sample.

The result of \Cref{f:massratio} has an impact on the total-to-stellar mass relation of central galaxies when centrals and satellites are not separated a priori, as done by \cite{vanuitert16}. As discussed by \cite{velander14}, the effect of this increased fraction of stripped material on the inferred halo masses is to reduce halo masses, by an amount that depends on both the level of stripping and the satellite fraction in the sample. Since the satellite fraction can be high at low stellar masses \citep[e.g., $>40$\percent\ at $m_\star<10^{10}\,\Msun$ in][]{velander14}, a stripping of 80\percent\ of the mass of subhaloes could have an appreciable effect. Exactly how much of an effect that is will also depend on the effect the stripping has on the density profile, however, and is not easy to quantify in advance. At the very least, our results should inform systematic uncertainty budgets for estimations of the total-to-stellar mass relation of central galaxies when the sample of lenses includes satellite galaxies as well.

Irrespective of what a comparison between present-day satellites and present-day centrals means, the solid lines shown in \Cref{f:massratio} represent a direct, quantitative prediction for hydrodynamical simulations. As shown in \Cref{f:massratio}, the same ratio measured in the EAGLE simulation yields significantly different results, with a ratio $m_\mathrm{bg}/M_\mathrm{200,central}\sim0.1$ for all stellar masses $m_\star<2\times10^{11}\,\Msun$. We discuss possible origins for this discrepancy in the next section.

\begin{table}
 \centering
\caption{Prior ranges and marginalized posterior estimates of best-fitting parameters of the satellite lensing model. Masses are in units of $\Msun$. Uncertainties correspond to 68\percent\ credible intervals. All parameters have flat priors in the quoted ranges. Note that priors are defined in linear, rather than logarithmic, space. The binning schemes are summarized in \Cref{t:bins}.}
\label{t:mcmc}
\begin{tabular}{c | c c c}
\hline\hline
Parameter & Prior &  $m_\star$ bins  &      $\Rsat$ bins     \\
          & range & (\Cref{s:shsmr}) & (\Cref{s:segregation}) \\[0.5ex]
\hline
$\log\langle m_\mathrm{bg,1} \rangle$ & $[7,14]$ & $10.64_{-0.53}^{+0.39}$ & $10.49_{-0.47}^{+0.35}$ \\[0.4ex]
$\log\langle m_\mathrm{bg,2} \rangle$ & $[7,14]$ & $11.41_{-0.21}^{+0.17}$ & $11.60_{-0.15}^{+0.16}$ \\[0.4ex]
$\log\langle m_\mathrm{bg,3} \rangle$ & $[7,14]$ & $11.71_{-0.17}^{+0.15}$ & $11.55_{-0.21}^{+0.21}$ \\[0.4ex]
$\log\langle m_\mathrm{bg,4} \rangle$ & $[7,14]$ & $11.84_{-0.15}^{+0.15}$ & $11.46_{-0.33}^{+0.25}$ \\[0.4ex]
$\log\langle m_\mathrm{bg,5} \rangle$ & $[7,14]$ & $12.15_{-0.20}^{+0.17}$ & \dots \\[0.4ex]
$a_{c,\mathrm{h}}$ & $[0,10]$ & $7.9_{-1.7}^{+1.4}$ & $7,2_{-2.0}^{+2.4}$ \\[0.4ex]
$b_{c,\mathrm{h}}$ & $[-1,1]$ & $-0.74_{-0.14}^{+0.17}$ & $-0.51_{-0.31}^{+0.38}$ \\[0.4ex]
$\log\langle M_\mathrm{h,1} \rangle$ & $[13,16]$ & $15.54_{-0.51}^{+0.30}$ & $15.31_{-0.34}^{+0.43}$ \\[0.4ex]
$\log\langle M_\mathrm{h,2} \rangle$ & $[13,16]$ & $15.54_{-0.37}^{+0.29}$ & $15.39_{-0.32}^{+0.36}$ \\[0.4ex]
$\log\langle M_\mathrm{h,3} \rangle$ & $[13,16]$ & $15.72_{-0.43}^{+0.20}$ & $15.69_{-0.76}^{+0.23}$ \\[0.4ex]
$\log\langle M_\mathrm{h,4} \rangle$ & $[13,16]$ & $15.72_{-0.49}^{+0.21}$ & $15.68_{-0.55}^{+0.24}$ \\[0.4ex]
$\log\langle M_\mathrm{h,5} \rangle$ & $[13,16]$ & $15.73_{-0.37}^{+0.20}$ & \dots \\[0.5ex]
\hline
\end{tabular}
\end{table}

\subsection{Comparison to other subhalo measurements and predictions}

In the right panel of \Cref{f:shsmr}, we compare our measurements to a previous measurement of subhalo mass as a function of stellar mass by \cite{li16}. We also compare to determinations of the subhalo-to-stellar mass relation in low-mass galaxy clusters ($\Mh\lesssim10^{14}\,\Msun$) from measurements in the EAGLE simulation \citep{crain15,schaye15} devised to match the satellite sample of \cite{sifon15_kids} by \citet{velliscig17}, and from a combination of galaxy clustering measurements and abundance matching predictions by \cite{rodriguez13}\footnote{\citet{rodriguez13} used their measurements to fit for $m_\star(m)$, which we invert by Monte Carlo-sampling their relation accounting for the subhalo mass function at the time of infall from \citet{vdbosch16}, including intrinsic scatter, and binning the data points by $m_\star$.}.

The measurements by \cite{li16} are consistent with our results within their comparatively large errorbars. It is worth noting that the high-stellar mass measurement by \cite{li16} supports our tentative detection of a difference between the SHSMR and the relation for central galaxies at the high mass end of subhaloes, $m\gtrsim10^{11}\,\Msun$. Their measurement is in fact consistent with a simple extrapolation of \Cref{eq:shsmr}. However, this comparison should be taken with care, as both the adopted density profiles and the mass definitions are different between us and \cite{li16}.

Combining our measurements with those of \cite{li16}, we find that satellite galaxies in EAGLE have a somewhat steeper SHSMR than the observations require. This may be due to the efficiency of tidal stripping implemented in the simulations, but it is likely that this is also influenced by technical differences such as assumptions about the density profiles and mass definitions, and potentially even the different halo mass regimes. The EAGLE simulation was calibrated to measurements of the stellar mass function at $z=0.1$ assuming the same \cite{chabrier03} IMF, and subhaloes and galaxies are identified, and their masses calculated, using \textsc{subfind}, so in terms of definitions the comparison with our work seems consistent. However, \cite{knebe11} have shown that the accuracy of subhalo masses estimated by \textsc{subfind} depends significantly on halo-centric distance. Given that more massive satellites are generally located further out than less massive satellites (conversely, the average stellar mass at $R_\mathrm{sat}\sim0.2\,\mathrm{Mpc}$ is rougly 60\percent\ that at $R_\mathrm{sat}\sim1.5\,\mathrm{Mpc}$ cf.\ \Cref{t:bins}), this bias in \textsc{subfind} might exacerbate real differences at low masses somewhat. This should not, however, strongly affect the high mass end. We also note that \cite{velliscig17} have shown that, on average, the excess surface density around subhaloes in EAGLE is consistent with lensing measurements of observed satellite galaxies, but the \emph{number} of subhaloes per host halo in EAGLE can appear inconsistent with the observations  if the selection function is not carefully accounted for. It is plausible that a combination of biases in \textsc{subfind} and inconsistent satellite fractions might explain the observed difference.

Similarly, the abundance matching-based measurement of \cite{rodriguez13} is significantly lower than our measurements at stellar masses $m_\star\gtrsim 5\times10^{9}\,\Msun$. At face value, this suggests that subhalo abundance matching does not capture the relation between total and stellar mass properly. The existence of `assembly bias'---the hypothesis that the correspondence between halo mass and stellar mass depends on halo formation history \citep{gao05}---would indeed mean that this is the case, but the extent of this problem is not well determined. Much work is devoted these days to the characterization and modelling of assembly bias \citep[e.g.,][]{hearin16}, and future work may be able to determine the role of assembly bias, or other effects, in this comparison.

\begin{figure*}
 \centerline{\includegraphics[width=\textwidth]{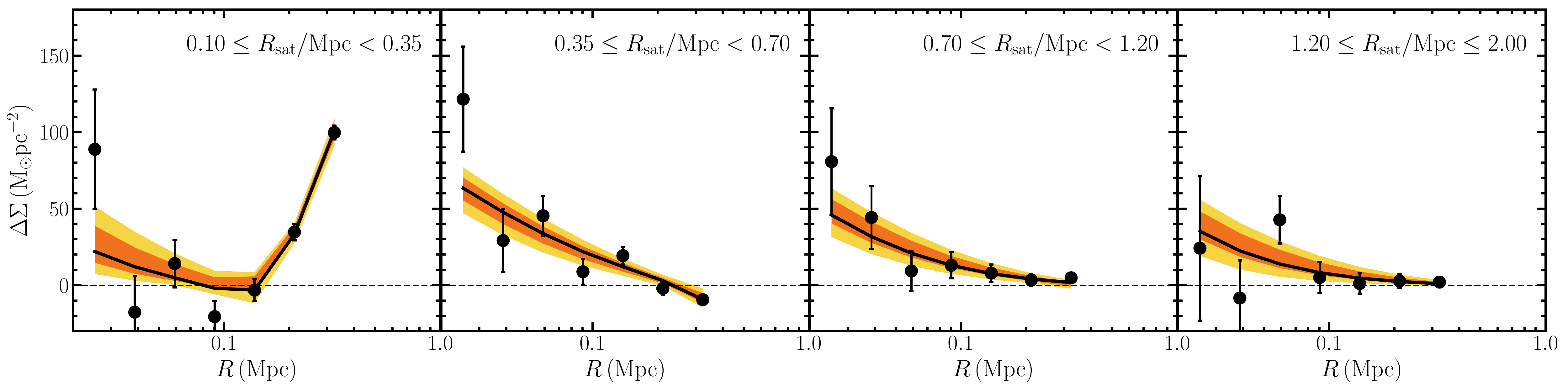}}
\caption{Excess surface mass density (black points with errorbars) and best-fit NFW model (black line) of satellites binned by cluster-centric distance. Black lines show best-fit models while orange and yellow regions show 68 and 95 per cent credible intervals. The model for the host clusters is not flexible enough because the small field of view (in physical units at $z\sim0.05$) biases our large-scale ($R\gtrsim0.5\,{\rm Mpc}$) measurements, but this has no impact on the modelling of the subhalo signal. See \Cref{s:segregation} for details.}
\label{f:esd_distBCG}
\end{figure*}

\begin{figure}
 \centerline{\includegraphics[width=3.4in]{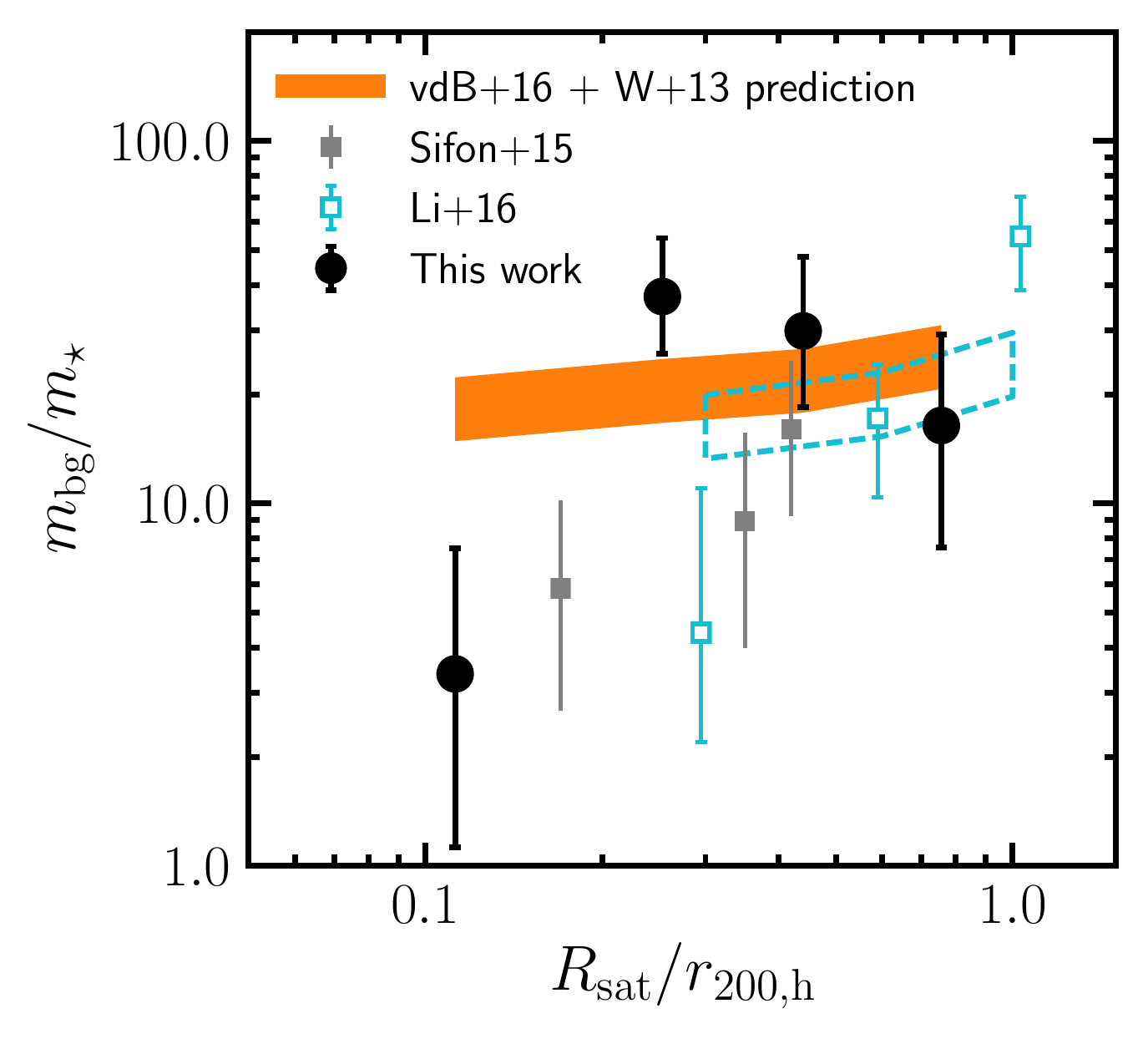}}
\caption{Best-fit subhalo-to-stellar mass ratio as a function of projected distance to the cluster centre, in units of $r_{200}$ of the host cluster assuming $M_\mathrm{200,h}=6\times10^{14}\,\Msun$ (see text). Uncertainties show 68\percent\ credible intervals. As in \Cref{f:shsmr}, subhalo masses refer to the mass within $r_\mathrm{bg}$. Also shown are previous measurements from \citet{sifon15_kids} and \citet{li16}. The orange band shows a prediction for the total-to-stellar mass ratio as a function of $R_\mathrm{sat}$ from numerical simulations from \citet{vdbosch16}, which give $m/m_\mathrm{acc}(R_\mathrm{sat})$ (where $m_\mathrm{acc}$ is the subhalo mass at the time at accretion), combined with the semi-analytic $m_\mathrm{acc}(m_\star)$ predictions from \citet{wang13} for the median stellar masses in the five $\Rsat$ bins (cf.\ \Cref{t:bins}). For illustration purposes, the orange band has a width of 20\percent. The cyan box outline shows the same predictions, also with a 20\percent\ width, for the mean stellar masses of \citet{li16}.
}
\label{f:mass_distBCG}
\end{figure}


\section{Subhalo mass segregation}\label{s:segregation}

In this section we explore the dependence of subhalo mass on the distance to the cluster centre. \cite{vdbosch16} have shown using \Nbody\ simulations that, although subhalo mass segregation is seen more strongly in three dimensions, the projected halo-centric distance still preserves some of the correlation of subhalo physical parameters with the binding energy, which is closely related to the time a subhalo has spent bound to the host halo. However, after multiple orbits the correlation is significantly reduced because at any particular (projected) distance from the halo centre there are subhaloes with a wide range of infall times. We might therefore expect satellites at similar $\Rsat$ to have been part of similar halo-subhalo interactions on average, but with a large scatter in their individual histories.

\Cref{f:esd_distBCG} shows the measured ESD and best-fit model when we split the satellite sample into four $\Rsat$ bins. The best-fit model has $\chi^2=28.8$ with 17 `nominal' degrees of freedom (the caveats discussed in \Cref{s:shsmr} also apply here). Because of the finite field of view of our observations, we cannot average galaxy shapes in full annuli with large lens-source separations around most lenses, so additive biases do not cancel out. For this reason, in this section we only use measurements out to lens-source separations $R=0.6$ Mpc. At larger separations the signal is dominated by the host clusters, with little to no contribution from the subhaloes, and we have verified that subhalo masses are not affected by this cut.

\subsection{Mass segregation of MENeaCS satellites}

We show the posterior subhalo masses, normalized by the median stellar mass in each bin, in \Cref{f:mass_distBCG}. In order to compare with literature measurements and predictions, most of which refer to lower halo masses, we normalize cluster-centric distances by cluster size $r_\mathrm{200,h}$. Following the discussion in \Cref{s:likelihood}, we adopt a mean cluster mass $M_\mathrm{200,h}=6\times10^{14}\,\Msun$ suggested by galaxy velocity dispersions and weak lensing measurements by \cite{sifon15_cccp} and Herbonnet et al.\ (in prep.), respectively, instead of the posterior masses reported in \Cref{t:mcmc}. 

There is mild evidence for the innermost satellites to have a lower total-to-stellar mass ratio. The weighted average of the three outermost data points\footnote{Because these are ratios of masses, we calculate the averages in logarithmic space, where the ratios correspond to differences and all operations are linear, and then revert the average quantities back to linear space.} is $\langle m_\mathrm{bg}/m_\star \rangle=30.7_{-7.2}^{+9.2}$ for satellites at $R_\mathrm{sat}>0.2r_\mathrm{200,h}$, while the ratio for the innermost satellites is $\langle m_\mathrm{sub}/m_\star \rangle = 3.3_{-2.2}^{+4.2}$, different at the $3\sigma$ level. At face value, this may be taken as evidence for tidal stripping manifest in satellites within roughly the scale radius of the hosts. However, we caution that the separation between the innermost bin and the other three have been made after the fact; even though $0.2r_\mathrm{200,h}$ corresponds roughly to the scale radius of massive clusters (i.e., they have a concentration of about 5), we did not have a reason to expect a discontinuity in the total-to-stellar mass ratio at this particular radius or any other \citep[as opposed, for instance, to the steady rise suggested by][]{li16}. Furthermore, we note that the mass ratio for the outermost data point is consistent with that of the innermost data point, suggesting that these variations may be due to statistical noise (or, e.g., different model biases applying to different data points, for instance due to the effect of tidal stripping). If we fit the four data points with a log-linear relation as a function of cluster-centric distance, we find $m_\mathrm{bg}/m_\star\propto(R_\mathrm{sat}/r_\mathrm{200,h})^{0.84\pm0.49}$, consistent with no dependence on cluster-centric distance within $2\sigma$, although this may not be a good description if there is indeed a single discontinuity between the first and second data points. We will explore the reality and nature of this feature in more detail in future work.

\subsection{Comparison to theoretical predictions}

\cite{vdbosch16} have shown that the parameter that correlates most strongly with both binding energy and halo-centric distance is the ratio $m/m_\mathrm{acc}$, where $m_\mathrm{acc}$ is the mass of the subhalo at the time of its accretion onto the main halo. This is because of the average relation between time a subhalo has spent in the host halo (or the accretion redshift, for a given redshift of observation) and the subhalo's distance to the halo centre, combined with the strong dependence of the mass ratio to the time since accretion as a result of tidal stripping.
We also show in \Cref{f:mass_distBCG} a prediction obtained by combining numerical simulations and a semi-analytic model, as follows. We use the average $m/m_\mathrm{acc}$ (that is, the ratio between present mass and mass at the time of accretion) as a function of projected distance measured by \cite{vdbosch16} for subhaloes in a set of N-body simulations. We combine these predictions with $m_\mathrm{acc}(m_\star)$ estimated by \cite{wang13}\footnote{Similar to the treatment of the $m_\star(m)$ relation of \citet{rodriguez13}, we Monte-Carlo sample the $m_\star(m_\mathrm{acc})$ relation of \citet{wang13}, convolved with the subhalo mass function and accounting for intrinsic scatter, in order to recover $m_\mathrm{acc}(m_\star)$.} by fitting predictions from semi-analytical models to the stellar mass function and the clustering of SDSS galaxies, adopting the median stellar masses for each cluster-centric bin, as quoted in \Cref{t:bins}. These predictions are in good agreement with our measurements, and show that we do not expect to see a dependence of the subhalo-to-stellar mass ratio with cluster-centric distance with the current uncertainty levels.

Note that the normalization of the predicted subhalo-to-stellar mass ratio is fixed by the $m_\star(m_\mathrm{acc})$ relation, and has not been adjusted to match our results, except for the use of the stellar masses of \Cref{t:bins} as input to the $m_\mathrm{acc}(m_\star)$ relation. The fact that the predicted normalization of $m/m_\star$ is consistent with our data lends credence to our definition of subhalo mass as $m_\mathrm{bg}$ or, at the very least, supports the idea that the comparison with theoretical predictions is internally consistent. In fact, \cite{vdbosch16} based their analysis on the \textsc{rockstar} phase-space halo finder \citep{behroozi13} which has been shown to accurately recover subhalo masses at all halo-centric distances; most other subhalo finders (including \textsc{subfind}) tend to underestimate subhalo masses closer to the halo centre \citep{knebe11}.

\subsection{Previous measurements of subhalo mass segregation}

Several previous observational studies have focused on the mass segregation of subhaloes. However, differences in the adopted density profiles, mass definitions, and the fact that some works did not report the masses of the host clusters (nor normalized cluster-centric distance by host cluster size), preclude a detailed comparison with our results. To contextualize our results, we nevertheless compare these studies to the present one in a qualitative sense.

\cite{okabe14} measured the lensing signal of galaxy- and subgroup-scale subhaloes in the Coma cluster. They found that, while subgroup-scale subhaloes (which they analyzed individually) are better fit by truncated profiles, a stack of individual luminous galaxies is well-fit by a simple NFW model like the one adopted in this work, with no discernible truncation radius. This suggests that, maybe, the stacking of subhaloes with varying truncation radii, produces an average signal in which a truncation radius is no longer discernible. However, this contrasts with the results of \cite{natarajan98,natarajan02,natarajan07,natarajan09} and \cite{limousin07}, who found evidence for galaxy truncation by interpreting the weak lensing signal of cluster galaxies using a maximum likelihood approach. Moreover, these studies found significant evidence for smaller truncation radii (or, equivalently, more compact cores) in galaxies closer to the cluster centres. It is unclear whether the methodology itself allowed the latter set of authors to detect a truncation radius while our methodology is more limiting in this respect, or if the parametrization of the subhalo mass density profile has any influence on this discrepancy, as argued by \cite{pastormira11}. 
 Since the papers above do not show the signal from which their results are derived, it is difficult to assess the origin of the different conclusions we reach compared to theirs. Our detailed assessment of shape measurements in \Cref{s:calibration} makes it unlikely that truncation radii of order 10--20 kpc can be detected directly with weak lensing measurements using ground-based observations \citep[as suggested by][]{limousin07}, unless perhaps if lens galaxies were subtracted from the images before the analysis, something we will explore in future work. On the other hand, by incorporating the spatial distribution of galaxies into the analysis, one may potentially be able to extract more information than our methodology allows. As it stands, this difference remains unresolved.

More recently, several authors have measured the stacked weak lensing signal as a function of cluster-centric distance, producing results that are more directly comparable to ours. As mentioned in the preceding section, \cite{sifon15_kids} also found no significant segregation of subhalo mass in GAMA groups using weak lensing measurements from KiDS. The results of \cite{niemiec17} also suggest no evidence for mass segregation.\footnote{\citet{niemiec17} interpret their measurements as evidence of tidal stripping, based on the claim that galaxies closer to the centre have smaller total-to-stellar mass ratios than galaxies further out. However, their results show that there is only a $1\sigma$ difference between galaxies closer in and further out, and we instead choose to interpret them as showing no evidence for mass segregation.} In contrast, \cite{li16} found a factor 10 increase in subhalo-to-stellar mass ratio going from $R_\mathrm{sat}\sim0.3r_\mathrm{200,h}$ to $R_\mathrm{sat}\sim r_\mathrm{200,h}$.\footnote{\citet{li16} do not report the masses of the host clusters; we adopt a mass $M_\mathrm{200,h}=10^{14}\,\Msun$ to estimate $r_\mathrm{200,h}$.} As shown in \Cref{f:mass_distBCG}, \cite{li16} probe cluster-centric distances larger than we do, and it is possible that the effect would be more apparent at larger distance, although as we show in \Cref{f:mass_distBCG}, this does not seem to be the theoretical expectation. We caution, however, that \cite{li16} used the \emph{photometric} redMaPPer cluster catalogue \citep{rykoff14} to construct their lens sample. \cite{sifon15_cccp} showed that even in the case of unbiased photometric redshifts, samples of galaxies selected to be at the cluster redshift are significantly contaminated at large distances. It is therefore possible that the trend observed by \cite{li16} may be due at least in part to contamination by unrelated galaxies although the fact that \cite{li16} use only high-probability cluster members (based on the redMaPPer definition) may somewhat mitigate this \citep[see][]{zu17}.


\section{Conclusions}\label{s:conclusions}

We present the average masses of satellite galaxies in massive galaxy clusters at $0.05<z<0.15$ using weak galaxy-galaxy lensing measurements. We use a combination of deep, wide-field observations of galaxy clusters and extensive archival spectroscopic data \citep{sifon15_cccp}. Using extensive image simulations of bright lenses in the foreground of a population of field galaxies resembling the source population in our data, we model and account for biases arising from (i) shape measurements, due to confusion of light from the lens with the faint sources, and (ii) contamination of the source sample by faint cluster members (\Cref{s:calibration}).

We model the lensing signal from subhaloes using an NFW profile and the subhalo mass-concentration relation measured from \Nbody\ simulations by \cite{moline17}, which depends on cluster-centric distance. We split the sample in bins of stellar mass and measure the subhalo-to-stellar mass relation (SHSMR) of galaxies in massive clusters. Fitting the resulting masses with a power-law relation, we find $\log\Msub=(11.54\pm0.05) + (\slope)\log\Mstar$ (\Cref{f:shsmr}). The slope of this relation is robust to both the adopted subhalo mass-concentration relation and the subhalo mass definition. We find that at a characteristic stellar mass of $\sim3\times10^{10}\,\Msun$, the ratio between subhalo mass and host halo mass is maximal, reaching a value of approximately 0.5, and dropping to 0.2 at the high-mass end (\Cref{f:massratio}). This behaviour is likely caused by a combination of tidal stripping and dynamical friction.

We also study the masses of subhaloes at different cluster-centric distances with the aim of studying the evolution of subhaloes within clusters. We cautiously point out the possibility of a discontinuity in the dependence of the total-to-stellar mass ratio at about the cluster scale radius, within which satellites seem to show a highly suppressed ratio. However, both more data and more accurate predictions are required to validate this result. Although direct comparison with the observational literature is complicated by the use of different definitions and conventions, our results are generally consistent within the overlapping stellar mass and cluster-centric distance, but the resulting picture is still unclear.

The halo model commonly employed in galaxy-galaxy lensing studies requires some assumptions about the density profiles of the subhaloes hosting satellite galaxies. At low stellar masses ($M_\star\lesssim10^{10}\Msun$), most red galaxies seem to be satellites \citep{mandelbaum06_ggl,velander14}. Therefore, host halo masses at low stellar masses are determined through the halo model based on observations of what are mostly satellite galaxies; degeneracies in the halo model dominate the resulting masses \citep{velander14}. While we are not able to constrain the density profiles of subhaloes at present, the subhalo-to-stellar mass relation is an important ingredient that could be incorporated in future galaxy-galaxy lensing analyses to inform these choices.


\section*{Acknowledgments}

We are indebted to Marcello Cacciato for valuable discussions and suggestions throughout the development of this work.
We thank David Sand and Melissa Graham for their participation in the MENeaCS proposal and observations.
CS, RH, HH and MV acknowledge support from the European Research Council under FP7 grant number 279396.
RFJvdB acknowledges support from the European Research Council under FP7 grant number 340519.

Based on observations obtained with MegaPrime/MegaCam, a joint project of CFHT and CEA/DAPNIA, at the Canada-France-Hawaii Telescope (CFHT) which is operated by the National Research Council (NRC) of Canada, the Institute National des Sciences de l'Univers of the Centre National de la Recherche Scientifique of France, and the University of Hawaii.

\bibliographystyle{mnras}
\bibliography{bibliography}

\begin{thebibliography}{}
\makeatletter
\relax
\def\mn@urlcharsother{\let\do\@makeother \do\$\do\&\do\#\do\^\do\_\do\%\do\~}
\def\mn@doi{\begingroup\mn@urlcharsother \@ifnextchar [ {\mn@doi@}
  {\mn@doi@[]}}
\def\mn@doi@[#1]#2{\def\@tempa{#1}\ifx\@tempa\@empty \href
  {http://dx.doi.org/#2} {doi:#2}\else \href {http://dx.doi.org/#2} {#1}\fi
  \endgroup}
\def\mn@eprint#1#2{\mn@eprint@#1:#2::\@nil}
\def\mn@eprint@arXiv#1{\href {http://arxiv.org/abs/#1} {{\tt arXiv:#1}}}
\def\mn@eprint@dblp#1{\href {http://dblp.uni-trier.de/rec/bibtex/#1.xml}
  {dblp:#1}}
\def\mn@eprint@#1:#2:#3:#4\@nil{\def\@tempa {#1}\def\@tempb {#2}\def\@tempc
  {#3}\ifx \@tempc \@empty \let \@tempc \@tempb \let \@tempb \@tempa \fi \ifx
  \@tempb \@empty \def\@tempb {arXiv}\fi \@ifundefined
  {mn@eprint@\@tempb}{\@tempb:\@tempc}{\expandafter \expandafter \csname
  mn@eprint@\@tempb\endcsname \expandafter{\@tempc}}}

\bibitem[\protect\citeauthoryear{{Akritas} \& {Bershady}}{{Akritas} \&
  {Bershady}}{1996}]{akritas96}
{Akritas} M.~G.,  {Bershady} M.~A.,  1996, \mn@doi [\apj] {10.1086/177901},
  \href {http://adsabs.harvard.edu/abs/1996ApJ...470..706A} {470, 706}

\bibitem[\protect\citeauthoryear{{Andrae}, {Schulze-Hartung}  \&
  {Melchior}}{{Andrae} et~al.}{2010}]{andrae10}
{Andrae} R.,  {Schulze-Hartung} T.,   {Melchior} P.,  2010, preprint, \href
  {http://adsabs.harvard.edu/abs/2010arXiv1012.3754A} {} (\mn@eprint {arXiv}
  {1012.3754})

\bibitem[\protect\citeauthoryear{{Applegate} et~al.,}{{Applegate}
  et~al.}{2014}]{applegate14}
{Applegate} D.~E.,  et~al., 2014, \mn@doi [\mnras] {10.1093/mnras/stt2129},
  \href {http://adsabs.harvard.edu/abs/2014MNRAS.439...48A} {439, 48}

\bibitem[\protect\citeauthoryear{{Baltz}, {Marshall}  \& {Oguri}}{{Baltz}
  et~al.}{2009}]{baltz09}
{Baltz} E.~A.,  {Marshall} P.,   {Oguri} M.,  2009, \mn@doi [\jcap]
  {10.1088/1475-7516/2009/01/015}, \href
  {http://adsabs.harvard.edu/abs/2009JCAP...01..015B} {1, 015}

\bibitem[\protect\citeauthoryear{{Behroozi}, {Wechsler}  \& {Wu}}{{Behroozi}
  et~al.}{2013}]{behroozi13}
{Behroozi} P.~S.,  {Wechsler} R.~H.,   {Wu} H.-Y.,  2013, \mn@doi [\apj]
  {10.1088/0004-637X/762/2/109}, \href
  {http://adsabs.harvard.edu/abs/2013ApJ...762..109B} {762, 109}

\bibitem[\protect\citeauthoryear{{Bertin} \& {Arnouts}}{{Bertin} \&
  {Arnouts}}{1996}]{bertin96}
{Bertin} E.,  {Arnouts} S.,  1996, \mn@doi [\aaps] {10.1051/aas:1996164}, \href
  {http://adsabs.harvard.edu/abs/1996A%26AS..117..393B} {117, 393}

\bibitem[\protect\citeauthoryear{{Blanton} et~al.,}{{Blanton}
  et~al.}{2005}]{blanton05_vagc}
{Blanton} M.~R.,  et~al., 2005, \mn@doi [\aj] {10.1086/429803}, \href
  {http://adsabs.harvard.edu/abs/2005AJ....129.2562B} {129, 2562}

\bibitem[\protect\citeauthoryear{{Bower}, {Benson}, {Malbon}, {Helly}, {Frenk},
  {Baugh}, {Cole}  \& {Lacey}}{{Bower} et~al.}{2006}]{bower06}
{Bower} R.~G.,  {Benson} A.~J.,  {Malbon} R.,  {Helly} J.~C.,  {Frenk} C.~S.,
  {Baugh} C.~M.,  {Cole} S.,   {Lacey} C.~G.,  2006, \mn@doi [\mnras]
  {10.1111/j.1365-2966.2006.10519.x}, \href
  {http://adsabs.harvard.edu/abs/2006MNRAS.370..645B} {370, 645}

\bibitem[\protect\citeauthoryear{{Boylan-Kolchin}, {Bullock}  \&
  {Kaplinghat}}{{Boylan-Kolchin} et~al.}{2011}]{boylan11}
{Boylan-Kolchin} M.,  {Bullock} J.~S.,   {Kaplinghat} M.,  2011, \mn@doi
  [\mnras] {10.1111/j.1745-3933.2011.01074.x}, \href
  {http://adsabs.harvard.edu/abs/2011MNRAS.415L..40B} {415, L40}

\bibitem[\protect\citeauthoryear{{Bruzual} \& {Charlot}}{{Bruzual} \&
  {Charlot}}{2003}]{bruzual03}
{Bruzual} G.,  {Charlot} S.,  2003, \mn@doi [\mnras]
  {10.1046/j.1365-8711.2003.06897.x}, \href
  {http://adsabs.harvard.edu/abs/2003MNRAS.344.1000B} {344, 1000}

\bibitem[\protect\citeauthoryear{{Bullock}, {Kolatt}, {Sigad}, {Somerville},
  {Kravtsov}, {Klypin}, {Primack}  \& {Dekel}}{{Bullock}
  et~al.}{2001}]{bullock01}
{Bullock} J.~S.,  {Kolatt} T.~S.,  {Sigad} Y.,  {Somerville} R.~S.,  {Kravtsov}
  A.~V.,  {Klypin} A.~A.,  {Primack} J.~R.,   {Dekel} A.,  2001, \mn@doi
  [\mnras] {10.1046/j.1365-8711.2001.04068.x}, \href
  {http://adsabs.harvard.edu/abs/2001MNRAS.321..559B} {321, 559}

\bibitem[\protect\citeauthoryear{{Cacciato}, {van den Bosch}, {More}, {Li},
  {Mo}  \& {Yang}}{{Cacciato} et~al.}{2009}]{cacciato09}
{Cacciato} M.,  {van den Bosch} F.~C.,  {More} S.,  {Li} R.,  {Mo} H.~J.,
  {Yang} X.,  2009, \mn@doi [\mnras] {10.1111/j.1365-2966.2008.14362.x}, \href
  {http://adsabs.harvard.edu/abs/2009MNRAS.394..929C} {394, 929}

\bibitem[\protect\citeauthoryear{{Chabrier}}{{Chabrier}}{2003}]{chabrier03}
{Chabrier} G.,  2003, \mn@doi [\pasp] {10.1086/376392}, \href
  {http://adsabs.harvard.edu/abs/2003PASP..115..763C} {115, 763}

\bibitem[\protect\citeauthoryear{{Chang}, {Macci{\`o}}  \& {Kang}}{{Chang}
  et~al.}{2013}]{chang13_stripping}
{Chang} J.,  {Macci{\`o}} A.~V.,   {Kang} X.,  2013, \mn@doi [\mnras]
  {10.1093/mnras/stt434}, \href
  {http://adsabs.harvard.edu/abs/2013MNRAS.431.3533C} {431, 3533}

\bibitem[\protect\citeauthoryear{{Clowe}, {Luppino}, {Kaiser}, {Henry}  \&
  {Gioia}}{{Clowe} et~al.}{1998}]{clowe98}
{Clowe} D.,  {Luppino} G.~A.,  {Kaiser} N.,  {Henry} J.~P.,   {Gioia} I.~M.,
  1998, \mn@doi [\apjl] {10.1086/311285}, \href
  {http://adsabs.harvard.edu/abs/1998ApJ...497L..61C} {497, L61}

\bibitem[\protect\citeauthoryear{{Coupon} et~al.,}{{Coupon}
  et~al.}{2015}]{coupon15}
{Coupon} J.,  et~al., 2015, \mn@doi [\mnras] {10.1093/mnras/stv276}, \href
  {http://adsabs.harvard.edu/abs/2015MNRAS.449.1352C} {449, 1352}

\bibitem[\protect\citeauthoryear{{Crain} et~al.,}{{Crain}
  et~al.}{2015}]{crain15}
{Crain} R.~A.,  et~al., 2015, \mn@doi [\mnras] {10.1093/mnras/stv725}, \href
  {http://adsabs.harvard.edu/abs/2015MNRAS.450.1937C} {450, 1937}

\bibitem[\protect\citeauthoryear{{Duffy}, {Schaye}, {Kay}  \& {Dalla
  Vecchia}}{{Duffy} et~al.}{2008}]{duffy08}
{Duffy} A.~R.,  {Schaye} J.,  {Kay} S.~T.,   {Dalla Vecchia} C.,  2008, \mn@doi
  [\mnras] {10.1111/j.1745-3933.2008.00537.x}, \href
  {http://adsabs.harvard.edu/abs/2008MNRAS.390L..64D} {390, L64}

\bibitem[\protect\citeauthoryear{{Dutton} \& {Macci{\`o}}}{{Dutton} \&
  {Macci{\`o}}}{2014}]{dutton14}
{Dutton} A.~A.,  {Macci{\`o}} A.~V.,  2014, \mn@doi [\mnras]
  {10.1093/mnras/stu742}, \href
  {http://adsabs.harvard.edu/abs/2014MNRAS.441.3359D} {441, 3359}

\bibitem[\protect\citeauthoryear{{Eichner} et~al.,}{{Eichner}
  et~al.}{2013}]{eichner13}
{Eichner} T.,  et~al., 2013, \mn@doi [\apj] {10.1088/0004-637X/774/2/124},
  \href {http://adsabs.harvard.edu/abs/2013ApJ...774..124E} {774, 124}

\bibitem[\protect\citeauthoryear{{Fahlman}, {Kaiser}, {Squires}  \&
  {Woods}}{{Fahlman} et~al.}{1994}]{fahlman94}
{Fahlman} G.,  {Kaiser} N.,  {Squires} G.,   {Woods} D.,  1994, \mn@doi [\apj]
  {10.1086/174974}, \href {http://adsabs.harvard.edu/abs/1994ApJ...437...56F}
  {437, 56}

\bibitem[\protect\citeauthoryear{{Fang}, {Clampitt}, {Dalal}, {Jain}, {Rozo},
  {Moustakas}  \& {Rykoff}}{{Fang} et~al.}{2016}]{fang16}
{Fang} Y.,  {Clampitt} J.,  {Dalal} N.,  {Jain} B.,  {Rozo} E.,  {Moustakas}
  J.,   {Rykoff} E.,  2016, \mn@doi [\mnras] {10.1093/mnras/stw2108}, \href
  {http://adsabs.harvard.edu/abs/2016MNRAS.463.1907F} {463, 1907}

\bibitem[\protect\citeauthoryear{{Foreman-Mackey}, {Hogg}, {Lang}  \&
  {Goodman}}{{Foreman-Mackey} et~al.}{2013}]{foreman13}
{Foreman-Mackey} D.,  {Hogg} D.~W.,  {Lang} D.,   {Goodman} J.,  2013, \mn@doi
  [\pasp] {10.1086/670067}, \href
  {http://adsabs.harvard.edu/abs/2013PASP..125..306F} {125, 306}

\bibitem[\protect\citeauthoryear{{Gabor} \& {Dav{\'e}}}{{Gabor} \&
  {Dav{\'e}}}{2015}]{gabor15}
{Gabor} J.~M.,  {Dav{\'e}} R.,  2015, \mn@doi [\mnras] {10.1093/mnras/stu2399},
  \href {http://adsabs.harvard.edu/abs/2015MNRAS.447..374G} {447, 374}

\bibitem[\protect\citeauthoryear{{Gao}, {Springel}  \& {White}}{{Gao}
  et~al.}{2005}]{gao05}
{Gao} L.,  {Springel} V.,   {White} S.~D.~M.,  2005, \mn@doi [\mnras]
  {10.1111/j.1745-3933.2005.00084.x}, \href
  {http://adsabs.harvard.edu/abs/2005MNRAS.363L..66G} {363, L66}

\bibitem[\protect\citeauthoryear{{Geiger} \& {Schneider}}{{Geiger} \&
  {Schneider}}{1998}]{geiger98}
{Geiger} B.,  {Schneider} P.,  1998, \mn@doi [\mnras]
  {10.1046/j.1365-8711.1998.01146.x}, \href
  {http://adsabs.harvard.edu/abs/1998MNRAS.295..497G} {295, 497}

\bibitem[\protect\citeauthoryear{{Ghigna}, {Moore}, {Governato}, {Lake},
  {Quinn}  \& {Stadel}}{{Ghigna} et~al.}{1998}]{ghigna98}
{Ghigna} S.,  {Moore} B.,  {Governato} F.,  {Lake} G.,  {Quinn} T.,   {Stadel}
  J.,  1998, \mn@doi [\mnras] {10.1046/j.1365-8711.1998.01918.x}, \href
  {http://adsabs.harvard.edu/abs/1998MNRAS.300..146G} {300, 146}

\bibitem[\protect\citeauthoryear{{Goodman} \& {Weare}}{{Goodman} \&
  {Weare}}{2010}]{goodman10}
{Goodman} J.,  {Weare} J.,  2010, \mn@doi [Communications in Applied
  Mathematics and Computational Science] {10.2140/camcos.2010.5.65}, 5, 65

\bibitem[\protect\citeauthoryear{{Gruen} \& {Brimioulle}}{{Gruen} \&
  {Brimioulle}}{2017}]{gruen17}
{Gruen} D.,  {Brimioulle} F.,  2017, \mn@doi [\mnras] {10.1093/mnras/stx471},
  \href {http://adsabs.harvard.edu/abs/2017MNRAS.468..769G} {468, 769}

\bibitem[\protect\citeauthoryear{{Gwyn}}{{Gwyn}}{2008}]{gwyn08}
{Gwyn} S.~D.~J.,  2008, \mn@doi [\pasp] {10.1086/526794}, \href
  {http://adsabs.harvard.edu/abs/2008PASP..120..212G} {120, 212}

\bibitem[\protect\citeauthoryear{{Haines} et~al.,}{{Haines}
  et~al.}{2015}]{haines15}
{Haines} C.~P.,  et~al., 2015, \mn@doi [\apj] {10.1088/0004-637X/806/1/101},
  \href {http://adsabs.harvard.edu/abs/2015ApJ...806..101H} {806, 101}

\bibitem[\protect\citeauthoryear{{Halkola}, {Seitz}  \& {Pannella}}{{Halkola}
  et~al.}{2007}]{halkola07}
{Halkola} A.,  {Seitz} S.,   {Pannella} M.,  2007, \mn@doi [\apj]
  {10.1086/510555}, \href {http://adsabs.harvard.edu/abs/2007ApJ...656..739H}
  {656, 739}

\bibitem[\protect\citeauthoryear{{Han}, {Cole}, {Frenk}  \& {Jing}}{{Han}
  et~al.}{2016}]{han16}
{Han} J.,  {Cole} S.,  {Frenk} C.~S.,   {Jing} Y.,  2016, \mn@doi [\mnras]
  {10.1093/mnras/stv2900}, \href
  {http://adsabs.harvard.edu/abs/2016MNRAS.457.1208H} {457, 1208}

\bibitem[\protect\citeauthoryear{{Hao} et~al.,}{{Hao} et~al.}{2009}]{hao09}
{Hao} J.,  et~al., 2009, \mn@doi [\apj] {10.1088/0004-637X/702/1/745}, \href
  {http://adsabs.harvard.edu/abs/2009ApJ...702..745H} {702, 745}

\bibitem[\protect\citeauthoryear{{Hayashi}, {Navarro}, {Taylor}, {Stadel}  \&
  {Quinn}}{{Hayashi} et~al.}{2003}]{hayashi03}
{Hayashi} E.,  {Navarro} J.~F.,  {Taylor} J.~E.,  {Stadel} J.,   {Quinn} T.,
  2003, \mn@doi [\apj] {10.1086/345788}, \href
  {http://adsabs.harvard.edu/abs/2003ApJ...584..541H} {584, 541}

\bibitem[\protect\citeauthoryear{{Hearin}, {Zentner}, {van den Bosch},
  {Campbell}  \& {Tollerud}}{{Hearin} et~al.}{2016}]{hearin16}
{Hearin} A.~P.,  {Zentner} A.~R.,  {van den Bosch} F.~C.,  {Campbell} D.,
  {Tollerud} E.,  2016, \mn@doi [\mnras] {10.1093/mnras/stw840}, \href
  {http://adsabs.harvard.edu/abs/2016MNRAS.460.2552H} {460, 2552}

\bibitem[\protect\citeauthoryear{{Heymans} et~al.,}{{Heymans}
  et~al.}{2006a}]{heymans06_step}
{Heymans} C.,  et~al., 2006a, \mn@doi [\mnras]
  {10.1111/j.1365-2966.2006.10198.x}, \href
  {http://adsabs.harvard.edu/abs/2006MNRAS.368.1323H} {368, 1323}

\bibitem[\protect\citeauthoryear{{Heymans} et~al.,}{{Heymans}
  et~al.}{2006b}]{heymans06_ggl}
{Heymans} C.,  et~al., 2006b, \mn@doi [\mnras]
  {10.1111/j.1745-3933.2006.00208.x}, \href
  {http://adsabs.harvard.edu/abs/2006MNRAS.371L..60H} {371, L60}

\bibitem[\protect\citeauthoryear{{Hoekstra}}{{Hoekstra}}{2001}]{hoekstra01}
{Hoekstra} H.,  2001, \mn@doi [\aap] {10.1051/0004-6361:20010293}, \href
  {http://adsabs.harvard.edu/abs/2001A%26A...370..743H} {370, 743}

\bibitem[\protect\citeauthoryear{{Hoekstra}}{{Hoekstra}}{2007}]{hoekstra07}
{Hoekstra} H.,  2007, \mn@doi [\mnras] {10.1111/j.1365-2966.2007.11951.x},
  \href {http://adsabs.harvard.edu/abs/2007MNRAS.379..317H} {379, 317}

\bibitem[\protect\citeauthoryear{{Hoekstra}, {Franx}, {Kuijken}  \&
  {Squires}}{{Hoekstra} et~al.}{1998}]{hoekstra98}
{Hoekstra} H.,  {Franx} M.,  {Kuijken} K.,   {Squires} G.,  1998, \mn@doi
  [\apj] {10.1086/306102}, \href
  {http://adsabs.harvard.edu/abs/1998ApJ...504..636H} {504, 636}

\bibitem[\protect\citeauthoryear{{Hoekstra}, {Franx}  \& {Kuijken}}{{Hoekstra}
  et~al.}{2000}]{hoekstra00}
{Hoekstra} H.,  {Franx} M.,   {Kuijken} K.,  2000, \mn@doi [\apj]
  {10.1086/308556}, \href {http://adsabs.harvard.edu/abs/2000ApJ...532...88H}
  {532, 88}

\bibitem[\protect\citeauthoryear{{Hoekstra}, {Hsieh}, {Yee}, {Lin}  \&
  {Gladders}}{{Hoekstra} et~al.}{2005}]{hoekstra05}
{Hoekstra} H.,  {Hsieh} B.~C.,  {Yee} H.~K.~C.,  {Lin} H.,   {Gladders} M.~D.,
  2005, \mn@doi [\apj] {10.1086/496913}, \href
  {http://adsabs.harvard.edu/abs/2005ApJ...635...73H} {635, 73}

\bibitem[\protect\citeauthoryear{{Hoekstra}, {Mahdavi}, {Babul}  \&
  {Bildfell}}{{Hoekstra} et~al.}{2012}]{hoekstra12}
{Hoekstra} H.,  {Mahdavi} A.,  {Babul} A.,   {Bildfell} C.,  2012, \mn@doi
  [\mnras] {10.1111/j.1365-2966.2012.22072.x}, \href
  {http://adsabs.harvard.edu/abs/2012MNRAS.427.1298H} {427, 1298}

\bibitem[\protect\citeauthoryear{{Hoekstra}, {Herbonnet}, {Muzzin}, {Babul},
  {Mahdavi}, {Viola}  \& {Cacciato}}{{Hoekstra} et~al.}{2015}]{hoekstra15}
{Hoekstra} H.,  {Herbonnet} R.,  {Muzzin} A.,  {Babul} A.,  {Mahdavi} A.,
  {Viola} M.,   {Cacciato} M.,  2015, \mn@doi [\mnras] {10.1093/mnras/stv275},
  \href {http://adsabs.harvard.edu/abs/2015MNRAS.449..685H} {449, 685}

\bibitem[\protect\citeauthoryear{{Kaiser}, {Squires}  \& {Broadhurst}}{{Kaiser}
  et~al.}{1995}]{kaiser95}
{Kaiser} N.,  {Squires} G.,   {Broadhurst} T.,  1995, \mn@doi [\apj]
  {10.1086/176071}, \href {http://adsabs.harvard.edu/abs/1995ApJ...449..460K}
  {449, 460}

\bibitem[\protect\citeauthoryear{{Kauffmann} et~al.,}{{Kauffmann}
  et~al.}{2003}]{kauffmann03_mstar}
{Kauffmann} G.,  et~al., 2003, \mn@doi [\mnras]
  {10.1046/j.1365-8711.2003.06291.x}, \href
  {http://adsabs.harvard.edu/abs/2003MNRAS.341...33K} {341, 33}

\bibitem[\protect\citeauthoryear{{Klypin}, {Gottl{\"o}ber}, {Kravtsov}  \&
  {Khokhlov}}{{Klypin} et~al.}{1999}]{klypin99}
{Klypin} A.,  {Gottl{\"o}ber} S.,  {Kravtsov} A.~V.,   {Khokhlov} A.~M.,  1999,
  \mn@doi [\apj] {10.1086/307122}, \href
  {http://adsabs.harvard.edu/abs/1999ApJ...516..530K} {516, 530}

\bibitem[\protect\citeauthoryear{{Knebe} et~al.,}{{Knebe}
  et~al.}{2011}]{knebe11}
{Knebe} A.,  et~al., 2011, \mn@doi [\mnras] {10.1111/j.1365-2966.2011.18858.x},
  \href {http://adsabs.harvard.edu/abs/2011MNRAS.415.2293K} {415, 2293}

\bibitem[\protect\citeauthoryear{{Kriek}, {van Dokkum}, {Labb{\'e}}, {Franx},
  {Illingworth}, {Marchesini}  \& {Quadri}}{{Kriek} et~al.}{2009}]{kriek09}
{Kriek} M.,  {van Dokkum} P.~G.,  {Labb{\'e}} I.,  {Franx} M.,  {Illingworth}
  G.~D.,  {Marchesini} D.,   {Quadri} R.~F.,  2009, \mn@doi [\apj]
  {10.1088/0004-637X/700/1/221}, \href
  {http://adsabs.harvard.edu/abs/2009ApJ...700..221K} {700, 221}

\bibitem[\protect\citeauthoryear{{Lacey} et~al.,}{{Lacey}
  et~al.}{2016}]{lacey16}
{Lacey} C.~G.,  et~al., 2016, \mn@doi [\mnras] {10.1093/mnras/stw1888}, \href
  {http://adsabs.harvard.edu/abs/2016MNRAS.462.3854L} {462, 3854}

\bibitem[\protect\citeauthoryear{{Laigle} et~al.,}{{Laigle}
  et~al.}{2016}]{laigle16}
{Laigle} C.,  et~al., 2016, \mn@doi [\apjs] {10.3847/0067-0049/224/2/24}, \href
  {http://adsabs.harvard.edu/abs/2016ApJS..224...24L} {224, 24}

\bibitem[\protect\citeauthoryear{{Leauthaud} et~al.,}{{Leauthaud}
  et~al.}{2012}]{leauthaud12}
{Leauthaud} A.,  et~al., 2012, \mn@doi [\apj] {10.1088/0004-637X/744/2/159},
  \href {http://adsabs.harvard.edu/abs/2012ApJ...744..159L} {744, 159}

\bibitem[\protect\citeauthoryear{{Li}, {Mo}, {Fan}, {Yang}  \& {Bosch}}{{Li}
  et~al.}{2013a}]{li13_ggl}
{Li} R.,  {Mo} H.~J.,  {Fan} Z.,  {Yang} X.,   {Bosch} F.~C.~v.~d.,  2013a,
  \mn@doi [\mnras] {10.1093/mnras/stt133}, \href
  {http://adsabs.harvard.edu/abs/2013MNRAS.430.3359L} {430, 3359}

\bibitem[\protect\citeauthoryear{{Li}, {Wang}  \& {Jing}}{{Li}
  et~al.}{2013b}]{li13_massproxy}
{Li} C.,  {Wang} L.,   {Jing} Y.~P.,  2013b, \mn@doi [\apjl]
  {10.1088/2041-8205/762/1/L7}, \href
  {http://adsabs.harvard.edu/abs/2013ApJ...762L...7L} {762, L7}

\bibitem[\protect\citeauthoryear{{Li} et~al.,}{{Li} et~al.}{2014}]{li14}
{Li} R.,  et~al., 2014, \mn@doi [\mnras] {10.1093/mnras/stt2395}, \href
  {http://adsabs.harvard.edu/abs/2014MNRAS.438.2864L} {438, 2864}

\bibitem[\protect\citeauthoryear{{Li} et~al.,}{{Li} et~al.}{2016}]{li16}
{Li} R.,  et~al., 2016, \mn@doi [\mnras] {10.1093/mnras/stw494}, \href
  {http://adsabs.harvard.edu/abs/2016MNRAS.458.2573L} {458, 2573}

\bibitem[\protect\citeauthoryear{{Limousin}, {Kneib}  \&
  {Natarajan}}{{Limousin} et~al.}{2005}]{limousin05}
{Limousin} M.,  {Kneib} J.-P.,   {Natarajan} P.,  2005, \mn@doi [\mnras]
  {10.1111/j.1365-2966.2004.08449.x}, \href
  {http://adsabs.harvard.edu/abs/2005MNRAS.356..309L} {356, 309}

\bibitem[\protect\citeauthoryear{{Limousin}, {Kneib}, {Bardeau}, {Natarajan},
  {Czoske}, {Smail}, {Ebeling}  \& {Smith}}{{Limousin}
  et~al.}{2007}]{limousin07}
{Limousin} M.,  {Kneib} J.~P.,  {Bardeau} S.,  {Natarajan} P.,  {Czoske} O.,
  {Smail} I.,  {Ebeling} H.,   {Smith} G.~P.,  2007, \mn@doi [\aap]
  {10.1051/0004-6361:20065543}, \href
  {http://adsabs.harvard.edu/abs/2007A%26A...461..881L} {461, 881}

\bibitem[\protect\citeauthoryear{{Luppino} \& {Kaiser}}{{Luppino} \&
  {Kaiser}}{1997}]{luppino97}
{Luppino} G.~A.,  {Kaiser} N.,  1997, \apj, \href
  {http://adsabs.harvard.edu/abs/1997ApJ...475...20L} {475, 20}

\bibitem[\protect\citeauthoryear{{Macci{\`o}}, {Dutton}  \& {van den
  Bosch}}{{Macci{\`o}} et~al.}{2008}]{maccio08}
{Macci{\`o}} A.~V.,  {Dutton} A.~A.,   {van den Bosch} F.~C.,  2008, \mn@doi
  [\mnras] {10.1111/j.1365-2966.2008.14029.x}, \href
  {http://adsabs.harvard.edu/abs/2008MNRAS.391.1940M} {391, 1940}

\bibitem[\protect\citeauthoryear{{Mancone} \& {Gonzalez}}{{Mancone} \&
  {Gonzalez}}{2012}]{mancone12}
{Mancone} C.~L.,  {Gonzalez} A.~H.,  2012, \mn@doi [\pasp] {10.1086/666502},
  \href {http://adsabs.harvard.edu/abs/2012PASP..124..606M} {124, 606}

\bibitem[\protect\citeauthoryear{{Mandelbaum} et~al.,}{{Mandelbaum}
  et~al.}{2005a}]{mandelbaum05_errors}
{Mandelbaum} R.,  et~al., 2005a, \mn@doi [\mnras]
  {10.1111/j.1365-2966.2005.09282.x}, \href
  {http://adsabs.harvard.edu/abs/2005MNRAS.361.1287M} {361, 1287}

\bibitem[\protect\citeauthoryear{{Mandelbaum}, {Tasitsiomi}, {Seljak},
  {Kravtsov}  \& {Wechsler}}{{Mandelbaum} et~al.}{2005b}]{mandelbaum05_ggl}
{Mandelbaum} R.,  {Tasitsiomi} A.,  {Seljak} U.,  {Kravtsov} A.~V.,
  {Wechsler} R.~H.,  2005b, \mn@doi [\mnras]
  {10.1111/j.1365-2966.2005.09417.x}, \href
  {http://adsabs.harvard.edu/abs/2005MNRAS.362.1451M} {362, 1451}

\bibitem[\protect\citeauthoryear{{Mandelbaum}, {Seljak}, {Kauffmann}, {Hirata}
  \& {Brinkmann}}{{Mandelbaum} et~al.}{2006}]{mandelbaum06_ggl}
{Mandelbaum} R.,  {Seljak} U.,  {Kauffmann} G.,  {Hirata} C.~M.,   {Brinkmann}
  J.,  2006, \mn@doi [\mnras] {10.1111/j.1365-2966.2006.10156.x}, \href
  {http://adsabs.harvard.edu/abs/2006MNRAS.368..715M} {368, 715}

\bibitem[\protect\citeauthoryear{{Mandelbaum}, {Wang}, {Zu}, {White},
  {Henriques}  \& {More}}{{Mandelbaum} et~al.}{2016}]{mandelbaum16}
{Mandelbaum} R.,  {Wang} W.,  {Zu} Y.,  {White} S.,  {Henriques} B.,   {More}
  S.,  2016, \mn@doi [\mnras] {10.1093/mnras/stw188}, \href
  {http://adsabs.harvard.edu/abs/2016MNRAS.457.3200M} {457, 3200}

\bibitem[\protect\citeauthoryear{{McGee}, {Balogh}, {Bower}, {Font}  \&
  {McCarthy}}{{McGee} et~al.}{2009}]{mcgee09}
{McGee} S.~L.,  {Balogh} M.~L.,  {Bower} R.~G.,  {Font} A.~S.,   {McCarthy}
  I.~G.,  2009, \mn@doi [\mnras] {10.1111/j.1365-2966.2009.15507.x}, \href
  {http://adsabs.harvard.edu/abs/2009MNRAS.400..937M} {400, 937}

\bibitem[\protect\citeauthoryear{{Mellier}}{{Mellier}}{1999}]{mellier99}
{Mellier} Y.,  1999, \mn@doi [\araa] {10.1146/annurev.astro.37.1.127}, \href
  {http://adsabs.harvard.edu/abs/1999ARA%26A..37..127M} {37, 127}

\bibitem[\protect\citeauthoryear{{Molin{\'e}}, {S{\'a}nchez-Conde},
  {Palomares-Ruiz}  \& {Prada}}{{Molin{\'e}} et~al.}{2017}]{moline17}
{Molin{\'e}} {\'A}.,  {S{\'a}nchez-Conde} M.~A.,  {Palomares-Ruiz} S.,
  {Prada} F.,  2017, \mn@doi [\mnras] {10.1093/mnras/stx026}, \href
  {http://adsabs.harvard.edu/abs/2017MNRAS.466.4974M} {466, 4974}

\bibitem[\protect\citeauthoryear{{Monna} et~al.,}{{Monna}
  et~al.}{2015}]{monna15}
{Monna} A.,  et~al., 2015, \mn@doi [\mnras] {10.1093/mnras/stu2534}, \href
  {http://adsabs.harvard.edu/abs/2015MNRAS.447.1224M} {447, 1224}

\bibitem[\protect\citeauthoryear{{Monna} et~al.,}{{Monna}
  et~al.}{2017}]{monna17_macs}
{Monna} A.,  et~al., 2017, \mn@doi [\mnras] {10.1093/mnras/stx015}, \href
  {http://adsabs.harvard.edu/abs/2017MNRAS.466.4094M} {466, 4094}

\bibitem[\protect\citeauthoryear{{Moore}, {Ghigna}, {Governato}, {Lake},
  {Quinn}, {Stadel}  \& {Tozzi}}{{Moore} et~al.}{1999}]{moore99}
{Moore} B.,  {Ghigna} S.,  {Governato} F.,  {Lake} G.,  {Quinn} T.,  {Stadel}
  J.,   {Tozzi} P.,  1999, \mn@doi [\apjl] {10.1086/312287}, \href
  {http://adsabs.harvard.edu/abs/1999ApJ...524L..19M} {524, L19}

\bibitem[\protect\citeauthoryear{{More}, {van den Bosch}, {Cacciato}, {Skibba},
  {Mo}  \& {Yang}}{{More} et~al.}{2011}]{more11}
{More} S.,  {van den Bosch} F.~C.,  {Cacciato} M.,  {Skibba} R.,  {Mo} H.~J.,
  {Yang} X.,  2011, \mn@doi [\mnras] {10.1111/j.1365-2966.2010.17436.x}, \href
  {http://adsabs.harvard.edu/abs/2011MNRAS.410..210M} {410, 210}

\bibitem[\protect\citeauthoryear{{Natarajan} \& {Kneib}}{{Natarajan} \&
  {Kneib}}{1997}]{natarajan97}
{Natarajan} P.,  {Kneib} J.-P.,  1997, \mn@doi [\mnras]
  {10.1093/mnras/287.4.833}, \href
  {http://adsabs.harvard.edu/abs/1997MNRAS.287..833N} {287, 833}

\bibitem[\protect\citeauthoryear{{Natarajan}, {Kneib}, {Smail}  \&
  {Ellis}}{{Natarajan} et~al.}{1998}]{natarajan98}
{Natarajan} P.,  {Kneib} J.-P.,  {Smail} I.,   {Ellis} R.~S.,  1998, \apj,
  \href {http://adsabs.harvard.edu/abs/1998ApJ...499..600N} {499, 600}

\bibitem[\protect\citeauthoryear{{Natarajan}, {Kneib}  \& {Smail}}{{Natarajan}
  et~al.}{2002}]{natarajan02}
{Natarajan} P.,  {Kneib} J.-P.,   {Smail} I.,  2002, \mn@doi [\apjl]
  {10.1086/345399}, \href {http://adsabs.harvard.edu/abs/2002ApJ...580L..11N}
  {580, L11}

\bibitem[\protect\citeauthoryear{{Natarajan}, {De Lucia}  \&
  {Springel}}{{Natarajan} et~al.}{2007}]{natarajan07}
{Natarajan} P.,  {De Lucia} G.,   {Springel} V.,  2007, \mn@doi [\mnras]
  {10.1111/j.1365-2966.2007.11399.x}, \href
  {http://adsabs.harvard.edu/abs/2007MNRAS.376..180N} {376, 180}

\bibitem[\protect\citeauthoryear{{Natarajan}, {Kneib}, {Smail}, {Treu},
  {Ellis}, {Moran}, {Limousin}  \& {Czoske}}{{Natarajan}
  et~al.}{2009}]{natarajan09}
{Natarajan} P.,  {Kneib} J.-P.,  {Smail} I.,  {Treu} T.,  {Ellis} R.,  {Moran}
  S.,  {Limousin} M.,   {Czoske} O.,  2009, \mn@doi [\apj]
  {10.1088/0004-637X/693/1/970}, \href
  {http://adsabs.harvard.edu/abs/2009ApJ...693..970N} {693, 970}

\bibitem[\protect\citeauthoryear{{Navarro}, {Frenk}  \& {White}}{{Navarro}
  et~al.}{1995}]{nfw95}
{Navarro} J.~F.,  {Frenk} C.~S.,   {White} S.~D.~M.,  1995, \mn@doi [\mnras]
  {10.1093/mnras/275.3.720}, \href
  {http://adsabs.harvard.edu/abs/1995MNRAS.275..720N} {275, 720}

\bibitem[\protect\citeauthoryear{{Niemiec} et~al.,}{{Niemiec}
  et~al.}{2017}]{niemiec17}
{Niemiec} A.,  et~al., 2017, \mn@doi [\mnras] {10.1093/mnras/stx1667}, \href
  {http://adsabs.harvard.edu/abs/2017MNRAS.471.1153N} {471, 1153}

\bibitem[\protect\citeauthoryear{{Oguri}, {Bayliss}, {Dahle}, {Sharon},
  {Gladders}, {Natarajan}, {Hennawi}  \& {Koester}}{{Oguri}
  et~al.}{2012}]{oguri12}
{Oguri} M.,  {Bayliss} M.~B.,  {Dahle} H.,  {Sharon} K.,  {Gladders} M.~D.,
  {Natarajan} P.,  {Hennawi} J.~F.,   {Koester} B.~P.,  2012, \mn@doi [\mnras]
  {10.1111/j.1365-2966.2011.20248.x}, \href
  {http://adsabs.harvard.edu/abs/2012MNRAS.420.3213O} {420, 3213}

\bibitem[\protect\citeauthoryear{{Okabe}, {Futamase}, {Kajisawa}  \&
  {Kuroshima}}{{Okabe} et~al.}{2014}]{okabe14}
{Okabe} N.,  {Futamase} T.,  {Kajisawa} M.,   {Kuroshima} R.,  2014, \mn@doi
  [\apj] {10.1088/0004-637X/784/2/90}, \href
  {http://adsabs.harvard.edu/abs/2014ApJ...784...90O} {784, 90}

\bibitem[\protect\citeauthoryear{{Old} et~al.,}{{Old} et~al.}{2015}]{old15}
{Old} L.,  et~al., 2015, \mn@doi [\mnras] {10.1093/mnras/stv421}, \href
  {http://adsabs.harvard.edu/abs/2015MNRAS.449.1897O} {449, 1897}

\bibitem[\protect\citeauthoryear{{Pastor Mira}, {Hilbert}, {Hartlap}  \&
  {Schneider}}{{Pastor Mira} et~al.}{2011}]{pastormira11}
{Pastor Mira} E.,  {Hilbert} S.,  {Hartlap} J.,   {Schneider} P.,  2011,
  \mn@doi [\aap] {10.1051/0004-6361/201116851}, \href
  {http://adsabs.harvard.edu/abs/2011A%26A...531A.169P} {531, A169}

\bibitem[\protect\citeauthoryear{{Peacock} \& {Smith}}{{Peacock} \&
  {Smith}}{2000}]{peacock00}
{Peacock} J.~A.,  {Smith} R.~E.,  2000, \mn@doi [\mnras]
  {10.1046/j.1365-8711.2000.03779.x}, \href
  {http://adsabs.harvard.edu/abs/2000MNRAS.318.1144P} {318, 1144}

\bibitem[\protect\citeauthoryear{{Peng}, {Ho}, {Impey}  \& {Rix}}{{Peng}
  et~al.}{2002}]{peng02}
{Peng} C.~Y.,  {Ho} L.~C.,  {Impey} C.~D.,   {Rix} H.-W.,  2002, \mn@doi [\aj]
  {10.1086/340952}, \href {http://adsabs.harvard.edu/abs/2002AJ....124..266P}
  {124, 266}

\bibitem[\protect\citeauthoryear{{Planck Collaboration} et~al.,}{{Planck
  Collaboration} et~al.}{2016}]{planck15xiii}
{Planck Collaboration} et~al., 2016, \mn@doi [\aap]
  {10.1051/0004-6361/201525830}, \href
  {http://adsabs.harvard.edu/abs/2016A%26A...594A..13P} {594, A13}

\bibitem[\protect\citeauthoryear{{Prada}, {Klypin}, {Cuesta}, {Betancort-Rijo}
  \& {Primack}}{{Prada} et~al.}{2012}]{prada12}
{Prada} F.,  {Klypin} A.~A.,  {Cuesta} A.~J.,  {Betancort-Rijo} J.~E.,
  {Primack} J.,  2012, \mn@doi [\mnras] {10.1111/j.1365-2966.2012.21007.x},
  \href {http://adsabs.harvard.edu/abs/2012MNRAS.423.3018P} {423, 3018}

\bibitem[\protect\citeauthoryear{{Rines}, {Geller}, {Diaferio}  \&
  {Hwang}}{{Rines} et~al.}{2016}]{rines16}
{Rines} K.~J.,  {Geller} M.~J.,  {Diaferio} A.,   {Hwang} H.~S.,  2016, \mn@doi
  [\apj] {10.3847/0004-637X/819/1/63}, \href
  {http://adsabs.harvard.edu/abs/2016ApJ...819...63R} {819, 63}

\bibitem[\protect\citeauthoryear{{Rix} et~al.,}{{Rix} et~al.}{2004}]{rix04}
{Rix} H.-W.,  et~al., 2004, \mn@doi [\apjs] {10.1086/420885}, \href
  {http://adsabs.harvard.edu/abs/2004ApJS..152..163R} {152, 163}

\bibitem[\protect\citeauthoryear{{Roberts}, {Parker}, {Joshi}  \&
  {Evans}}{{Roberts} et~al.}{2015}]{roberts15}
{Roberts} I.~D.,  {Parker} L.~C.,  {Joshi} G.~D.,   {Evans} F.~A.,  2015,
  \mn@doi [\mnras] {10.1093/mnrasl/slu188}, \href
  {http://adsabs.harvard.edu/abs/2015MNRAS.448L...1R} {448, L1}

\bibitem[\protect\citeauthoryear{{Rodr{\'{\i}}guez-Puebla}, {Drory}  \&
  {Avila-Reese}}{{Rodr{\'{\i}}guez-Puebla} et~al.}{2012}]{rodriguez12}
{Rodr{\'{\i}}guez-Puebla} A.,  {Drory} N.,   {Avila-Reese} V.,  2012, \mn@doi
  [\apj] {10.1088/0004-637X/756/1/2}, \href
  {http://adsabs.harvard.edu/abs/2012ApJ...756....2R} {756, 2}

\bibitem[\protect\citeauthoryear{{Rodr{\'{\i}}guez-Puebla}, {Avila-Reese}  \&
  {Drory}}{{Rodr{\'{\i}}guez-Puebla} et~al.}{2013}]{rodriguez13}
{Rodr{\'{\i}}guez-Puebla} A.,  {Avila-Reese} V.,   {Drory} N.,  2013, \mn@doi
  [\apj] {10.1088/0004-637X/767/1/92}, \href
  {http://adsabs.harvard.edu/abs/2013ApJ...767...92R} {767, 92}

\bibitem[\protect\citeauthoryear{{Rowe} et~al.,}{{Rowe} et~al.}{2015}]{rowe15}
{Rowe} B.~T.~P.,  et~al., 2015, \mn@doi [Astronomy and Computing]
  {10.1016/j.ascom.2015.02.002}, \href
  {http://adsabs.harvard.edu/abs/2015A%26C....10..121R} {10, 121}

\bibitem[\protect\citeauthoryear{{Rykoff} et~al.,}{{Rykoff}
  et~al.}{2014}]{rykoff14}
{Rykoff} E.~S.,  et~al., 2014, \mn@doi [\apj] {10.1088/0004-637X/785/2/104},
  \href {http://adsabs.harvard.edu/abs/2014ApJ...785..104R} {785, 104}

\bibitem[\protect\citeauthoryear{{Sand} et~al.,}{{Sand} et~al.}{2012}]{sand12}
{Sand} D.~J.,  et~al., 2012, \mn@doi [\apj] {10.1088/0004-637X/746/2/163},
  \href {http://adsabs.harvard.edu/abs/2012ApJ...746..163S} {746, 163}

\bibitem[\protect\citeauthoryear{{Schaye} et~al.,}{{Schaye}
  et~al.}{2015}]{schaye15}
{Schaye} J.,  et~al., 2015, \mn@doi [\mnras] {10.1093/mnras/stu2058}, \href
  {http://adsabs.harvard.edu/abs/2015MNRAS.446..521S} {446, 521}

\bibitem[\protect\citeauthoryear{{Schechter}}{{Schechter}}{1976}]{schechter76}
{Schechter} P.,  1976, \mn@doi [\apj] {10.1086/154079}, \href
  {http://adsabs.harvard.edu/abs/1976ApJ...203..297S} {203, 297}

\bibitem[\protect\citeauthoryear{{Schlafly} \& {Finkbeiner}}{{Schlafly} \&
  {Finkbeiner}}{2011}]{schlafly11}
{Schlafly} E.~F.,  {Finkbeiner} D.~P.,  2011, \mn@doi [\apj]
  {10.1088/0004-637X/737/2/103}, \href
  {http://adsabs.harvard.edu/abs/2011ApJ...737..103S} {737, 103}

\bibitem[\protect\citeauthoryear{{Schlegel}, {Finkbeiner}  \&
  {Davis}}{{Schlegel} et~al.}{1998}]{schlegel98}
{Schlegel} D.~J.,  {Finkbeiner} D.~P.,   {Davis} M.,  1998, \mn@doi [\apj]
  {10.1086/305772}, \href {http://adsabs.harvard.edu/abs/1998ApJ...500..525S}
  {500, 525}

\bibitem[\protect\citeauthoryear{{Schneider}}{{Schneider}}{2003}]{schneider03}
{Schneider} P.,  2003, \mn@doi [\aap] {10.1051/0004-6361:20031035}, \href
  {http://adsabs.harvard.edu/abs/2003A%26A...408..829S} {408, 829}

\bibitem[\protect\citeauthoryear{{Schrabback} et~al.,}{{Schrabback}
  et~al.}{2018}]{schrabback18}
{Schrabback} T.,  et~al., 2018, \mn@doi [\mnras] {10.1093/mnras/stx2666}, \href
  {http://adsabs.harvard.edu/abs/2018MNRAS.474.2635S} {474, 2635}

\bibitem[\protect\citeauthoryear{{Seljak}}{{Seljak}}{2000}]{seljak00}
{Seljak} U.,  2000, \mn@doi [\mnras] {10.1046/j.1365-8711.2000.03715.x}, \href
  {http://adsabs.harvard.edu/abs/2000MNRAS.318..203S} {318, 203}

\bibitem[\protect\citeauthoryear{{S\'{e}rsic}}{{S\'{e}rsic}}{1968}]{sersic68}
{S\'{e}rsic} J.~L.,  1968, {Atlas de galaxias australes}

\bibitem[\protect\citeauthoryear{{Sif{\'o}n} et~al.,}{{Sif{\'o}n}
  et~al.}{2015a}]{sifon15_kids}
{Sif{\'o}n} C.,  et~al., 2015a, \mn@doi [\mnras] {10.1093/mnras/stv2051}, \href
  {http://adsabs.harvard.edu/abs/2015MNRAS.454.3938S} {454, 3938}

\bibitem[\protect\citeauthoryear{{Sif{\'o}n}, {Hoekstra}, {Cacciato}, {Viola},
  {K{\"o}hlinger}, {van der Burg}, {Sand}  \& {Graham}}{{Sif{\'o}n}
  et~al.}{2015b}]{sifon15_cccp}
{Sif{\'o}n} C.,  {Hoekstra} H.,  {Cacciato} M.,  {Viola} M.,  {K{\"o}hlinger}
  F.,  {van der Burg} R.~F.~J.,  {Sand} D.~J.,   {Graham} M.~L.,  2015b,
  \mn@doi [\aap] {10.1051/0004-6361/201424435}, \href
  {http://adsabs.harvard.edu/abs/2015A%26A...575A..48S} {575, A48}

\bibitem[\protect\citeauthoryear{{Sif{\'o}n}, {van der Burg}, {Hoekstra},
  {Muzzin}  \& {Herbonnet}}{{Sif{\'o}n} et~al.}{2018}]{sifon18_udg}
{Sif{\'o}n} C.,  {van der Burg} R.~F.~J.,  {Hoekstra} H.,  {Muzzin} A.,
  {Herbonnet} R.,  2018, \mn@doi [\mnras] {10.1093/mnras/stx2648}, \href
  {http://adsabs.harvard.edu/abs/2018MNRAS.473.3747S} {473, 3747}

\bibitem[\protect\citeauthoryear{{Springel}, {White}, {Tormen}  \&
  {Kauffmann}}{{Springel} et~al.}{2001}]{springel01_cluster}
{Springel} V.,  {White} S.~D.~M.,  {Tormen} G.,   {Kauffmann} G.,  2001,
  \mn@doi [\mnras] {10.1046/j.1365-8711.2001.04912.x}, \href
  {http://adsabs.harvard.edu/abs/2001MNRAS.328..726S} {328, 726}

\bibitem[\protect\citeauthoryear{{Springel} et~al.,}{{Springel}
  et~al.}{2005}]{springel05_millenium}
{Springel} V.,  et~al., 2005, \mn@doi [\nat] {10.1038/nature03597}, \href
  {http://adsabs.harvard.edu/abs/2005Natur.435..629S} {435, 629}

\bibitem[\protect\citeauthoryear{{Springel} et~al.,}{{Springel}
  et~al.}{2008}]{springel08}
{Springel} V.,  et~al., 2008, \mn@doi [\mnras]
  {10.1111/j.1365-2966.2008.14066.x}, \href
  {http://adsabs.harvard.edu/abs/2008MNRAS.391.1685S} {391, 1685}

\bibitem[\protect\citeauthoryear{{Suyu} \& {Halkola}}{{Suyu} \&
  {Halkola}}{2010}]{suyu10}
{Suyu} S.~H.,  {Halkola} A.,  2010, \mn@doi [\aap]
  {10.1051/0004-6361/201015481}, \href
  {http://adsabs.harvard.edu/abs/2010A%26A...524A..94S} {524, A94}

\bibitem[\protect\citeauthoryear{{Taffoni}, {Mayer}, {Colpi}  \&
  {Governato}}{{Taffoni} et~al.}{2003}]{taffoni03}
{Taffoni} G.,  {Mayer} L.,  {Colpi} M.,   {Governato} F.,  2003, \mn@doi
  [\mnras] {10.1046/j.1365-8711.2003.06395.x}, \href
  {http://adsabs.harvard.edu/abs/2003MNRAS.341..434T} {341, 434}

\bibitem[\protect\citeauthoryear{{Taylor} \& {Babul}}{{Taylor} \&
  {Babul}}{2005}]{taylor05}
{Taylor} J.~E.,  {Babul} A.,  2005, \mn@doi [\mnras]
  {10.1111/j.1365-2966.2005.09581.x}, \href
  {http://adsabs.harvard.edu/abs/2005MNRAS.364..535T} {364, 535}

\bibitem[\protect\citeauthoryear{{Taylor} et~al.,}{{Taylor}
  et~al.}{2011}]{taylor11}
{Taylor} E.~N.,  et~al., 2011, \mn@doi [\mnras]
  {10.1111/j.1365-2966.2011.19536.x}, \href
  {http://adsabs.harvard.edu/abs/2011MNRAS.418.1587T} {418, 1587}

\bibitem[\protect\citeauthoryear{{Tormen}, {Diaferio}  \& {Syer}}{{Tormen}
  et~al.}{1998}]{tormen98}
{Tormen} G.,  {Diaferio} A.,   {Syer} D.,  1998, \mn@doi [\mnras]
  {10.1046/j.1365-8711.1998.01775.x}, \href
  {http://adsabs.harvard.edu/abs/1998MNRAS.299..728T} {299, 728}

\bibitem[\protect\citeauthoryear{{Umetsu} et~al.,}{{Umetsu}
  et~al.}{2014}]{umetsu14}
{Umetsu} K.,  et~al., 2014, \mn@doi [\apj] {10.1088/0004-637X/795/2/163}, \href
  {http://adsabs.harvard.edu/abs/2014ApJ...795..163U} {795, 163}

\bibitem[\protect\citeauthoryear{{Umetsu}, {Zitrin}, {Gruen}, {Merten},
  {Donahue}  \& {Postman}}{{Umetsu} et~al.}{2016}]{umetsu16}
{Umetsu} K.,  {Zitrin} A.,  {Gruen} D.,  {Merten} J.,  {Donahue} M.,
  {Postman} M.,  2016, \mn@doi [\apj] {10.3847/0004-637X/821/2/116}, \href
  {http://adsabs.harvard.edu/abs/2016ApJ...821..116U} {821, 116}

\bibitem[\protect\citeauthoryear{{Velander} et~al.,}{{Velander}
  et~al.}{2014}]{velander14}
{Velander} M.,  et~al., 2014, \mn@doi [\mnras] {10.1093/mnras/stt2013}, \href
  {http://adsabs.harvard.edu/abs/2014MNRAS.437.2111V} {437, 2111}

\bibitem[\protect\citeauthoryear{{Velliscig} et~al.,}{{Velliscig}
  et~al.}{2017}]{velliscig17}
{Velliscig} M.,  et~al., 2017, \mn@doi [\mnras] {10.1093/mnras/stx1789}, \href
  {http://adsabs.harvard.edu/abs/2017MNRAS.471.2856V} {471, 2856}

\bibitem[\protect\citeauthoryear{{Viola} et~al.,}{{Viola}
  et~al.}{2015}]{viola15}
{Viola} M.,  et~al., 2015, \mn@doi [\mnras] {10.1093/mnras/stv1447}, \href
  {http://adsabs.harvard.edu/abs/2015MNRAS.452.3529V} {452, 3529}

\bibitem[\protect\citeauthoryear{{Wang}, {De Lucia}  \& {Weinmann}}{{Wang}
  et~al.}{2013}]{wang13}
{Wang} L.,  {De Lucia} G.,   {Weinmann} S.~M.,  2013, \mn@doi [\mnras]
  {10.1093/mnras/stt188}, \href
  {http://adsabs.harvard.edu/abs/2013MNRAS.431..600W} {431, 600}

\bibitem[\protect\citeauthoryear{{Yang}, {Mo}, {van den Bosch}, {Jing},
  {Weinmann}  \& {Meneghetti}}{{Yang} et~al.}{2006}]{yang06}
{Yang} X.,  {Mo} H.~J.,  {van den Bosch} F.~C.,  {Jing} Y.~P.,  {Weinmann}
  S.~M.,   {Meneghetti} M.,  2006, \mn@doi [\mnras]
  {10.1111/j.1365-2966.2006.11091.x}, \href
  {http://adsabs.harvard.edu/abs/2006MNRAS.373.1159Y} {373, 1159}

\bibitem[\protect\citeauthoryear{{Yang}, {Mo}  \& {van den Bosch}}{{Yang}
  et~al.}{2009}]{yang09}
{Yang} X.,  {Mo} H.~J.,   {van den Bosch} F.~C.,  2009, \mn@doi [\apj]
  {10.1088/0004-637X/693/1/830}, \href
  {http://adsabs.harvard.edu/abs/2009ApJ...693..830Y} {693, 830}

\bibitem[\protect\citeauthoryear{{Zolotov} et~al.,}{{Zolotov}
  et~al.}{2012}]{zolotov12}
{Zolotov} A.,  et~al., 2012, \mn@doi [\apj] {10.1088/0004-637X/761/1/71}, \href
  {http://adsabs.harvard.edu/abs/2012ApJ...761...71Z} {761, 71}

\bibitem[\protect\citeauthoryear{{Zu} \& {Mandelbaum}}{{Zu} \&
  {Mandelbaum}}{2015}]{zu15}
{Zu} Y.,  {Mandelbaum} R.,  2015, \mn@doi [\mnras] {10.1093/mnras/stv2062},
  \href {http://adsabs.harvard.edu/abs/2015MNRAS.454.1161Z} {454, 1161}

\bibitem[\protect\citeauthoryear{{Zu}, {Mandelbaum}, {Simet}, {Rozo}  \&
  {Rykoff}}{{Zu} et~al.}{2017}]{zu17}
{Zu} Y.,  {Mandelbaum} R.,  {Simet} M.,  {Rozo} E.,   {Rykoff} E.~S.,  2017,
  \mn@doi [\mnras] {10.1093/mnras/stx1264}, \href
  {http://adsabs.harvard.edu/abs/2017MNRAS.470..551Z} {470, 551}

\bibitem[\protect\citeauthoryear{{van Uitert}, {Hoekstra}, {Velander},
  {Gilbank}, {Gladders}  \& {Yee}}{{van Uitert} et~al.}{2011}]{vanuitert11}
{van Uitert} E.,  {Hoekstra} H.,  {Velander} M.,  {Gilbank} D.~G.,  {Gladders}
  M.~D.,   {Yee} H.~K.~C.,  2011, \mn@doi [\aap] {10.1051/0004-6361/201117308},
  \href {http://adsabs.harvard.edu/abs/2011A%26A...534A..14V} {534, A14}

\bibitem[\protect\citeauthoryear{{van Uitert} et~al.,}{{van Uitert}
  et~al.}{2016}]{vanuitert16}
{van Uitert} E.,  et~al., 2016, \mn@doi [\mnras] {10.1093/mnras/stw747}, \href
  {http://adsabs.harvard.edu/abs/2016MNRAS.tmp..601V} {}

\bibitem[\protect\citeauthoryear{{van Uitert} et~al.,}{{van Uitert}
  et~al.}{2017}]{vanuitert17}
{van Uitert} E.,  et~al., 2017, \mn@doi [\mnras] {10.1093/mnras/stx344}, \href
  {http://adsabs.harvard.edu/abs/2017MNRAS.467.4131V} {467, 4131}

\bibitem[\protect\citeauthoryear{{van den Bosch}}{{van den
  Bosch}}{2017}]{vdbosch17}
{van den Bosch} F.~C.,  2017, \mn@doi [\mnras] {10.1093/mnras/stx520}, \href
  {http://adsabs.harvard.edu/abs/2017MNRAS.468..885V} {468, 885}

\bibitem[\protect\citeauthoryear{{van den Bosch}, {More}, {Cacciato}, {Mo}  \&
  {Yang}}{{van den Bosch} et~al.}{2013}]{vdbosch13}
{van den Bosch} F.~C.,  {More} S.,  {Cacciato} M.,  {Mo} H.,   {Yang} X.,
  2013, \mn@doi [\mnras] {10.1093/mnras/sts006}, \href
  {http://adsabs.harvard.edu/abs/2013MNRAS.430..725V} {430, 725}

\bibitem[\protect\citeauthoryear{{van den Bosch}, {Jiang}, {Campbell}  \&
  {Behroozi}}{{van den Bosch} et~al.}{2016}]{vdbosch16}
{van den Bosch} F.~C.,  {Jiang} F.,  {Campbell} D.,   {Behroozi} P.,  2016,
  \mn@doi [\mnras] {10.1093/mnras/stv2338}, \href
  {http://adsabs.harvard.edu/abs/2016MNRAS.455..158V} {455, 158}

\bibitem[\protect\citeauthoryear{{van der Burg} et~al.,}{{van der Burg}
  et~al.}{2013}]{vdburg13}
{van der Burg} R.~F.~J.,  et~al., 2013, \mn@doi [\aap]
  {10.1051/0004-6361/201321237}, \href
  {http://adsabs.harvard.edu/abs/2013A%26A...557A..15V} {557, A15}

\bibitem[\protect\citeauthoryear{{van der Burg}, {Hoekstra}, {Muzzin},
  {Sif{\'o}n}, {Balogh}  \& {McGee}}{{van der Burg} et~al.}{2015}]{vdburg15}
{van der Burg} R.~F.~J.,  {Hoekstra} H.,  {Muzzin} A.,  {Sif{\'o}n} C.,
  {Balogh} M.~L.,   {McGee} S.~L.,  2015, \mn@doi [\aap]
  {10.1051/0004-6361/201425460}, \href
  {http://adsabs.harvard.edu/abs/2015A%26A...577A..19V} {577, A19}

\makeatother
\end{thebibliography}


\appendix

\section{Lens-induced biases on the shape measurements}
\label{ap:ct}

\begin{figure}
 \centerline{\includegraphics[width=3.3in]{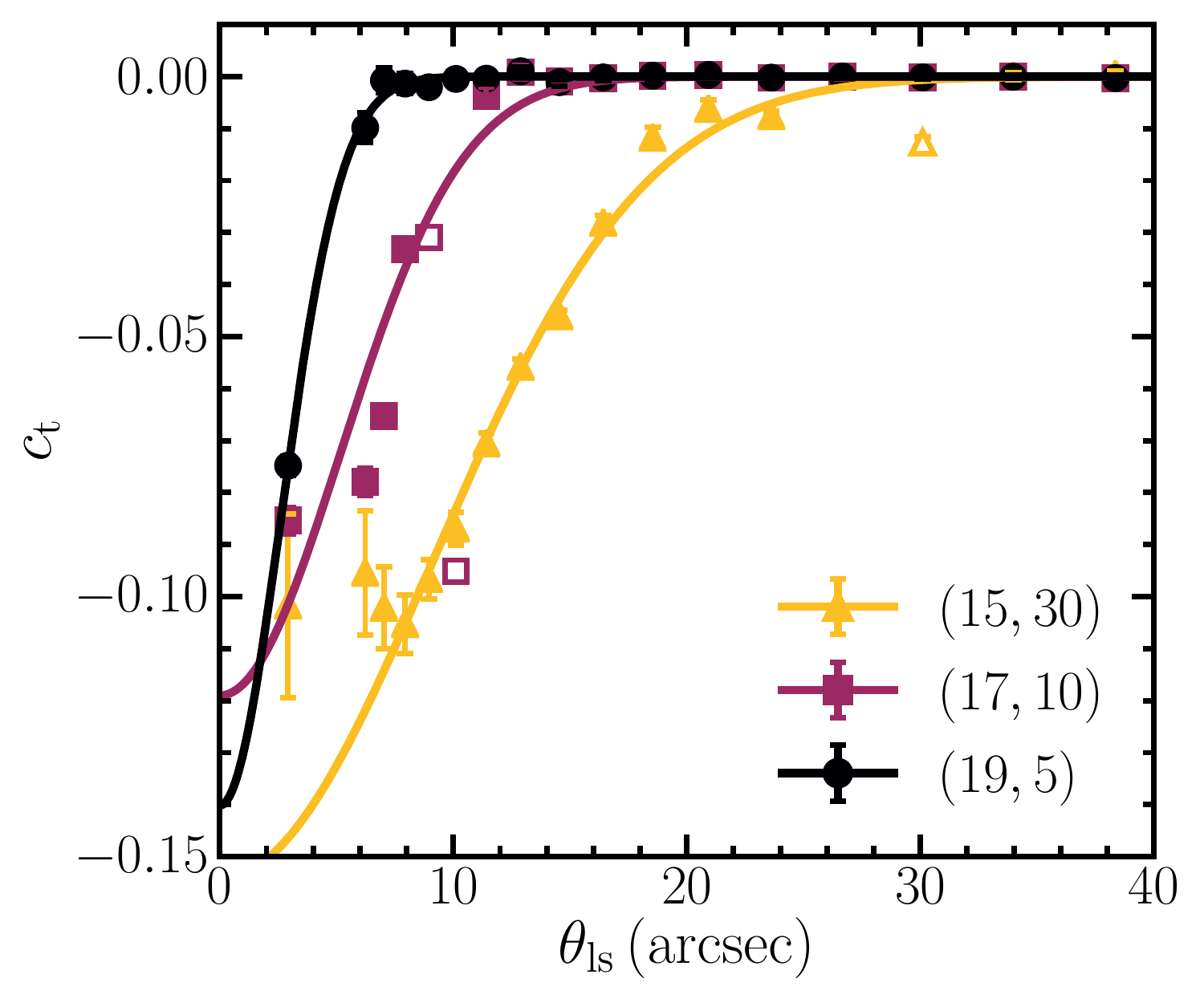}}
\caption{Additive tangential shear biases measured in three sets of image simulations as shown in the legend, which shows the magnitude and size (in pixels $=0.185$ arcseconds) of each set. The three examples correspond to big bright (yellow triangles), average (purple squares), and small faint (black circles) simulated lenses, and illustrate the range of biases. The relevance of each set with respect to the real satellite galaxies can be seen in \Cref{f:magsize}: both extremes are very rare, while the purple set corresponds to the most common (mag,size) configuration. Data points with errorbars show measured tangential shear and solid lines show Gaussian fits to each set of simulations. Empty points are biased because they are adjacent the chosen truncation radius of the lenses, and are excluded from the fits.}
\label{f:ct_sims}
\end{figure}

\begin{figure}
 \centerline{\includegraphics[width=3.3in]{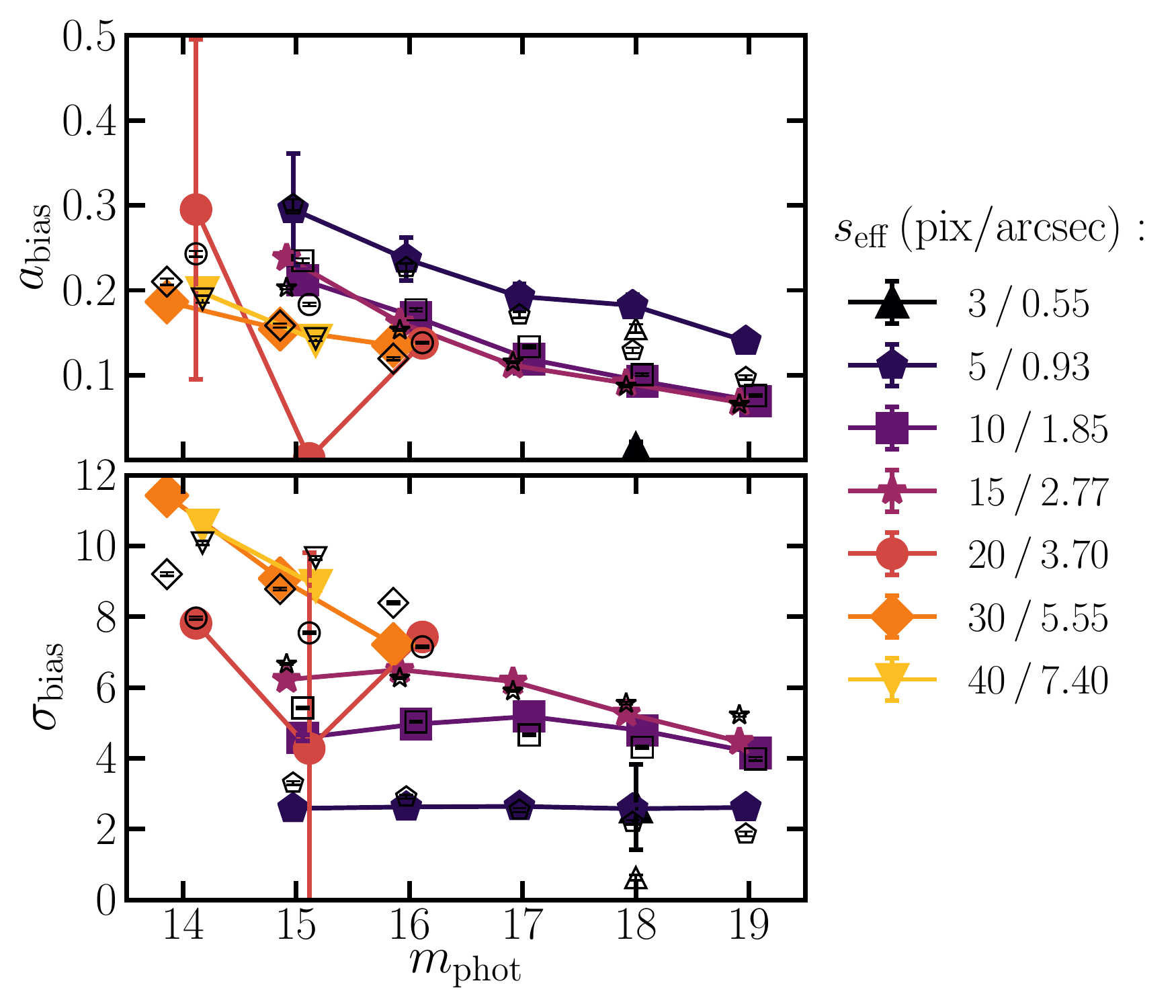}}
\caption{Amplitude and width of Gaussian fits to the additive bias $\ct$ (solid symbols), and the results from an overall fit to each panel given in \Cref{eq:ct_params} (empty symbols). Solid lines connect simulation sets with the same half-light radius as shown in the legend.}
\label{f:ct_params}
\end{figure}

Extended light from bright lens galaxies affects measurements of sources, such that their shapes are estimated to be more radially elongated than they really are. This induces a negative additive bias \emph{in the coordinate frame of the lens galaxy}, which we label $\ct$.

In order to account for this bias we measure the shapes of galaxies in the image simulations of \cite{hoekstra15}, after adding bright lens galaxies in a grid pattern (separated by 1 arcmin from each other). These injected lenses are modelled as a circular \cite{sersic68} profile (i.e., $I(r) \propto r^{1/n}$) using \textsc{galsim}, with a power-law index $n=4$. A S\'{e}rsic profile with a high index has very extended wings and to avoid the surface brightness profiles of different lenses to overlap we truncate the lens profiles at 5 $s_\mathrm{eff}$, where $s_\mathrm{eff}$ is the effective, or half-light, radius of the S\'{e}rsic profile. The source galaxies in the image simulations have a constant shear applied to them, which cancels out when we average over an isotropic grid of shears. Therefore any measured shear in the tangential frame can be attributed to a bias induced by extended light from the lenses. The lenses we inject into the simulations span the ranges $14 \leq \rmag \leq 20$ and $3 \leq \size/{\rm pix} \leq 40$ (corresponding to $0.\!''55\leq\size\leq7.\!''40$), and are compared to the magnitude and size distribution \citep[as measured by \galfit, see][]{sifon15_cccp} in the \meneacs\ data in \Cref{f:magsize}.

We show the measured $\ct$ for three sample sets of simulations in \Cref{f:ct_sims}. We find that the bias profiles can be well modelled in each bin as a Gaussian centred at $\thetals=0$,
\begin{equation}\label{eq:ct_profile}
 \ct(\thetals) = 
      a_\mathrm{bias} \exp\left[\frac{-\theta^2}{2\sigma_\mathrm{bias}^2}\right].
\end{equation}
We then fit the best-fit parameters $a_\mathrm{bias}$ and $\sigma_\mathrm{bias}$ as functions of lens magnitude and size,
\begin{equation}\label{eq:ct_params}
 \begin{split}
 a_\mathrm{bias} &= -0.15 - 0.023(\rmag-16) - 0.066\log(\size/15\,\mathrm{pix}), \\
 \sigma_\mathrm{bias} &= 6.27 -14.01\log(\rmag/16) + 7.04\log(\size/15\,\mathrm{pix}).
 \end{split}
\end{equation}

\Cref{f:ct_params} shows the best-fit individual values of $a_\mathrm{bias}$ and $\sigma_\mathrm{bias}$ and the values predicted by \Cref{eq:ct_params}. While at face value \Cref{eq:ct_params} is not a good description of the measurements in the simulations for the full ($\rmag$,$\size$) space (and especially for $\sigma_\mathrm{bias}$), the discrepancy is limited to the extremes of this space. One notable discrepancy is roughly a 25\percent\ difference in the prediction of $\sigma_\mathrm{bias}$ for $(\rmag,\size)=(14,30)$ (here, sizes are given in pixels). However, as shown in \Cref{f:magsize}, this combination of magnitude and size accounts for much less than 1\percent\ of the lenses in our sample. The other notable difference happens at $(\rmag,\size)=(18,3)$, but the bias introduced by such small, faint galaxies is negligible to start with. Moreover, as can be seen in \Cref{f:ct_params}, the difference arises because of the degeneracy between the amplitude and width of the Gaussian, such that the predicted bias is negligible as well.

\section{Boost correction on the satellite lensing signal}
\label{ap:boost}

\begin{figure*}
  \centerline{\includegraphics[height=1.4in]{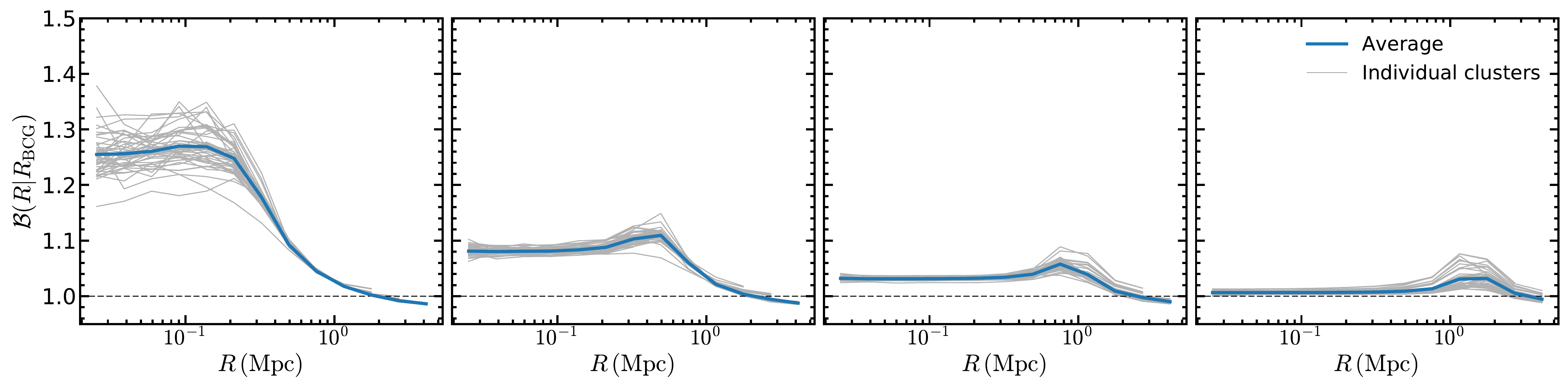}}
  \centerline{\includegraphics[height=1.4in]{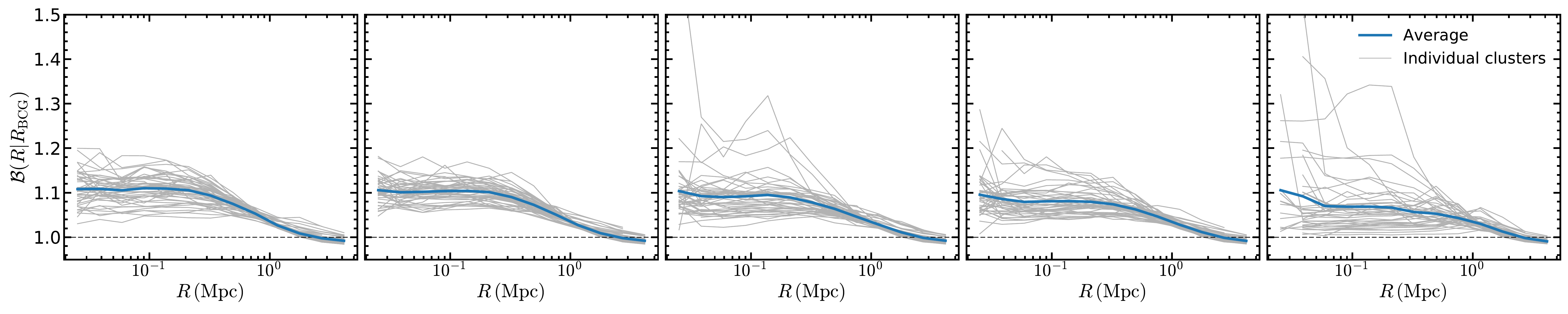}}
\caption{Boost corrections applied to the satellite lensing measurements binned by satellite cluster-centric distance, $R_\mathrm{sat}$ (top), and by stellar mass. Each grey line shows the correction for a single cluster; these are the corrections applied in our analysis. We also show for reference the mean correction in each of the bins. The dashed horizontal line shows a boost correction of 1 (i.e., no bias). Note the increase in $\mathcal{B}$ at the typical $R_\mathrm{sat}$ in each bin in the top plot, because these data points include sources approaching the cluster center. When binning by stellar mass, on the other hand, the profile is much smoother.
}
\label{f:boost_avg_clusters}
\end{figure*}

\begin{figure*}
  \centerline{\includegraphics[width=\linewidth]{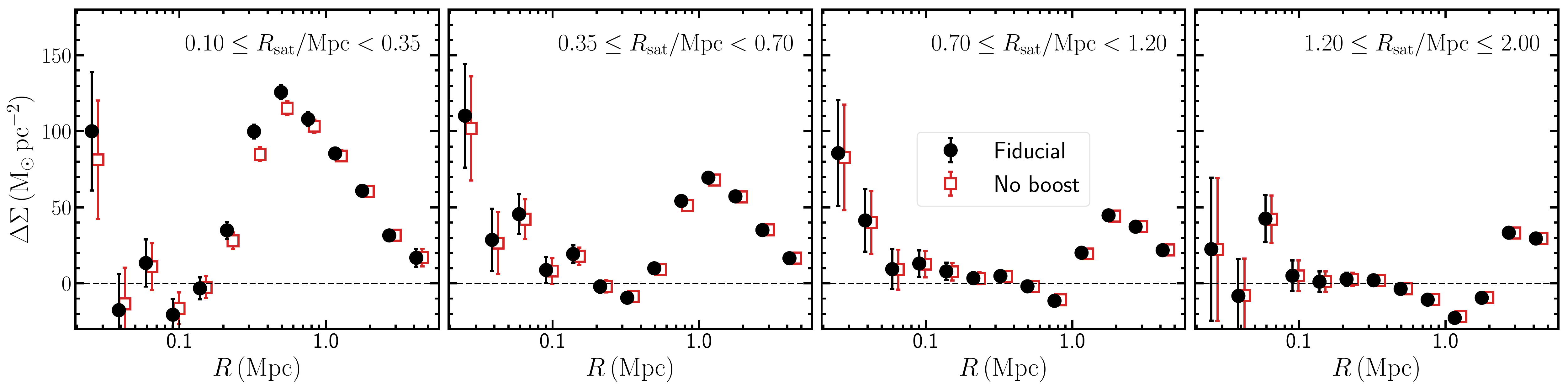}}
\caption{Satellite lensing signal before (red squares) and after (black circles) applying the boost correction, when the sample is binned by cluster-centric distance. The conversion from the red to the black points is shown by the blue thick lines in the top panels of \Cref{f:boost_avg_clusters}.}
\label{f:boost_effect}
\end{figure*}

In this section we show explicitly the impact of lenses entering our source sample---the boost correction---on our satellite lensing measurements. In \Cref{s:calibration} we discussed at length the methodology used to measure the biases introduced by cluster members, and presented a fitting function which depends on $R_\mathrm{BCG}$, since the cluster-centric distance is the factor determining the amount of contamination by cluster members. We use the same fitting function to calculate the boost correction for all clusters. However, the satellite lensing signal is not calculated as a function of cluster-centric distance but as a function of satellite-source separation, $R$, and therefore the boost correction applied to the satellite lensing signal varies from cluster to cluster, as it depends on the radial distribution of lenses within each cluster. \Cref{f:boost_avg_clusters} shows the boost factors that were applied to our measurements to each cluster, for our measurements in both \Cref{s:shsmr,s:segregation}.

The boost correction for each $\Rsat$ bin is constant for $R<\langle\Rsat\rangle$ and then falls rather rapidly to a value of 1 (i.e., no correction). The bump seen in all except the first panel around the average $\Rsat$ is analogous to the negative value of $\Delta\Sigma$ at the same lens-source separations. The boost correction in different stellar mass bins, on the other hand, is averaged over a large range in $\Rsat$, and therefore is smooth and rarely exceeds 20\percent, except for high-stellar mass objects in a few clusters.

Note that at large lens-source separations, $R>2$ Mpc, some boost corrections have values less than unity. This is, however, a 1\percent\ effect, and we make no attempts to correct for it. Furthermore, these lens-source separations have no influence on the subhalo masses and in fact we excluded them from our analysis in \Cref{s:shsmr,s:segregation}. In \Cref{f:boost_effect} we show the effect of the boost correction in our measurements by directly comparing the lensing signal before and after applying the boost factors, when binning by cluster-centric distance. The almost constant $\sim$10\percent\ boost correction for all stellar mass bins makes the data before and after the boost correction essentially indistinguishable, especially given the statistical uncertainties, and we therefore do not show the signal binned by stellar mass.

\section{Mass definition and density profile}
\label{ap:models}

As discussed in the main text, other studies of satellite lensing have made different assumptions about subhalo density profiles, as well as used different definitions of mass. For instance, \cite{li16,niemiec17} adopted different versions of a truncated NFW profile, while \cite{sifon15_kids} assumed a full NFW profile and quoted $m_{200}$ as the subhalo mass, after arguing that statistical uncertainties dominated uncertainties from the mass profile. In our main analysis we adopted yet another mass definition, based on a full NFW profile but only including the mass within the region where the subhalo is overdense compared to the parent cluster, which we denoted $m_\mathrm{bg}$. Furthermore, each of these works adopted different mass-concentration relations \citep[or, in the case of][marginalized over $r_\mathrm{s}$]{li16}, which may also impact the results and preclude a meaningful comparison.

In this section we show posterior mass estimates based on different models for the subhalo density profile, for the case of satellites binned by cluster-centric distance. In addition to our fiducial model described in \Cref{s:subhalo}, we implement two models which we briefly describe below.

Our first alternative model consists of a full NFW density profile but with a concentration that depends on cluster-centric distance through
\begin{equation}\label{eq:cRsat}
  c(\Rsat) = A \Rsat^B\,,
\end{equation}
but both $A$ and $B$ remain largely unconstrained by our data. This is a reflection of the fact that weak lensing alone simply cannot constrain the concentration because the signal cannot be measured at radii below the scale radius in ground-based data (see \Cref{s:obscuration}). This model is therefore effectively a model with a full NFW profile where the concentration of subhaloes is marginalized out.

Our second alternative model consists of a truncated NFW profile which falls of as $\rho \propto r^7$ beyond the truncation radius,
\begin{equation}\label{eq:tnfw}
  \rho(r) = \frac{\delta_\mathrm{c}\rho_\mathrm{m}}
                 {r/r_\mathrm{s}(1+r/r_\mathrm{s})^2}
      \left(\frac{r_\mathrm{t}}{r^2+r_\mathrm{t}^2}\right)^2\,,
\end{equation}
with both the scale radius, $r_\mathrm{s}$, and the trunaction radius, $r_\mathrm{t}$, in addition to subhalo mass, left as free parameters for each $\Rsat$ bin. This model is identical to the truncated NFW model used by \cite{li16} and therefore results from this model can be directly compared to those of \cite{li16}. Analytic expressions for the excess surface density of \Cref{eq:tnfw} were derived by \cite{baltz09}. Like \cite{li16}, we cannot constrain neither the scale nor truncation radii.

\begin{figure}
  \centerline{\includegraphics[width=\linewidth]{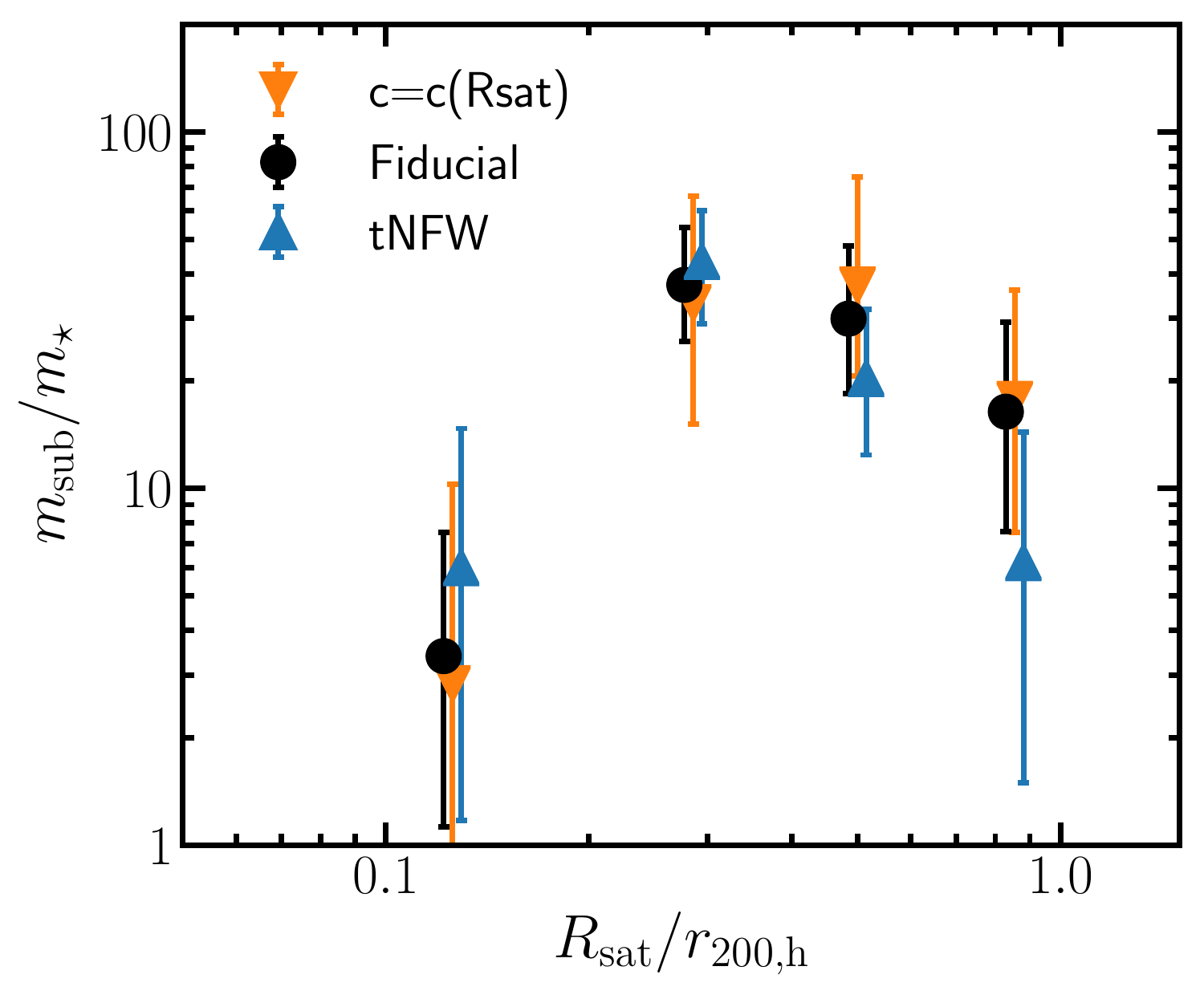}}
\caption{Posterior ratios of total-to-stellar masses as a function of cluster-centric distance for the two alternative models described in \Cref{ap:models}, compared to our fiducial model. Points have been slightly shifted horizontally for visual clarity.}
\label{f:masses_alternative}
\end{figure}

We show the posterior mass estimates of these models, along with our fiducial model from \Cref{s:subhalo}, in \Cref{f:masses_alternative}. The results of these two alternative models are fully consistent with our fiducial analysis; the increased errorbars are a reflection of the additional free parameters in these models. Like previous work \citep[e.g.,][]{li16}, we conclude that systematic uncertainties due to the choice of density profile are subdominant to statistical uncertainties.

\end{document}